\documentclass[fleqn,12pt,a4paper,
]{article}
\usepackage[a4paper, 
top=1in, left=1in, right=1in, bottom=1in]{geometry}

\usepackage{amsthm}
\usepackage{amsmath}
\usepackage{amssymb}
\usepackage{amsopn}
\usepackage{eurosym}
\usepackage{latexsym}
\usepackage{setspace}
\usepackage{dsfont}
\usepackage{datetime}
\usepackage[allnumber,warning,easyold,math]{easyeqn}
\usepackage[12pt]{moresize}
\usepackage{natbib}
\usepackage[bookmarks=true,bookmarksnumbered=true,
pdfpagemode=None,pdfstartview=FitH,hidelinks]{hyperref}
\usepackage{fix-cm}
\usepackage[calcwidth]{titlesec}
\usepackage{graphicx}
\usepackage[utf8]{inputenc}
\usepackage[english]{babel}
\usepackage{amssymb}
\usepackage{braket}
\usepackage{upgreek}
\usepackage[font=footnotesize,skip=3pt]{caption}
\usepackage[font=footnotesize]{subcaption}
\usepackage{float}
\usepackage{multirow}
\usepackage{soul}
\usepackage{lscape}
\usepackage{nicematrix}
\usepackage{fixltx2e}
\usepackage{tikz}
\usetikzlibrary{arrows, decorations.markings}
\usetikzlibrary{arrows.meta, positioning, shadows, shapes.symbols}
\usetikzlibrary{shapes, shadows, arrows}
\usetikzlibrary{positioning}
\usepackage{circuitikz}
\usepackage[colorinlistoftodos]{todonotes}
\usepackage{longtable}
\usepackage{svg,svg-extract,svgcolor}

\newtheorem{proposition}{Proposition}

\newtheorem{highlight}{Highlight}

\newcommand{\Prests}{\raisebox{-0.7pt}{%
\includegraphics[height=0.35cm]{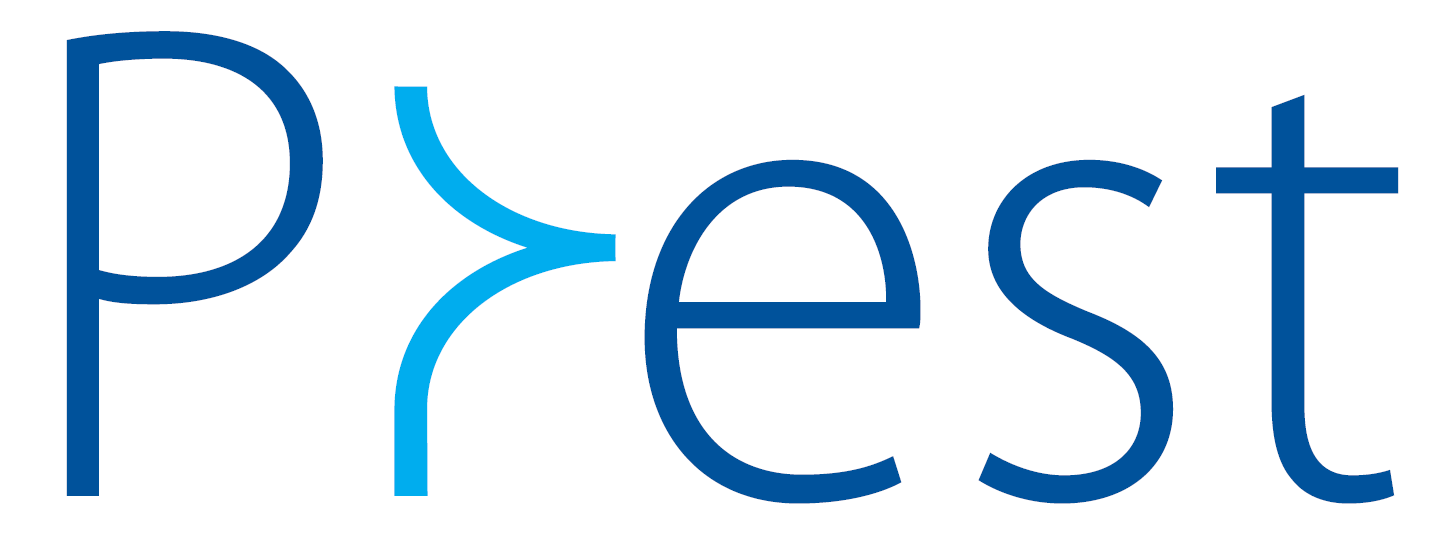}\;}%
}

\newcommand{\PrestsFootnote}{\raisebox{-0.7pt}{%
\includegraphics[height=0.30cm]{images/prest.png}\;}%
}

\NiceMatrixOptions
{
custom-line =
{
letter = - ,
command = dashedline ,
tikz = dashed
}
}

\titleformat{\section}[hang]{\rmfamily\bfseries}{\color{black}\fontsize{14}{14}
\selectfont\thesection}{30pt}{\fontsize{14}{14}\selectfont}
\titleformat{\subsection}[hang]{\rmfamily\bfseries}{\color{black}\fontsize{13}{13}
\selectfont\thesubsection}{30pt}{\fontsize{13}{13}\selectfont}
\titleformat{\subsubsection}[hang]{\rmfamily\bfseries}{\color{black}\fontsize{12}{12}
\selectfont\thesubsubsection}{30pt}{\fontsize{12}{12}\selectfont}

\graphicspath{{./images/}}

\begin{document}

\parskip=2pt

\author{\vspace{-10pt} \normalsize
Thomas Dohmen\thanks{tdohmen@uni-bonn.de} \\
{\scriptsize University of Bonn}
\and \vspace{-10pt} \normalsize
Georgios Gerasimou\thanks{georgios.gerasimou@glasgow.ac.uk} \\
{\scriptsize University of Glasgow}}

\title{\Large \textbf{Convergence to 
Utility Maximization\linebreak and the Indifference 
Hypothesis}\thanks{
We thank
Carlos Al\'{o}s-Ferrer, Junaid Arshad, Andrew Caplin,
David Gill, Ido Erev, John List, Antonio Penta,
Jakub Steiner, Jianying Qiu and audiences at
Manchester Economic Theory Conference 2025,
Frontiers of Measurement \& Formation of Skills (Jinan University, 2025),
ESA 2024 (Ohio State), Psychonomics 2024 (Lancaster),
EEA-ESEM 2024 (Rotterdam), BSE Bounded Rationality 2024 (Barcelona),
SAET 2023 (Paris), Sydney,
Nottingham, Radboud/Nijmegen, Glasgow and St Andrews
for useful comments. Funding by
the Deutsche Forschungsgemeinschaft
(DFG, German Research Foundation) through CRC TR 224 (Project
A05) and Germany's Excellence Strategy - EXC 2126/1-390838866
as well as through NSF Award 2201888 is gratefully acknowledged.
The experiment was approved by the University of St Andrews Research Ethics
Committee (EC14020).
Max Jahn provided excellent research assistance with computations.
Any errors are our own.}}

\parindent=0pt

\date{\footnotesize \today}

\maketitle

\vspace{-30pt}

\begin{abstract}
\noindent We ask if participants in a choice experiment with repeated presentation
of the same menus and no feedback provision:
(i) exhibit overall behaviour that is consistent with ordinal and expected utility
theory under \textit{weak} preferences; 
(ii) become more consistent with the predictions of
these theories under \textit{strict} preferences.
To answer these questions we designed and implemented
a free-choice lab experiment with 15 distinct menus. Each menu
contained two, three and four lotteries with three monetary outcomes,
and was shown five times.
Subjects could avoid/defer making active choices at a positive expected cost.
Among our 308 subjects from the UK and Germany, significantly more
were compatible with ordinal and expected utility
maximization in their last 15 than in their
first 15 identical decision problems.
Around a quarter and a fifth of all subjects, moreover,
were compatible with those modes \textit{throughout} the experiment,
with nearly half of these under weak preferences.
Choice consistency is positively correlated with cognitive ability,
and  subjects who converged to utility-maximizing behaviour were
more cognitively able than those who did not.
We discuss potential implications of our study's
novel set of findings.

\vspace{10pt}

\end{abstract}

\setcounter{page}{0}

\thispagestyle{empty}

\vfill

\pagebreak

\parindent=15pt


\section{Introduction}\label{s1}

A large body of experimental work in economics and psychology shows
that individuals often make different choices under risk when
confronted with the same decision problems repeatedly.
Such patterns are often interpreted as evidence
against the hypothesis of maximization of stable, complete and transitive preferences
that constitutes the rationality cornerstone of neoclassical economic analysis.
Six commonly invoked explanations for the occurrence of such patterns include
the possibilities of: (i) ``noisy'' utility maximization; (ii)
systematically bounded-rational behaviour with stable or context-dependent preferences;
(iii) imprecise/incomplete preferences;
(iv) utility maximization with costly information acquisition;
(v) limited attention; and (vi) deliberate randomization.\footnote{See, for example,
(i) \cite{gillen-snowberg-yariv19,apesteguia-ballester18,harless-camerer94,hey-orme94};
(ii) \cite{cerreia-vioglio-dillenberger-ortoleva15,bordalo-gennaioli-shleifer12};
(iii) \cite{cubitt&navarro-martinez&starmer15};
(iv) \cite{dean-neligh23};
(v) \cite{barseghyan-molinari-thirkettle21};
(vi) \cite{machina85,cerreia-vioglio-dillenberger-ortoleva-riella19,agranov&ortoleva};
and references therein.}
In this study, we focus on the potential role of two basic, complementary
explanations by asking the following questions, which have received less attention
in the literature:
\begin{enumerate}
	\item Could some of the observed `volatility' in behaviour
	across different presentations of the same menus
	be attributable to subjects' \textit{rational indifferences}
	between the relevant alternatives?
\end{enumerate}
This intuitive possibility is in line with conventional
economic interpretations of the concept of indifference.
In one of the early studies that presented results
from an experiment on stochastic choice, for example,
\citet[p. 233]{davidson-marschak59}	remarked that
\textit{``One may interpret every case of inconsistency
	as a case of indifference: if the subject has chosen
	`$a$' rather than `$b$' but soon afterwards chooses `$b$'
	rather than `$a$', this is interpreted as indifference between those
	two objects; if he chooses `$a$' rather than `$b$', `$b$'
	rather than `$c$', and `$c$' rather than `$a$', this is interpreted 
    as indifference between those three objects.''}
The authors concluded with their warning that
\textit{``In empirical application, this approach would
	probably make indifference all-pervasive.''}
However, there appears to be no systematic attempt
that examines whether rational choice with
\textit{weak} preferences can account for some of the observed---and 
seemingly non-rational---choice reversals.
Indeed, it is unclear to what extent
Davidson and Marschak's (\citeyear{davidson-marschak59})
warning is justified once subjects' indifferences and strict preferences
are recovered from such data in a theory-guided way, using
modern computational tools.\footnote{For a quick
	preview of this task's computational demands in our environment
	we note that there are $7!=5,040$ strict orders
	and 47,293 weak orders over a set of 7 choice alternatives
	(\href{https://oeis.org/A000670}{\color{blue}OEIS: A000670}).}
\begin{enumerate}
\item[2.] Do subjects become more rational over the course of the experiment
\textit{without} receiving any feedback, interventions or
other opportunities to acquire new information,
and without being forced to always make an active choice at each menu?
\end{enumerate}
This question is important for at least two reasons.
First, if such convergence does occur in experiments featuring repeated presentation
of the same decision problems, without subjects
receiving any new information, and in a free-choice environment where
avoiding/deferring is also possible, then the targeted design and use of such experiments
should be promoted further for more accurate theoretical tests and preference recovery.
Second, under this convergence hypothesis an analyst could be justified in focusing
on subjects' behaviour at the later stages of the experiment and analyse
it through the lens of ordinal or expected utility theory.
With the exception of a few studies that we discuss in Section \ref{s7}---which
were considerably more limited in their scope and sample sizes---here too 
it appears to be no systematic attempt to answer this question.

The decision environment in our experiment that aimed to shed light on these 
questions was structured around seven money lotteries and fifteen menus derived from them.
Each lottery assigned a positive probability to three prizes.
The fifteen menus included nine binary, four ternary and two quaternary ones.
In particular, contrary to the vast majority of experimental
studies on choice under risk, which focus only on behaviour at binary menus
and implicitly assume subjects' compliance with ordinal utility maximization
at \textit{all} menus, this design allows for testing
various implications of both ordinal and expected utility
theory \textit{simultaneously}.

Specifically, we aimed to assess, among others, the Transitivity,
Contraction Consistency and Weak Axiom of
Revealed Preference implications of ordinal utility theory,
as well as the Independence and First-Order-Stochastic Dominance
implications of expected utility theory.
Our test of the Independence axiom is new in that,
although invoking Allais-style
(Allais, \citeyear{allais}; Kahneman and Tversky, \citeyear{kahneman-tverksy79})
lottery mixtures, neither the prizes nor the probabilities are as extreme as
in the original Allais paradox or its ``common-ratio'', ``common-consequence''
and ``certainty-effect'' variations and decompositions that have been studied
extensively in the literature.\footnote{See \cite{mcgranaghanetal2024}
and references therein.} Instead, in each of the two binary menus that were
designed to test Independence in our study, the two lotteries within each of
these menus had the same expected value and were unrelated by Second-Order Stochastic
Dominance---hence presented relatively more difficult decisions.
In addition, the expected value in these two pairs differed by a factor of two
(12 and 6 Pounds Sterling/Euros, respectively).
Thus, our test of Independence differs from standard ones in that it
aims to shed light on humans' conformity with this axiom when the two
underlying decisions are not as intuitive as in those situations where
it is well-known to be violated often.

In addition to these tests, we formally introduce and empirically assess
the \textit{Stability of Attitudes to Risk} (StAR) implication that
expected utility theory has in our environment.
As we show, in our environment this amounts to a subject always choosing either 
the dominant lottery (risk-averse agent) or dominated one (risk-seeking agent) 
in each of our three distinct binary menus where a Second-Order Stochastic Dominance 
relation exists.

Importantly, moreover, incorporating a free-/non-forced-choice approach into our
experimental design allows us to also investigate the Decisiveness
implication of rational choice theory, according
to which the decision maker always chooses
one of the available lotteries, in line with the Completeness/Comparability
axiom on preferences under utility theory.
Implicitly, Decisiveness also assumes that decision makers
are not subject to fatigue
or cognitive overload at any decision problem.
An essential aspect of our design is that it grants subjects 
the freedom to avoid or delay, at a small expected cost,
making an active choice at menus where, for any reason,
they felt uncomfortable to do so.
This approach allows us to test all of the above-mentioned
distinct six implications of the theory
while eliminating the confounds arising from the interactions
between standard forced-choice experimental designs and
the potential incompleteness of subject's preferences or reluctance
to engage with the decision problem at hand.\footnote{As far as testing
Transitivity is concerned, early warnings to that effect appeared in
\cite{luce&raiffa} and, citing these authors,
\cite{aumann_1962}. Motivated by these, and also by
experimental findings in psychology suggesting
that hard decisions lead to choice paralysis
\citep{tversky_shafir92,iyengar_lepper00,dhar97,dhar-simonson03},
\cite{gerasimou18} proposed models of fully consistent
active choices in general non-forced decision environments.
Some predictions of one of these models were subsequently tested
experimentally in \cite{CCGT22}, \cite{gerasimou21} and,
less directly, \cite{nielsen-rigotti22}.
These three studies share the finding that free/non-forced choices
are significantly more consistent than
forced-choice ones, in line with Luce and Raiffa's
(\citeyear{luce&raiffa}) intuition that \textit{``intransitivities
often occur when a subject forces choices between inherently
incomparable alternatives''}.
Our study does not feature a forced-choice treatment.
Instead, it allows testing, for
the first time, a rich set of specific implications of expected utility theory
without forcing subjects to always make active choices,
thereby extending the crux
of Luce and Raiffa's insights to that domain also.}

Deploying new computational tools on data collected from 308 subjects 
in the UK and in Germany, we find that:
\begin{enumerate}
\item About a quarter and a fifth of all subjects
consistently exhibited
rational behaviour in these two respects
\textit{throughout} their 75 decisions, with about half of them revealing
at least one indifference between distinct lotteries.
\item Nearly twice as many subjects conformed with ordinal- (57.5\%)
and expected-utility (40\%)
maximization with strict preferences at the end of the experiment
than at the beginning, accompanied by significant reductions
in decision times.
\item The number of subjects violating either one or all of Transitivity,
Contraction Consistency, Weak Axiom of Revealed Preference,
Independence, First-Order Stochastic Dominance and Stability
of Attitudes to Risk decreases steadily over the course of the experiment,
and significantly so between the first and last round.
\item Deferring/avoiding behaviour is generally infrequent
and stable throughout, mainly occurring
at menus with increased decision difficulty,
i.e., those without a stochastically dominant lottery
and/or where the feasible lotteries are relatively complicated.
\item Cognitive ability, particularly in verbal reasoning,
is positively correlated with choice consistency
and with ``early-onset'' rationality.
\item Subjects who deviated from rationality initially
but complied with it by the end
of the experiment were significantly more
cognitive able than those who did not.
\end{enumerate}

\noindent These findings have implications for testing
existing theories of choice under risk and interpreting the results of
such tests, as well as for experimental design, preference elicitation,
the development of new theories, and for efforts to improve real-world
decision-making under risk.
We discuss these implications in the last section. 

Overall, we view this paper's contribution as documenting and discussing the relevance
of these novel empirical findings as the outcome
of an analysis that features
the following methodological innovations which, we hope, add meaningfully
to the currently available toolkit in the literatures of choice under risk,
revealed preference analysis, behavioural economics, as well as the emerging
field of cognitive economics \citep{caplin25}:
\begin{enumerate}
\item[i.] Decisions made from a collection of 15 binary and
non-binary menus from 7 lotteries with 3 outcomes, whose construction
was theory-motivated and guided by the presence
or absence of first/second stochastic dominance relations. 
This allows for targeted systematic tests of both ordinal- and expected-utility
maximization in a richer environment than those
typically seen in experiments on choice under risk. To our knowledge, this is
the first such design that presents repeated choices from both binary
\textit{and} non-binary menus of lotteries.
This is particularly important because it provides the basis
for a very general test of these theories and allows for
a detailed assessment of the ``extra rationality steps''
that separate conformity with ordinal versus expected-utility maximization.
\item[ii.] The design of a free-choice experiment
that combines repetition of the distinct decision problems with
allowing subjects to avoid/defer making active choices
at a small expected cost, and with the non-provision of any kind of feedback.
\item[iii.] A carefully administered presentation of the 5 repetition blocks
for each subject, which combines homogeneity within and across subjects
in the first and last rounds with complete randomness---also within
and across subjects---in the middle three rounds, motivated
by the specific goal of testing whether feedback-less learning occurs.
\item[iv.] The analyses of deterministic ordinal and
expected-utility maximization through the lens of
both strict and---new with this paper---\textit{weak}
preferences, and the
recovery by means of recently developed combinatorial-optimization techniques
of the best-matching such relations for each subject.

\end{enumerate}

The next section introduces the relevant theoretical framework
and then provides a detailed description of the experimental design
and its relation to that framework.
Our empirical results are presented in Sections \ref{s4}-\ref{s6}.
The penultimate section discusses the related literature
and how this paper is placed in it.
The final section discusses implications of our findings.
Additional information about the experiment, including screenshots,
is provided in Online Appendices (O.A.) \ref{a3}--\ref{a5}.

\section{Design of the Experiment}\label{s3}

Our experimental data can be analysed
from the point of view of (expected-)utility-maximizing behaviour
with strict preferences that are revealed by single-valued choices,
either per individual decision round or overall/across rounds.
Yet they can also be investigated for such behaviour under
possibly weak preferences that are revealed by subjects' constructed
\textit{multi-valued} choices.
Each of these two modes of analysis has well-known behavioural implications,
many of which our experiment was specifically designed to test.
We proceed with reviewing those before turning to our experiment's design.

\subsection{Theoretical Background}\label{s2-1}

We consider a general choice domain of finitely many menus
containing lotteries that are defined over a finite set of
monetary outcomes $Z\subset \mathbb{R}_+$.
Denoting by $X$ the finite set of lotteries over $Z$ that we consider,
and denoting by $\mathcal{M}$ the collection of such menus,
a decision maker's behaviour is described by a choice correspondence
$C:\mathcal{M}\twoheadrightarrow X$,
i.e., a mapping that satisfies $C(A)\subseteq A$ for every menu $A$ in $\mathcal{M}$.
Unlike a single-valued choice \textit{function},
the value of a choice correspondence at menu
$A$ could contain multiple alternatives, which are typically
interpreted as those that the decision maker
\textit{might} choose from $A$.\footnote{See, for example,
	Chapter 1 in \cite{mwg} or Chapter 1 in \cite{kreps12}.}
Clearly, since the empty set is a subset of every
set, this basic definition allows for the possibility of $C(A)=\emptyset$.
Because the ultimate decision outcomes in our free-/non-forced-choice
experimental design are monetary amounts---with
higher clearly preferred to lower, and all non-zero amounts being
desirable---the potential unattractiveness
of the items in $A$ is not a likely explanation for
observing $C(A)=\emptyset$ in our setting.
Hence, in our environment, this notation will be used to represent
situations where the decision maker
opts to avoid or delay choice at $A$ because they find
it difficult to make an active choice at that menu.

The four basic choice axioms on the observable values of $C$
listed below are implied
by every deterministic model of utility maximization over lotteries over $X$
and not just those that belong to the expected-utility class
of \cite{neumann_morgenstern}.

\vspace{3pt}

\noindent \textbf{Decisiveness}\\
\textit{$C(A)\neq\emptyset$ for every menu $A$.}

\vspace{3pt}

\noindent \textbf{Transitivity}\\
\textit{If $p\in C(\{p,q\})$ and $q\in C(\{q,r\})$, then $p\in C(\{p,r\})$.}

\vspace{3pt}

\noindent \textbf{Contraction Consistency / Independence of Irrelevant Alternatives}\\
\textit{If $p\in C(A)$ and $p\in B\subset A$, then $p\in C(B)$.}

\vspace{3pt}

\noindent \textbf{Weak Axiom of Revealed Preference (WARP)}\\
\textit{If $p\in C(A)$, $q\in A\setminus C(A)$ and $q\in C(B)$, then $p\not\in B$.}

\vspace{3pt}

\noindent \textit{Decisiveness} is typically assumed
in most choice-theoretic analyses as part of the definition
of a choice correspondence. As was pointed out above, however,
it is in fact an additional restriction
that has behavioural meaning.
Relaxing Decisiveness when the analyst suspects that decision makers
may avoid/delay making an active choice because
of decision difficulty is potentially fruitful theoretically
\citep{hurwicz86,kreps90,kreps12,gerasimou18} and relevant empirically
\citep{tversky_shafir92,iyengar_lepper00,dhar97,CCGT22}.
We stress that, because our environment involves non-forced choices
where Decisiveness is not a priori assumed to hold,
in addition to ruling out \textit{cyclic} preferences,
Transitivity also rules out acyclic
but still non-transitive preferences.
For example, $x\in C(\{x,y\})$, $y\in C(\{y,z\})$,
$\emptyset=C(\{x,z\})$ reveal acyclic but intransitive and incomplete preferences,
where $x\succsim y\succsim z$ and $x\not\succsim z\not\succsim x$.
Hence, testing Transitivity with choices that arise
from such a free-/non-forced choice
decision environment amounts to testing for transitive preferences in the absence
of any potential confounds that the exogenous imposition
of Decisiveness may not be able to account for
\citep{luce&raiffa,aumann_1962}.

WARP is a fundamental rationality property, requiring that there be no
direct choice reversals between any two lotteries.
\textit{Contraction Consistency},
also known as \textit{Independence of Irrelevant Alternatives},
the \textit{Chernoff axiom}, and
\textit{Property} $\alpha$ \citep{sen97},
is implied by WARP under Decisiveness but not in general;
for example, $x\in B\subset A$, $x\in C(A)$, $C(B)=\emptyset$
violates this axiom but not WARP. In the baseline case where
$C$ is always non-empty-valued,
this axiom rules out a large class of context-dependent choice reversals
that are driven by the presence or absence of
irrelevant alternatives. Specifically, it requires that when an alternative
is declared choosable at a menu, removing other alternatives
from that menu should not alter this status. That is, the absence of those
``irrelevant'' alternatives in the smaller menu should not
make the agent choose something else or, in our more general environment,
avoid/delay choice.

The next two revealed-preference axioms are relevant for finite
choice datasets that are either obtained from general environments
of choice under risk or from more specific environments of choice
over money lotteries.

\vspace{3pt}

\noindent \textbf{Independence}\\
\textit{For any $p,q,r$ and $\alpha\in(0,1)$,}\\
\centerline{$p\in C(\{p,q\})$ $\Longrightarrow$
	$\alpha p+(1-\alpha)r\in C(\{\alpha p +(1-\alpha)r,\alpha q + (1-\alpha)r\})$.}
	
\vspace{3pt}

\noindent \textbf{First-Order Stochastic Dominance (FOSD)}\\
\textit{If $p$ \textit{FOSD} $q$,
	then $C(\{p,q\})=\{p\}$.}

\vspace{3pt}

\noindent
Both these axioms are well-known
and extensively studied implications of
expected utility theory.
Yet the remarks made earlier
about the generality of testing
Transitivity in our free-choice environment also carry
over to Independence and FOSD.
Namely, by not imposing forced-choice ex ante, we are allowing
for testing each of these axioms independently of Completeness.

\subsection{Lotteries and Choice Menus}\label{s3-1}

We constructed 7 lotteries (Table \ref{tab:lotteries};
Figure \ref{fig:lotteries} in O.A.\ref{a3}),
each with three monetary outcomes
from the set $\{0,9,10,20,24\}$,
where the numbers denote either Pounds Sterling, \pounds,
or Euros, {\small \euro}.
Out of the 127 possible menus that are derivable
from this grand choice set we selected
15 that contained either two lotteries (9 menus),
three lotteries (4 menus) or four lotteries (2 menus)
(Table \ref{tab:menus}; Figure \ref{fig:menus} in O.A.\ref{a3}).
All menus were presented 5 times, resulting in a total of 75 decision problems.
This decision environment is therefore richer than those
usually seen in experiments on choice under risk where subjects
are presented with only binary menus, often over
two monetary outcomes.\footnote{That said, we note that
\cite{apesteguia-ballester21} also analysed an experimental dataset
on choice under risk where some non-binary
menus were shown. Unlike our experiment where each menu was presented five times,
subjects in that experiment saw each menu once.}
This domain richness is relevant because it allows testing an
analogously rich set of consistency principles that pertain both to
binary and larger menus.

In each decision, subjects could decide to avoid/defer
making an active choice by opting for the option \textit{``I'm not choosing now''}
(it is in this sense that choices were not forced). 
If a choice at a menu was deferred, and if that menu was randomly drawn for payment
at the end of the experiment, subjects would be asked to choose a lottery from 
this menu then.
We stress at this point that no new information about any of the
lotteries was given to subjects after the main part of the experiment.
In particular, opting for \textit{``I'm not choosing now''}
was not associated with any informational gains.

\begin{table}[!htbp]
\centering
\footnotesize
\caption{\centering The 7 lotteries.}
\setlength{\tabcolsep}{6pt} 
\renewcommand{\arraystretch}{1.5} 
\makebox[\textwidth][c]{
\begin{tabular}{|c|c|c|c|c|c|c|}
\hline
\multicolumn{1}{|r|}{\multirow{1.5}{18pt}{Prize}} 	
& \multirow{2}{20pt}{\centering {\scriptsize \euro/\pounds} \textbf{0}}
& \multirow{2}{20pt}{\centering {\scriptsize \euro/\pounds} \textbf{9}}
& \multirow{2}{25pt}{\centering {\scriptsize \euro/\pounds} \textbf{10}}
& \multirow{2}{25pt}{\centering {\scriptsize \euro/\pounds} \textbf{20}}
& \multirow{2}{25pt}{\centering {\scriptsize \euro/\pounds} \textbf{24}}
& \multirow{2}{45pt}{\centering \textbf{Expected}}\\
\multicolumn{1}{|c|}{\multirow{1}{*}{Lottery}} & & & & &
& \multirow{1.3}{*}{\centering \textbf{value}}\\
\hline
\textbf{A1} & $\frac{10}{100}$ & -- & $\frac{60}{100}$ & $\frac{30}{100}$ & --
& {\scriptsize \euro/\pounds} \textbf{12}\\
\hline
\textbf{A2} & $\frac{20}{100}$ & -- & $\frac{50}{100}$ & $\frac{30}{100}$ & --
& {\scriptsize \euro/\pounds} \textbf{11}\\
\hline
\textbf{B1} & $\frac{25}{100}$ & -- & $\frac{30}{100}$ & $\frac{45}{100}$ & --
& {\scriptsize \euro/\pounds} \textbf{12}\\
\hline
\textbf{B2} & $\frac{25}{100}$ & $\frac{40}{100}$ & -- & -- & $\frac{35}{100}$
& {\scriptsize \euro/\pounds} \textbf{12}\\
\hline
\textbf{C1} & $\frac{625}{1000}$ & -- & $\frac{150}{1000}$ & $\frac{225}{1000}$ & --
& {\scriptsize \euro/\pounds} \textbf{6}\\
\hline
\textbf{C2} & $\frac{625}{1000}$ & $\frac{200}{1000}$ & -- & -- & $\frac{175}{1000}$
& {\scriptsize \euro/\pounds} \textbf{6}\\
\hline
\textbf{D}  & $\frac{15}{100}$ & -- & $\frac{50}{100}$ & $\frac{35}{100}$ & --
& {\scriptsize \euro/\pounds} \textbf{12}\\
\hline
\end{tabular}
}
\label{tab:lotteries}
\end{table}

\begin{table}[!htbp]
\centering
\footnotesize
\caption{\centering The 15 lottery menus and some of the axioms
they were designed to test.}
\setlength{\tabcolsep}{6pt} 
\renewcommand{\arraystretch}{1.3} 
\makebox[\textwidth][c]{
\begin{tabular}{|c|cccc|l|l|}
\cline{1-7}
\textbf{Menu \#}
& \multicolumn{4}{c|}{\textbf{Lotteries in Menu}}
& \multicolumn{1}{c|}{\textbf{(Non-)Dominance structure}}
& \multicolumn{1}{c|}{\tiny \textbf{Additional remarks}}\\
\cline{1-6}
1& A1 & A2 &  &  & A1 $FOSD$ A2 & \\
\cline{1-6}
2&  B1 & B2 &  &  & No $SOSD$dominance
& \multirow{-1.5}{157pt}{\tiny \hspace{-4pt} Menus  2, 3
	jointly test \textbf{\textit{Independence}}:}\\
\cline{1-6}
3& C1 & C2&  &  & No $SOSD$; `hard' probabilities
& \multirow{-2}{157pt}{\tiny $Ci=\frac{1}{2} Bi + \frac{1}{2} (1,0,0,0,0)$,{ }
	$i=1,2$}\\
\cline{1-6}
4& B1 & D &  &  &  D $SOSD$ B1
& \multirow{2}{157pt}{\tiny \hspace{-4pt} Menus 2, 3, 5, 7, 13
	feature \textit{`hard decisions'} \vspace{-1pt} \\
	\hspace{-4pt} \& test \textit{\textbf{Decisiveness}}
	via the \textit{(in)Completeness} channel}\\
\cline{1-6}
5& B2 & D &  &  &  No $SOSD$ominance &
\multirow{3.75}{157pt}{\hspace{-1pt}\tiny Pairs {$\{4,6\}$, $\{4,9\}$, $\{6,9\}$
		test}} \\
\cline{1-6}
6& A1 & B1 &  &  &  A1 $SOSD$ B1 & \multirow{3}{157pt}{\hspace{-1pt}\tiny
\textbf{\textit{Stable Attitudes to Risk}} (Proposition \ref{prp:StAR})}\\
\cline{1-6}	
7& A1 & B2 &  &  &  No $SOSD$ominance & \\
\cline{1-6}
8& A2 & D &  &  &  D $FOSD$ A2
& \multirow{2.5}{155pt}{\tiny \hspace{-4pt}
	Triples $\{1,8,9\}$, $\{2,5,4\}$, $\{6,4,9\}$,}\\
\cline{1-6}
9& A1 & D &  &  &  A1 $SOSD$ D
& \multirow{1.8}{157pt}{\tiny \hspace{-4pt} $\{7,2,6\}$, $\{7,5,9\}$
	test \textit{\textbf{Transitivity}}}\\
\cline{1-6}
10& A1 & A2 & C1 &  & A1 is $FOSD$ominant
& \multirow{3.75}{157pt}{\hspace{-4pt} \tiny
	Pairs $\{1,3\}\times \{10\}$, $\{1,3\}\times \{11\}$,}\\
\cline{1-6}
11& A1 & A2 & C2 &  & A1 is ``nearly''* $FOSD$ominant
& \multirow{3}{157pt}{\hspace{-5pt} \tiny
	$\{2,4,5\}\times \{13\}$, $\{$2, 4, 5, 9, 12, 13$\}\times\{$14$\}$,}\\
\cline{1-6}
12& A1 & B1 & B2 &  & A1 is ``nearly''** $SOSD$ominant
& \multirow{2.25}{157pt}{\hspace{-2pt}\tiny
	$\{2,6,7\}\times \{12\}$, $\{$1, 3, 10, 11$\}\times\{$15$\}$}\\
\cline{1-6}
13& B1 & B2 & D &  & D is ``nearly''** $SOSD$ominant
& \multirow{1.25}{157pt}{\hspace{-1pt}\tiny test
	\textbf{\textit{Contraction Consistency}}}\\
\cline{1-6}
14& A1 & B1 & B2 & D & A1 is ``nearly''** $SOSD$ominant
& \multirow{2}{157pt}{\tiny \hspace{-4pt} Menus 14, 15
	are relatively \textit{`complex'} and test\\
	\textit{\textbf{Decisiveness}} via the \textit{overload} channel} \\
\cline{1-6}
15& A1 & A2 & C1 & C2 & A1 is ``nearly''* $FOSD$ominant & \\
\cline{1-7}
\end{tabular}
}
\label{tab:menus}
\vspace{1pt}
\makebox[\textwidth][l]{\scriptsize

*A1 first-order stochastically dominates A2 \& C1
and ``almost'' dominates C2 (see Table \ref{tab:fosd}).}
\makebox[\textwidth][l]{\scriptsize
**A1 second-order stochastically dominates
B1 \& D and ``almost'' dominates B2;}
\makebox[\textwidth][l]{\scriptsize
D second-order stochastically dominates B1
and ``almost'' dominates B2 (see Table \ref{tab:sosd}).}
\end{table}

Two of the binary menus, \{\textnormal{A1, A2}\}
and \{\textnormal{A2, D}\},
featured a FOSD relation, hence an easy decision.
Another three such menus,
\{\textnormal{B1, D}\}, \{\textnormal{A1, B1}\} and
\{\textnormal{A1, D}\}, featured a SOSD relation,
hence an easy decision for any risk-averse (more likely)
or risk-seeking expected-utility maximizing subject,
who would \textit{always} choose the SOSDominant and SOSDominated
such options, respectively.
These lotteries and menus were designed to test
	what we highlight below (and prove in O.A.\ref{a8})
	as an implication of expected-utility theory
	in this environment.

\begin{proposition}[Stability of Attitudes to Risk---StAR]
	\label{prp:StAR}
	Exactly one of the following holds for every non-risk-neutral
	expected-utility maximizer:\vspace{-10pt}
	\begin{eqnarray}
		\label{star-ra}
		C(\{\textnormal{A1, D}\})=\{\textnormal{A1}\}  &\Longleftrightarrow &
		C(\{\textnormal{A1, B1}\})=\{\textnormal{A1}\}\\
		& \Longleftrightarrow &
		C(\{\textnormal{B1, D}\})=\{\textnormal{D}\}\\
		\label{star-rs}
		C(\{\textnormal{A1, D}\})=\{\textnormal{D}\}  &\Longleftrightarrow &
		C(\{\textnormal{A1, B1}\})=\{\textnormal{B1}\}\\
		&\Longleftrightarrow &
		C(\{\textnormal{B1, D}\})=\{\textnormal{B1}\}
	\end{eqnarray}
	
	\vspace{-15pt}
	
	\noindent
	The following holds for every risk-neutral expected-utility maximizer:
	\vspace{-10pt}
	\begin{eqnarray}
		\label{star-rn}
		C(\{\textnormal{A1, D}\})=\{\textnormal{A1, D}\} & \Longleftrightarrow &
		C(\{\textnormal{A1, B1}\})=\{\textnormal{A1, B1}\}\\
		& \Longleftrightarrow &
		C(\{\textnormal{B1, D}\})=\{\textnormal{B1, D}\}
	\end{eqnarray}
\end{proposition}

It is intuitive and well-known that both safer
(SOSDominant) and riskier (dominated)
lotteries with the same expected value
might be preferred by the same expected-utility maximizer
at low and high wealth levels,
respectively \citep{friedman-savage48}.
What the StAR implication of Proposition \ref{prp:StAR} clarifies
is that, in our experiment's decision environment,
an expected-utility agent's general attitude toward risk---as measured 
by their preference for SOSDominant (risk-averse case),
SOSDominated (risk-seeking) or either type of lotteries (risk-neutral)---is 
consistent throughout, irrespective of the agent's
(unmeasured) background wealth level.
We can therefore test a potentially expected-utility maximizing
subject's stability of risk attitudes by checking whether
they consistently opted for the same type
of lottery at menus \{A1, B1\},
\{B1, D\} and \{A1, D\}.\footnote{Our motivation for introducing and testing StAR
is in the same vein as Alaoui and Penta's (\citeyear{alaoui-penta25})
motivation for introducing---also
within the Expected-Utility framework---their
\textit{Same-Side No Reversal} (SSNR) property.
Unlike StAR, however, SSNR is defined in terms of an 
exogenous threshold wealth $z_0\in \mathbb{R}_+$ and the 
following requirements:
(i) a stable risk attitude with respect to lotteries that
are supported on the left of $z_0$; (ii) a stable
and distinct risk attitude with respect to lotteries that are supported
on the right of $z_0$;
(iii) risk-attitude reversals with respect to some lotteries
that are supported both on the left and right of $z_0$.}

By contrast, the remaining four binary menus---\{\textnormal{B1, B2}\},
\{\textnormal{C1, C2}\},
\{\textnormal{A1, B2}\}
and \{\textnormal{B2, D}\}---contained lotteries that were
unrelated by SOSD, thereby resulting in potentially challenging
decision problems that, as we hypothesized,
could lead some subjects to opt for
the costly choice avoidance/deferral option.
For menu \{\textnormal{C1, C2}\}, in particular, the absence
of a dominant alternative was coupled
by the complexity associated with the non-trivial
probabilities in the definition of these lotteries,
which included three significant decimal points.
On the other hand, at menus \{\textnormal{A1, B2}\}
and \{\textnormal{B2, D}\},
lotteries A1 and D are ``almost'' second-order
stochastically dominating their respective counterparts,
in the sense that such a relationship exists for most of
the money range under consideration
(see Table \ref{tab:sosd} in O.A.\ref{a6} for more details).
As such, deciding at these two menus is relatively easier
than at the other two, at least for risk-averse
expected-utility maximizers.
In general, these four menus,
and especially \{\textnormal{B2, B2}\} \&
\{\textnormal{C1, C2}\}, invite a natural
targeted test of Decisiveness via the
(in)complete preferences channel.

The lotteries at the two difficult binary menus
\{\textnormal{B1, B2}\} and
\{\textnormal{C1, C2}\}, moreover, were constructed
so as to also allow for testing Independence.
Indeed, letting $R:=(1,0,0,0,0)$ be the fictitious lottery that assigns
probability 1 to the zero prize and probability
0 to prizes 9, 10, 20 and 24, we have C1 $= \dfrac{1}{2}$B1 $+ \dfrac{1}{2}$R
and C2 $= \dfrac{1}{2}$B2 $+ \dfrac{1}{2}$R.
Thus, any expected-utility maximizing subject
weakly prefers B1 to B2 if and only if
they weakly prefer C1 to C2.
This test for Independence is clearly different from
existing and well-studied, Allais-type tests of that
axiom such as the ``common-ratio'', ``common consequence''
and ``certainty'' effects in two respects:
(1) both the monetary outcomes and probabilities
are less extreme here; (2) within each pair,
the two lotteries have the same expected value
(12 and 6, respectively) and are unrelated by SOSD.
Hence, the behavioural trade-offs in our
test of Independence are different from those
found in the typical tests of that axiom.

Finally, our collection of 9 binary menus was also designed
to test for Transitivity at five
distinct triples of lotteries:
A1,D,A2; A1,B2,D; A1,B1,B2; A1,D,B1; and D,B2,B1.
The 3 pairs of lotteries within each triple feature different
combinations of dominance/no-dominance
relations, thereby leading to varying levels of ``difficulty''
across triples for subjects to ``pass''
the Transitivity test. We discuss these in the next section.

We now turn to menus that contain more than two lotteries.
From those with three such items, menu \{A1, A2, C1\}
has a FOSDominant lottery (A1) and therefore presents
a relatively easy decision.
A1 is ``nearly'' a FOSDominant lottery at
menu \{A1, A2, C2\} as well, dominating C2 in most of the
relevant monetary range, and featuring twice as high an
expected value (see Table \ref{tab:fosd} in O.A.\ref{a6} for
more details).
On the other hand, menus \{A1, B1, B2\} and \{B1, B2, D\}
do not have a dominant lottery under either the FOSD or SOSD criterion,
although A1 and D, respectively, properly SOSDdominate B1
and ``nearly'' dominate B2 as well, doing so for all $x\in [0,22.143]$
(see Table \ref{tab:sosd} in O.A.\ref{a6} for more details).
Finally, for these reasons,
lottery A1 was ``nearly'' FOSDominant at menu \{A1, A2, C1, C2\},
whereas no such option was available at \{A1, B1, B2, D\}, although
A1 was ``nearly'' SOSDominant at the latter menu.
Together with the nine binary menus above and their respective tests,
the presence of these 3- and 4-lottery menus further allows for testing
WARP and Contraction Consistency.
In addition, the two quaternary menus invite tests for potential
\textit{``choice-overload''} effects \citep{iyengar_lepper00}
whereby avoiding/deferring choice is more likely
at larger menus.
In particular, they allow us to test whether any such effect
is influenced by the presence/absence of a dominant alternative, as
suggested, for example, by the meta-analyses
conducted by \cite{scheibehenne-greifeneder-todd}
and \cite{chernevetal15}, and as predicted by decision processes that
are based on dominant choice with
incomplete preferences \citep[Section 2]{gerasimou18}.
If such a mechanism is present in our data,
then we should intuitively observe more violations of
Decisiveness at menu \{A1, B1, B2, D\}, which lacks a
FOSDominant lottery, than at menu \{A1, A2, C1, C2\},
which ``nearly does'' have such a clearly superior option.

\subsection{Sequence of Choices, Tasks and Payments}\label{s3-2}

After having received and been quizzed on the experiment's
instructions (see O.A.\ref{a5} for details), 
subjects were sequentially presented with 75
decisions, each from one of the 15 menus presented in Table \ref{tab:menus}.
Each menu was presented five times
(no practice rounds were involved).
In the set of the first 15 and in the set of the last 15 choices
we presented each of the 15 menus once.
The order of presentation in these two rounds of 15 choices
was identical and common to all subjects,
and coincides with the order that menus appear
in Table \ref{tab:menus}.
In the remaining 45 decision problems, i.e., from the 16th to the 60th,
each menu was presented three times and the order was randomized for each
participant. The order of presentation within a given menu
was not randomized across rounds.
Once subjects had gone through the 75 decision problems,
and before the payout procedure commenced, they were asked to
complete a series of questionnaires.
These included questions on basic demographic characteristics
as well as the ICAR-16 test of cognitive ability
\citep{condon-revelle14} and the 22 indecisiveness-scale items
in \cite{germeijs&deboeck}.\footnote{This was done in order to test
for possible correlations between subjects' scores in
that scale and the frequency of hesitation-revealing behaviours
such as deferrals and/or long response times.
No notable such correlations were found in our data.}
ICAR-16 contains four questions on each of the
following four types of cognitive tasks: (i) letter and number series;
(ii) verbal reasoning; (iii) three-dimensional rotation;
(iv) matrix reasoning.\footnote{A different set of items from
the ICAR database of questions was also used, for example,
by \cite*{chapman_etal23} to measure subjects' cognitive ability.}
We report on these data in Section \ref{s-ca}.

\subsection{Incentives}\label{s3-3}

Choices in our experiment were incentivized.
Subjects were informed at the start of the experiment that
15 distinct decision problems
would be shown to them 5 times\footnote{This was communicated to
them from the outset in order to prevent disruption during the experiment
that might be caused by subjects'
expecting to see only different problems and hence mistakenly thinking
that there were errors in the experiment's implementation.
This kind of early disclosure is in line with other experimental studies
featuring repeated choices, e.g. \cite{agranov&ortoleva}.} and
in random order, although they were not told what
the menu-size distribution was or that
all menus were constructed from just 7 distinct lotteries. Thus,
they could not anticipate that the lotteries and binary menus
would appear frequently. We also informed them that
one of the 75 decision problems would be randomly drawn at
the end of the experiment and that the lottery they had chosen in that
decision would be played out for
them and paid out accordingly. If they had previously selected
\textit{``I'm not choosing now''} at that decision problem, they
would be asked to choose a lottery from that menu at that point, and
this would then
be played out for them. Subjects received the lottery’s prize minus a fee of
{\footnotesize \euro/\pounds} 0.5 for having deferred the decision.
All subjects also received an additional {\footnotesize \euro/\pounds}
5 flat monetary fee.
As an extra incentive for subjects to make deliberated and non-rushed decisions,
they were told from the beginning that no participant would be able
to receive their rewards and leave the lab in the first 60 minutes of
the session.

Motivated by intuition and previous research
\citep{tversky_shafir92,danan&ziegelmeyer06,CCGT22}, we hypothesized that
the option of not choosing could be selected if subjects found a
decision problem difficult enough that they would
be willing to risk a small deduction
(\euro/\pounds {} 0.5) from their total monetary earnings
in order to avoid/delay making an active choice there,
either because they did not have a most preferred lottery
at the relevant menu or because they considered the task of
identifying their most preferred lottery too cognitively
costly. Although \cite{tversky_shafir92} did not use this terminology,
the indecisiveness-based motivation
for allowing choice avoidance/deferral follows their work.
Making such deferral costly to subjects on the other hand---and hence 
embedding it in the design's incentivization---follows
\cite{danan&ziegelmeyer06} and \cite{CCGT22}.
Unlike the design of this paper, however, the one in the latter
study allowed subjects to switch their active choice
at their randomly selected menu, at an even higher cost than
the cost associated with avoidance/deferral.
No such reversal was possible here. Furthermore, unlike the
design in the working paper of \cite{danan&ziegelmeyer06}, ours allows for
binary as well as non-binary menus; does not frame the decisions
as choices between menus of lotteries; and does not involve
a week's delay between when deferrals were made and when subjects
were asked to make an active choice at their randomly selected menu.

\subsection{Implementation and Procedural Details}\label{s3-4}

The experiment was conducted in two locations:
(i) the University of St Andrews Experimental Economics Lab
on 17-18th January 2022 ($N=86$) and on 8-9th May 2023 ($N=115$);
(ii) the University of Bonn
Laboratory for Experimental Economics (BonnEconLab) on
20th December 2022/11-12th January 2023 ($N=107$).
Subject recruitment was done with ORSEE \citep{ORSEE15}
in St Andrews, and with `hroot' \citep{hroot} in Bonn.
The experimental interface was programmed in Qualtrics.
All instructions were translated from English into German using that
platform's built-in translation tool, with manual adjustments
made when necessary.
Because we wanted to run the study in two countries and replicate 
the findings from the initial UK sample, we aimed for roughly 100 
participants per sample---a size that is common in risky-choice experiments. 
The only procedural difference between the first sample and the two subsequent 
samples was an additional payment of \euro/\pounds {} 2 for completing the 
cognitive-ability questionnaire, which participants learned about only after 
the main part of the experiment had concluded. Further details are provided 
in O.A.\ref{a7}.

\section{Preference Maximization With or Without Indifferences}\label{s4}

We start by investigating the extent to which subjects behaved as if they
were ordinal or expected-utility maximizers across \textit{all} 75 decisions.
We examine the possibility that choice reversals at different occurrences 
of the same menu are rational manifestations of subjects' indifference 
between lotteries, and that their overall behaviour is compatible with 
utility maximization when such alternating choices are viewed as the 
rational outcome of subjects' indifference.

To answer this question, we first sliced every subject's data into five regions,
each corresponding to one ``round'' where the 15 distinct menus displayed in
Table \ref{tab:menus} were presented.
In the following, we refer to the ``$i$-th round'' as the grouping of the set
of decisions at those menus which subjects
made the $i$-th time they saw those menus.\footnote{Recall that the order of choice
menus was identical and
common to all subjects in the first 15 and last 15 decisions,
but that the 15 menus appeared three times
in a subject-specific random order in the 45 decisions in between.
Hence, in the sequence of decision problems numbered 16 to 60 in the
Qualtrics survey,
the same menu might be displayed consecutively in those 45 decisions.
Furthermore, a decision problem whose
Qualtrics-survey number was between 46 and 60 could be presented
before a menu numbered between 16 and 30 or 31 and 45.}
Following that, we accounted for the possibility that subjects
are \textit{indifferent} between distinct
lotteries by \textit{merging} their decisions at each menu across the 5 rounds,
thereby creating a choice correspondence for each of them.
Although this intuitive approach
is endorsed by, among others, \citet[p.10]{mwg}
and has been supported by relevant computational tools
\citep[\Prests\!\!]{prest}, apparently it has not been followed
in experimental studies where subjects were
repeatedly presented with the same menus.\footnote{That said,
we remark that \cite{balakrishnan-ok-ortoleva}
is a recent, primarily theoretical
study that refines this approach by introducing
a choice-probability threshold rule into
the merging process, extending \cite{fishburn78}.
The authors apply their method on the binary
forced-choice data from \cite{tversky69}
to construct choice correspondences,
and find that more than half of these are transitive
under certain threshold values.
Unlike that study, here we do not impose a threshold 
in the analysis of our data, and do not require the 
primitive or merged choices to be non-empty.
Furthermore, we apply this
model-free choice-merging approach to our richer and novel
experimental dataset to carry out
a more extensive test of subjects' conformity with rational choice under risk.}

More specifically, letting $C_i^n(\{\textnormal{A1, A2}\})$ denote the
(possibly empty) choice at menu \{A1, A2\} that subject $n$ made
the $i$-th time they saw that menu, for $i\leq 5$,
this merging process is illustrated with the following hypothetical situation:
\begin{eqnarray*}
\left.
\begin{array}{l}
\overbrace{C^{n}_1(\{\textnormal{A1, A2}\}) = \emptyset}^{\text{round-1 choice}}
\vspace{5pt}\\
C^{n}_2(\{\textnormal{A1, A2}\})=\{A2\} \vspace{5pt}\\
C^{n}_3(\{\textnormal{A1, A2}\})=\{A1\} \vspace{5pt}\\
C^{n}_4(\{\textnormal{A1, A2}\})=\{A2\} \vspace{5pt}\\
\underbrace{C^{n}_5(\{\textnormal{A1, A2}\})=\{A1\}}_{\text{round-5 choice}}
\end{array}
\right\}
& \Longrightarrow &
\underbrace{C^{n}(\{\textnormal{A1, A2}\})=\{\textnormal{A1, A2}\}}_{\text{merged choice}}
\end{eqnarray*}
Similarly, if $C^n_i(\{\text{A1, A2}\})=\{\text{A1}\}$
for all $i\leq 5$, then
$C^n(\{\text{A1, A2}\})=\{\text{A1}\}$.\footnote{One may argue
that one cannot rule out that a subject is indifferent between A1 and A2
even though they might have chosen A1 in all 5 rounds.
Although possible, this is unlikely: if someone who is indifferent between
A1 and A2 chooses either one with probability $\frac{1}{2}$ each time
this menu is presented, then
the probability of choosing the same option all 5 times is
$\left(\frac{1}{2}\right)^5=0.03125$.}
The correspondence $C^{n}$ thus defined satisfies
$$\emptyset \subseteq C^{n}(A) \subseteq A\quad\text{for every menu }A,$$
with $C^{n}(A)=\emptyset$ iff $C^{n}_i(A)=\emptyset$ for all $i\leq 5$.

For every subject $n$ we used $C^n$ to
test for conformity with ordinal and expected-utility
maximization of that subject's behaviour
across their 75 decisions, with strict or
weak preferences.
For ordinal utility maximization, we used the Houtman-Maks
(\citeyear{houtman&maks}) (HM) method to find how close subjects' choices are to
this mode of behaviour.
This method assigns a non-negative integer, the HM score, 
to each subject's set of 15 menus. The HM score reflects the minimum number 
of decisions that must be dropped in order for the remaining ones to be
compatible with utility maximization.
Perfect compatibility with utility maximization is associated with an 
HM score of zero. 
An HM score of one, for example, means that removing a single choice 
(out of 15) is sufficient for the subject to be consistent with utility 
maximization in the remaining 14 choices. Note, however, that even in this case 
the individual may violate more than one rationality axiom across 
all 15 choices.
While computing the HM score is a standard---though often
computationally demanding---revealed-preference test,
we apply it to our data using a relatively new computational
technique that also allows for testing a subject's behavioural proximity to
maximization of a preference relation with
indifferences.\footnote{\cite{gerasimou21}
is the only other paper that we are aware of which applies the same
technique on possibly multi-valued choice correspondences. The experimental
design, (riskless) choice domain, focus, research questions
and results in that paper, however, are very different from those
in the present study.}
To illustrate with an example
the idea of this indifference-permitting HM test, suppose
$C^n(\{\textnormal{A1, A2}\})=\{\textnormal{A1, A2}\}
=C^n(\{\textnormal{A1, A2, C2}\})$
and $C^n(\{\textnormal{A1, A2, C1, C2}\})=\{\textnormal{A1}\}$.
The first two choices
are consistent with maximization of a preference relation
where subject $n$ is indifferent between A1 and A2,
and prefers either of them to C2. The third choice,
however, contradicts this by suggesting that A1 is
preferred to A2.
Thus, treating the absence of A2 from the optimal choices
at that menu as a mistake and ``dropping'' it from
the dataset leads to an HM score of 1 in this example.

For our expected-utility analysis, furthermore, we construct
a binary variable that indicates whether a subject's behaviour
is compatible with this model or not. Clearly, since every
expected-utility subject is also an ordinal-utility maximizer,
this group will be a subset of those subjects with an HM score
of zero. Given the structure of the 7 lotteries in our experiment
and of the 15 distinct menus presented to subjects,
the additional tests that must be carried out on those subjects'
choices pertain to FOSD, Independence and StAR.
In line with the discussion of Section \ref{s3-1}, FOSD is satisfied
if both $C^n(\{\textnormal{A1, A2}\})=\{\textnormal{A1}\}$ and
$C^n(\{\textnormal{A2, D}\})=\{D\}$ are true.
Additionally, Independence is satisfied if and only if one of the following
is true: (i) $C^n(\{\textnormal{B1, B2}\})=\emptyset=C^n(\{\textnormal{C1, C2}\})$;
(ii) $C^n(\{\textnormal{B1, B2}\})=\{\textnormal{B1}\}$ and
$C^n(\{\textnormal{C1, C2}\})=\{\textnormal{C1}\}$;
(iii) $C^n(\{\textnormal{B1, B2}\})=\{\textnormal{B2}\}$ and
$C^n(\{\textnormal{C1, C2}\})=\{\textnormal{C2}\}$;
or (iv) $C^n(\{\textnormal{B1, B2}\})=\{\textnormal{B1, B2}\}$
and $C^n(\{\textnormal{C1, C2}\})=\{\textnormal{C1, C2}\}$.
Finally, StAR is satisfied if and only if
one of the following holds at \textit{every} pair of menus
of lotteries $\{P,Q\}$ and $\{P',Q'\}$ where $P$ SOSD $Q$ and
$P'$ SOSD $Q'$: (i) $C^n(\{P,Q\})=\emptyset=C^n(\{P',Q'\})$;
(ii) $C^n(\{P,Q\})=\{P\}$ and $C^n(\{P',Q'\})=\{P'\}$;
(iii)  $C^n(\{P,Q\})=\{Q\}$ and $C^n(\{P',Q'\})=\{Q'\}$;
(iv) $C^n(\{P,Q\})=\{P,Q\}$ and $C^n(\{P',Q'\})=\{P',Q'\}$.
Assuming that the other requirements of expected utility are satisfied,
the choice patterns in cases (ii), (iii) and (iv), respectively, are
necessary and sufficient for the agent to be revealed
risk-averse, risk-seeking and risk-neutral.

The main results from this analysis are summarized in Table \ref{tab:merged}.
Perhaps surprisingly given the relatively large number
and difficulty of the experiment's decision environment,
approximately 26\% of all subjects behaved as if they consistently
maximized a stable, complete and transitive preference
relation over the 7 lotteries across \textit{all} 75 decisions.\footnote{In 
a previous version of this paper we also analysed the data across the 5 rounds 
through the lens of stochastic choice theory for potential conformity with the 
\textit{random utility model}. We found a significant subject overlap between 
deterministic utility-maximizing behaviour with indifferences and 
random utility maximization. We refer the interested reader to 
\href{https://arxiv.org/abs/2402.16538v1}{https://arxiv.org/abs/2402.16538v1}
for more details.}
Importantly, moreover, for more than half of those subjects, this
conclusion could only be
reached because we specifically tested for the possibility
that subjects' different choices at the same menus, across repeated 
appearances of these menus, could reflect subjects' rational indifference
rather than than violations of rationality.

\begin{table}[!htbp]
\centering
\footnotesize
\setlength{\tabcolsep}{6pt} 
\renewcommand{\arraystretch}{1.5} 
\caption{\centering Subjects who behaved as utility maximizers across all 75 decisions,
with or without strict preferences.}
\label{tab:merged}
\makebox[\textwidth][c]{
\begin{tabular}{|r|rl|rl|}
\hline
&\multicolumn{2}{l|}{\multirow{3}{80pt}{\centering\textbf{Ordinal-Utility
			maximizers}}}
&\multicolumn{2}{l|}{\multirow{3}{85pt}{\centering\textbf{Expected-Utility
			maximizers}}}\\
&&&&\\
&&&&\\
\hline
\multirow{1}{140pt}{\raggedleft\textbf{Revealing strict preferences}}
& 34
& (11\%)
& 32
& (10\%)\\
\hline
\multirow{2}{140pt}{\raggedleft\textbf{Revealing strict preferences
		and indifferences}}	
& \multirow{2}{30pt}{\raggedleft46}
& \multirow{2}{30pt}{(15\%)}
& \multirow{2}{30pt}{\raggedleft21} & \multirow{2}{30pt}{(7\%)}\\
&
&
&
&  \\
\hline
\color{black} \multirow{1}{140pt}{\raggedleft\textbf{Total}}
& \color{black} 80
&\color{black} (26\%)
& \color{black}53
& (17\%)\\
\hline		
\end{tabular}
}
\label{tab:merged-b}
\end{table}

The expected-utility types in our experiment are those subjects who are both 
ordinal utility maximizers across all 75 decisions and expected-utility 
maximizers on the binary menus. In our sample, 53 of 308 subjects (17\%) 
belong to this category: 21 revealed a stable weak order including 
some indifferences, while 32 exhibited a stable strict preference relation. 
All but one of these subjects were revealed to be risk-averse.
Looking specifically at the nine binary menus 
(five repetitions each, for a total of 45 decisions), 
we find that roughly 21\% of all subjects behaved as expected-utility 
maximizers on these choices. Among those binary-menu EUT maximizers, 
just under half exhibited at least one non-trivial indifference 
between distinct lotteries.

We summarise this information as follows:

\begin{highlight}
Twenty-six percent of all subjects were perfectly compatible with
ordinal utility maximization throughout the experiment,
with more than half being so under some
indifferences.
Moreover, 65\% of those subjects were perfectly compatible with
expected utility maximization throughout, with nearly 40\% of them under
some indifferences.
Fifty two of the 53 EUM-complying subjects
were risk-averse and one was risk-neutral.
\end{highlight}

\begin{table}[!htbp]
\centering
\footnotesize
\caption{\centering Subjects whose 15 indifference-permitting merged choices
deviate from the predictions \newline of deterministic ordinal/expected utility theory.}
\label{tab:axiomsDet}
\makebox[\textwidth][c]{
\setlength{\tabcolsep}{6pt} 
\renewcommand{\arraystretch}{1.5} 
\begin{tabular}{|l|rl|}
\hline
\textbf{Decisiveness}						
& 22
& (7\%) \\
\hline
\textbf{First-Order Stochastic Dominance}	
& 26
& (8.5\%) \\
\hline
\textbf{Independence}						
& 114
& (37\%) \\
\hline
\textbf{Stability of Attitudes to Risk}		
& 130
& (42\%) \\
\hline
\textbf{Contraction Consistency}			
& 155
& (50\%)\\
\hline
\textbf{Transitivity} 						
& 156
& (51.5\%) \\
\hline			
\textbf{Weak Axiom of Revealed Preference}	
& 220
& (71.5\%)\\
\hline
\end{tabular}
}
\end{table}

\begin{figure}[!htbp]
\centering
\caption[]{\centering The most frequent
weak revealed preferences that are compatible with
EUM \linebreak (merged choices).\footnotemark[16]
\footnotetext[16]{The displayed directed graphs in Figures 
	\ref{fig:prefs-weak} and \ref{fig:pref-strict} 
	correspond to \Prests-computed optimal orders
	and were produced with a GraphViz add-on
	\citep{graphviz}.}}
\begin{minipage}{0.3\textwidth}
\centering
\includegraphics[width=0.06\textheight]{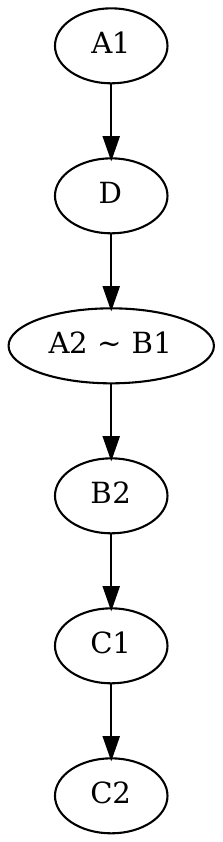}
\vspace{7pt}
\end{minipage}
\begin{minipage}{0.3\textwidth}
\centering
\includegraphics[width=0.06\textheight]{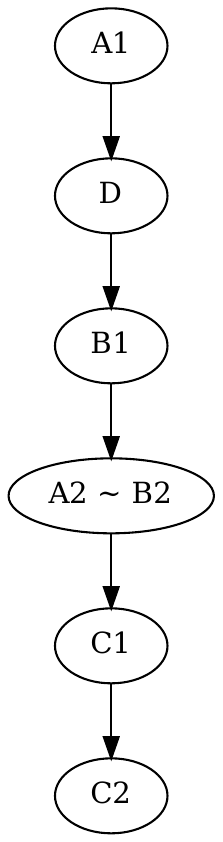}
\end{minipage}
\begin{minipage}{0.3\textwidth}
\centering
\includegraphics[width=0.06\textheight]{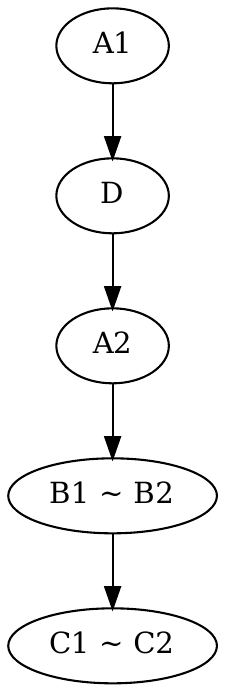}
\vspace{7pt}
\end{minipage}
\label{fig:prefs-weak}
\end{figure}

The three most frequent revealed preference orderings that are recoverable from each
subject's merged choices and, in addition, are compatible with expected-utility maximization
apply to 49 of the 53 subjects. All three entail 
A1 $\succ$ D $\succ$ B1 (Figure \ref{fig:prefs-weak}).\footnote{To
arrive at these preference relations, we first used \PrestsFootnote
to find each  subject's set of weak orders over the 7 lotteries that were compatible
with that person's 15 merged choices. The cardinality of this
set ranged from 3 to 173, with a mean (median) of 92 (63). 
This reflects the fact that
only 15 out of the 120 non-singleton menus
that are derivable from a set of 7 alternatives were shown to subjects,
implying that the revealed preferences that
are compatible with ordinal and/or expected utility maximization are
not uniquely identified in general.
Following this, we pinned down the EU-compatible orders with the highest and
second-highest frequency.}
Recalling that A1 \textit{SOSD} $D$ \textit{SOSD} B1, this is a reflection
of the strict risk aversion revealed by 52 of these 53 subjects' overall behaviour.

Retaining our focus on the subjects' overall behaviour
across their 75 decisions through
the resulting merged choices at the 15 distinct menus,
we proceed next to an analysis of
the main factors behind subjects' deviations from
indifference-permitting rational choice in the ordinal and/or
expected-utility sense. A summary of this analysis is
presented in Table \ref{tab:axiomsDet}.	
This clarifies that around half of all subjects' merged
choices violated Contraction Consistency and Transitivity, 
while over 70\% exhibited choice reversals, in violation of WARP.
Among subjects violating Transitivity, however, no subject
exhibited strict binary choice cycles of the form
$\{p\}=C(\{p,q\})$, $\{q\}=C(\{q,r\})$ and $\{r\}=C(\{p,r\})$.
Similarly, among the 8\% of subjects
who violated FOSD, none did so strictly in the sense
of always choosing the dominated lottery.
Twenty-two subjects (7\%), moreover, violated Decisiveness
by consistently avoiding/delaying making an active choice
in at least one of the 15 distinct menus (we discuss later
some patterns in those violations).
Finally, 42\% and 37\% of subjects, respectively, deviated from
StAR and Independence in this analysis.
While the proportion of Independence violators is lower than the one 
corresponding to some of the other consistency principles,
we recall that---unlike the latter---there was only one pair
of menus here where this axiom could have been violated. 
By contrast, there were 3 pairs of menus where StAR could be violated, 
5 triples of for Transitivity, 20 pairs for Contraction Consistency,
and even more for WARP. Cast in this light, our finding here
that 37\% of subjects violated Independence in this
merged-choice analysis cannot by itself be interpreted as
evidence suggesting that it is easier or harder
to comply with this axiom than the other axioms.
We return to this point in Section \ref{s5-1}.

\section{Convergence to Rationality, One Round at a Time}\label{s5}

We now turn to our other main question: \textit{Do subjects come
closer to maximizing utility with strict preferences
as they make choices at the same decision problems repeatedly,
without receiving any new information in the process?}
Table \ref{tab:first_last_15} summarizes the relevant findings
from this investigation, which is based on the
round-per-round behaviour of every subject according to the following criteria:
\begin{enumerate}
\item How many subjects' decisions in each round are perfectly
compatible with ordinal and expected-utility maximization
with strict preferences, and how many violate the
seven axioms of rational choice under risk that were discussed in Section \ref{s3}?
\item How many total and \textit{active-choice}
(i.e., excluding deferrals) decisions would have to be changed
on average for each subject in each round to make decisions
consistent with utility maximization with strict preferences?
\item How long does it take to make a decision on average?
\end{enumerate}

\begin{table}[!htbp]
\centering
\caption{\centering Utility maximization and axiom violations in each round.}
\label{tab:first_last_15}
\footnotesize
\setlength{\tabcolsep}{6pt} 
\renewcommand{\arraystretch}{1.3} 
\makebox[\textwidth][c]{
\begin{tabular}{|l|c|c|c|c|c|c|}
\multicolumn{7}{c}{(a) Utility and Expected-Utility maximization.} \\
\hline
& \multicolumn{1}{c}{\multirow{1}{30pt}{\centering\textbf{1st}}}
& \multicolumn{1}{c}{\multirow{1}{30pt}{\centering\textbf{2nd}}}
& \multicolumn{1}{c}{\multirow{1}{30pt}{\centering\textbf{3rd}}}
& \multicolumn{1}{c}{\multirow{1}{30pt}{\centering\textbf{4th}}}
& \multicolumn{1}{c|}{\multirow{1}{30pt}{\centering\textbf{5th}}}
& \multirow{1}{62pt}{\centering \scriptsize First vs Last 15: 2-sided test $p$-values}\\
& \multicolumn{5}{c|}{\multirow{1}{*}{\centering\textbf{round of 15 decisions*}}}
& \multirow{-1}{60pt}{\centering \scriptsize }\\
\hline
\multirow{2}{175pt}{\raggedleft \textbf{Utility Maximizers$^\natural$}}
& \multirow{2}{30pt}{\centering106\\ (34\%)}
& \multirow{2}{30pt}{\centering129\\ (42\%)}
& \multirow{2}{30pt}{\centering143\\ (46.5\%)}
& \multirow{2}{30pt}{\centering161\\ (52\%)}
& \multirow{2}{30pt}{\centering177\\ (57.5\%)}
& \multirow{2}{60pt}{\centering $<0.001$\\
	\tiny (Fisher's exact)}\\
& & & & & &\\
\hline
\multirow{2}{175pt}{\raggedleft \textbf{Approximate$^\flat$\\ Utility Maximizers}}	
& \multirow{2}{30pt}{\centering173\\ (56\%)}	
& \multirow{2}{30pt}{\centering202\\ (65.5\%)}
& \multirow{2}{30pt}{\centering211\\ (68.5\%)}
& \multirow{2}{30pt}{\centering225\\ (73\%)}
& \multirow{2}{30pt}{\centering242\\ (78.5\%)}
& \multirow{2}{60pt}{\centering $<0.001$\\
	\tiny (Fisher's exact)}\\
& & & & & & \\
\hline
\multirow{2}{175pt}{\raggedleft \textbf{\textit{Active-Choice$^\sharp$}\\
		Utility Maximizers}}	
& \multirow{2}{30pt}{\centering121\\ (39\%)}	
& \multirow{2}{30pt}{\centering156\\ (50.5\%)}
& \multirow{2}{30pt}{\centering171\\ (55.5\%)}
& \multirow{2}{30pt}{\centering186\\ (60\%)}
& \multirow{2}{30pt}{\centering211\\ (68.5\%)}
& \multirow{2}{60pt}{\centering $<0.001$\\
	\tiny (Fisher's exact)}\\
& & & & & & \\
\hline
\multirow{2}{175pt}{\raggedleft \textbf{Expected-Utility Maximizers\\
		at all menus}}	
& \multirow{2}{30pt}{\centering74\\ (24\%)}
& \multirow{2}{30pt}{\centering85\\ (27.5\%)}
& \multirow{2}{30pt}{\centering99\\ (32\%)}
& \multirow{2}{30pt}{\centering111\\ (34\%)}
& \multirow{2}{30pt}{\centering123\\ (40\%)}
& \multirow{2}{60pt}{\centering $<0.001$\\
	\tiny (Fisher's exact)}\\
& & & & & & \\
\hline
\multirow{2}{175pt}{\raggedleft \textbf{Average (median) decisions
		away from Utility Maximization {\tiny (HM)}}}
& \multirow{2}{47pt}{\centering1.45 (1)}
& \multirow{2}{47pt}{\centering1.19 (1)}
& \multirow{2}{47pt}{\centering1.1 (1)}
& \multirow{2}{47pt}{\centering0.97 (0)}
& \multirow{2}{47pt}{\centering0.86 (0)}
& \multirow{2}{60pt}{\centering $<0.001$\\
	\tiny (Mann-Whitney)}\\
& & & & & & \\
\hline
\multirow{2}{175pt}{\raggedleft \textbf{Average (median)
\textit{active} decisions away from Utility Maximization}}
& \multirow{2}{47pt}{\centering1.13 (1)}
& \multirow{2}{47pt}{\centering0.88 (0)}
& \multirow{2}{47pt}{\centering0.79 (0)}
& \multirow{2}{47pt}{\centering0.65 (0)}
& \multirow{2}{47pt}{\centering0.55 (0)}
& \multirow{2}{60pt}{\centering$<0.001$\\
	\tiny (Mann-Whitney)}\\
& & & & & & \\
\hline
\multirow{2}{175pt}{\raggedleft \textbf{Average (median) response time}\\
	\scriptsize (in seconds)}
& \multirow{2}{47pt}{\centering20.7 (16.2)}
& \multirow{2}{47pt}{\centering13.3 (10.3)}
& \multirow{2}{47pt}{\centering10.7 (8.2)}
& \multirow{2}{47pt}{\centering9.4 (7.2)}
& \multirow{2}{47pt}{\centering8.1 (6.3)}
& \multirow{2}{60pt}{\centering $<0.001$\\ 	\tiny (Mann-Whitney)}\\
& & & & & & \\
\hline
\multicolumn{7}{c}{(b) Axiom violations.} \\
\hline
\multirow{2}{175pt}{\raggedleft \textbf{Violating FOSD}}
& \multirow{2}{47pt}{\centering21\\ (7\%)}
& \multirow{2}{47pt}{\centering12\\ (4\%)}
& \multirow{2}{47pt}{\centering12\\ (4\%)}
& \multirow{2}{47pt}{\centering8\\ (2.5\%)}
& \multirow{2}{47pt}{\centering9\\ (3\%)}
& \multirow{2}{60pt}{\centering 0.041\\ \tiny (Fisher's exact)}\\
& & & & & & \\
\hline
\multirow{2}{175pt}{\raggedleft \textbf{Violating Independence}}
& \multirow{2}{47pt}{\centering98\\ (32\%)}
& \multirow{2}{47pt}{\centering84\\ (27\%)}
& \multirow{2}{47pt}{\centering87\\ (28\%)}
& \multirow{2}{47pt}{\centering85\\ (27.5\%)}
& \multirow{2}{47pt}{\centering76\\ (25\%)}
& \multirow{2}{60pt}{\centering 0.060\\ \tiny (Fisher's exact)}\\
& & & & & & \\
\hline
\multirow{2}{175pt}{\raggedleft \textbf{Violating\\
		Stability of Attitudes to Risk}}
& \multirow{2}{47pt}{\centering112\\ (36\%)}
& \multirow{2}{47pt}{\centering106\\ (34\%)}
& \multirow{2}{47pt}{\centering96\\ (31\%)}
& \multirow{2}{47pt}{\centering101\\ (33\%)}
& \multirow{2}{47pt}{\centering80\\ (26	\%)}
& \multirow{2}{60pt}{\centering 0.025\\ \tiny (Fisher's exact)}\\
& & & & & & \\
\hline
\multirow{2}{175pt}{\raggedleft \textbf{Violating WARP}}
& \multirow{2}{30pt}{\centering186\\ (60\%)}
& \multirow{2}{30pt}{\centering151\\ (49\%)}
& \multirow{2}{30pt}{\centering137\\ (44.5\%)}
& \multirow{2}{30pt}{\centering119\\ (38.5\%)}
& \multirow{2}{30pt}{\centering97\\ (31.5\%)}
& \multirow{2}{60pt}{\centering $<0.001$\\
	\tiny (Fisher's exact)}\\
& & & & & & \\
\hline		
\multirow{2}{175pt}{\raggedleft \textbf{Violating\\ Contraction Consistency}}
& \multirow{2}{30pt}{\centering190\\ (62\%)}
& \multirow{2}{30pt}{\centering153\\ (50\%)}
& \multirow{2}{30pt}{\centering146\\ (47\%)}
& \multirow{2}{30pt}{\centering122\\ (39.5\%)}
& \multirow{2}{30pt}{\centering101\\ (33\%)}
& \multirow{2}{60pt}{\centering $<0.001$\\
	\tiny (Fisher's exact)}\\
& & & & & & \\
\hline
\multirow{2}{175pt}{\raggedleft {\textbf{Violating Transitivity}}}
& \multirow{2}{47pt}{\centering77\\ (25\%)}
& \multirow{2}{47pt}{\centering67\\ (22\%)}
& \multirow{2}{47pt}{\centering64\\ (21\%)}
& \multirow{2}{47pt}{\centering56\\ (18\%)}
& \multirow{2}{47pt}{\centering36\\ (12\%)}
& \multirow{2}{60pt}{\centering $<0.001$\\ \tiny (Fisher's exact)}\\
& & & & & & \\
\hline		
\multirow{2}{175pt}{\raggedleft \textbf{Violating Decisiveness}}
& \multirow{2}{30pt}{\centering50\\ (16\%)}
& \multirow{2}{30pt}{\centering46\\ (15\%)}
& \multirow{2}{30pt}{\centering47 (15\%)}
& \multirow{2}{30pt}{\centering48 (15.5\%)}
& \multirow{2}{30pt}{\centering46\\ (15\%)}
& \multirow{2}{60pt}{\centering 0.739\\
	\tiny (Fisher's exact)}\\
&
&
&
&
&
& \\
\hline
\end{tabular}
}
\raggedright \footnotesize *The order of menu presentation
	was identical and common across subjects in rounds 1, 5 and subject-specific
	in rounds 2, 3, 4. The reported statistics in round $n\in\{2,3,4\}$
	account for this and pertain to the $n$-th appearance
	of each of the 15 distinct menus for each subject. 
	$^\natural$Unless the ``active-choice''
	qualification is present, both active-choice and deferral decisions
	are accounted for and, where relevant, penalized.
	$^\flat$Up to one decision away from Utility Maximization 
	(HM score $\leq 1$). $^\sharp$Ignoring/not penalizing deferrals.
\end{table}

With the exception of subjects violating Decisiveness, whose proportion stayed
in the 15\%-16\% range throughout, our findings unambiguously suggest
that subjects learned to be more consistent with the above
rationality criteria between the first and fifth rounds.
Furthermore, for virtually all of these criteria,
such learning occurred in a strictly
monotonic way; that is, for almost every aspect of rationality that
we consider, there is strictly higher conformity in the sample
as we move from one round to the next.

Before discussing those findings in more detail, it is worth pointing out that
the random and subject-specific order of appearance of the
middle 45 decision problems in our design alleviates potential concerns
that such learning might be driven by the particular order of presentation.
At the same time, the commonality of the presentation order between the
first and fifth round and between subjects allows us to conduct
a like-for-like comparison of behaviour at the beginning and at the end
of the experiment, and hence a targeted test of our hypothesis.

\begin{highlight}\label{find:1}
By the last round, 57.5\% of all subjects' behaviour converges to
utility maximization with strict preferences.
For 69.5\% of those subjects such convergence is to
\textnormal{expected} utility maximization.
\end{highlight}

Indeed, the relevant proportions nearly doubled from 34\% to 57.5\% and
from 24\% to 40\%, respectively,
between the first and fifth rounds ($p<0.001$ in both cases).
Notably, the proportion of strict-preference ordinal
utility maximizers who are also
expected-utility maximizers is relatively stable across rounds,
ranging from 61\% to 70\%.
This suggests that subjects' ability to learn to comply with
the general principles
of rational choice is positively associated with their ability to comply
with the more specialised principles of rational choice under risk.
In addition, the proportions of \textit{approximate}
ordinal utility maximizers,
defined as those who are
at most one decision away from perfect conformity with that model
(i.e., with an HM score less than or equal
to one), are relatively high and also increasing throughout,
from 56\% initially to 78.5\% finally ($p<0.001$).\footnote{In simulations with
uniform-random behaving subjects
on this collection of menus, the 2.5th percentile in the HM score
distribution is 2 decisions. This suggests that our approximation
threshold of one decision
is unlikely to have been reached by human subjects who behaved randomly.}
Furthermore, the distribution of subjects' HM scores is also shifted
significantly to the left in the last 15 compared to the first 15 decisions,
down from 1.45 to 0.86 decisions away from rationality, on average ($p<0.001$).

\subsection{Axiom Violations}\label{s5-1}

We now focus on the evolution of subjects' conformity
(or lack thereof) with each of the deterministic choice axioms
that were presented in Section \ref{s2-1}.

\begin{highlight}
	The proportions of subjects violating each of Transitivity,
	Contraction Consistency, WARP, Independence, FOSD and
	StAR in the last round are significantly
	lower than in the first.
	The proportion of those violating Decisiveness is stable.
\end{highlight}

Despite this significant improving trend, the most persistently violated
axioms were Contraction Consistency
(typically, but---in our free-choice environment---not always,
associated with WARP violations as well;
see Section \ref{s2-1} for more details), with 101
subjects (33\%) still deviating from this consistency principle
in their last 15 decisions.
At the same time, however, Contraction Consistency and WARP
are the two principles with the largest gains in compliance
(approximately 29 percentage points), and with nearly halved
proportions of subjects violating them in the last compared
to the first round.
On the other hand, the smallest gains in compliance were seen
with the Independence axiom, where the proportion
of violators fell from 32\% to 25\% ($p=0.060$).
The lack of an SOSD ranking in the two
menus involved in the Independence test, along with the
relatively high degree of decision difficulty,
may partly explain the slower pace of learning with respect to
that axiom. FOSD, on the other hand, is violated by very few subjects
(down from 7\% to 3\% by the fifth round).
This is in line with findings in \cite{levy08},
where, as in this study, subjects were shown lotteries without
any budget-constraint considerations.

\begin{table}[!htbp]
\centering
\footnotesize
\caption{\centering Frequencies of Decisiveness violations
at different menus.}
\setlength{\tabcolsep}{6pt} 
\renewcommand{\arraystretch}{1.3} 
\makebox[\textwidth][c]{
\begin{tabular}{|cccc|c|c|c|c|c|}
\cline{1-9}
\multicolumn{4}{|c|}{\multirow{3}{*}{\textbf{Menu}}}
& 	\multicolumn{1}{c|}{\multirow{1.3}{*}{\textbf{Deferrals}}}
&	\multicolumn{1}{c|}{\multirow{1.3}{*}{\textbf{Deferrals}}}
&  \multirow{1.3}{*}{\textbf{1st-round}}
& 	\multirow{1.3}{*}{\textbf{Average}}
& \multirow{1.3}{*}{\textbf{Average}}
\\
\multicolumn{4}{|c|}{\multirow{2}{*}{}}
& 	\multicolumn{1}{c|}{\multirow{1}{*}{\textbf{in 75}}}
& 	\multicolumn{1}{c|}{\multirow{1}{*}{\textbf{in merged}}}
&  \multirow{1}{*}{\textbf{presentation}}
&  \multirow{1}{*}{\textbf{1st-round}}
& \multirow{1}{*}{\textbf{response time}}
\\
\multicolumn{4}{|c|}{\multirow{2}{*}{}}
& 	\multicolumn{1}{c|}{\multirow{0.7}{*}{\textbf{decisions}}}
& 	\multicolumn{1}{c|}{\multirow{0.7}{*}{\textbf{15 decisions}}}
& \multirow{0.7}{*}{\textbf{order}}
& \multirow{0.7}{*}{\textbf{response time}}
& \multirow{0.7}{*}{\textbf{overall}}\\
\cline{1-9}
A1
& A2
&
&  									
& 13		
& 0
& 1
& 30.20
& 12.01\\
\cline{1-9}
\textbf{B1}
&\textbf{B2}
&
&  				
&\textbf{59}
&\textbf{2}
&\textbf{2}
&\textbf{29.85}
&\textbf{15.10}\\
\cline{1-9}
\textbf{C1}
& \textbf{C2}
&
&   			
&\textbf{192}
&\textbf{20}
&\textbf{3}
&\textbf{30.23}
&\textbf{16.69}\\
\cline{1-9}
B1
& D
&
&   									
&19 		
&1
& 4
& 21.76
& 11.71\\
\cline{1-9}
\textbf{B2}
&\textbf{D}
&
&   				
&\textbf{26}
&\textbf{1}
&\textbf{5}
&\textbf{16.00}
&\textbf{10.22}\\
\cline{1-9}
A1
& B1
&
&  									
&16 		
&1
& 6
& 16.33
& 10.22\\
\cline{1-9}
\textbf{A1} & \textbf{B2}
&
&  			
&\textbf{16} 	
&\textbf{1}
& \textbf{7}
& \textbf{13.01}
& \textbf{8.95}\\
\cline{1-9}
A2
& D
&
&  									
&19 		
&1
& 8
& 14.20
& 9.57\\
\cline{1-9}
A1
& D
&
&  									
&13 		
&1
& 9
& 13.53
& 10.05\\
\cline{1-9}
A1
& A2
& C1
&  								
&14 		
&1
& 10
& 24.80
& 12.98\\
\cline{1-9}
A1
& A2
& C2
&  								
&16 		
&1
& 11
& 15.80
& 10.94\\
\cline{1-9}
\textbf{A1}
& \textbf{B1}
& \textbf{B2}
&  		
&\textbf{22}
&\textbf{1}
&\textbf{12}
&\textbf{21.08}
&\textbf{14.27}\\
\cline{1-9}
\textbf{B1}
&\textbf{B2}
&\textbf{D}
&  			
&\textbf{23}
&\textbf{1}
&\textbf{13}
&\textbf{19.05}
&\textbf{14.50}\\
\cline{1-9}
\textbf{A1}
&\textbf{B1}
&\textbf{B2}
&\textbf{D}	
&\textbf{25}
&\textbf{3}
&\textbf{14}
&\textbf{26.64}
&\textbf{17.44}\\
\cline{1-9}
A1
&A2
&C1
&C2 								
&13 		
&2
&15
&17.29
&11.98\\
\cline{1-9}
\end{tabular}
}
\makebox[\linewidth][c]{\scriptsize { } }
\makebox[\linewidth][l]{\scriptsize
Note: in bold are ``hard'' menus that were hypothesised
ex ante to have a higher deferral frequency
(cf. Table \ref{tab:menus}).}
\makebox[\linewidth][l]{\scriptsize
First-round and overall mean/median response times between
``hard'' \& ``non-hard'' menus are, respectively,}
\makebox[\linewidth][l]{\scriptsize
22.3/16.6 vs 19.2/15.9 ($p=0.002$; 2-sided Mann-Whitney test) and
12.93/9.13 vs 12.02/8.89 ($p=0.004$).}
\label{tab:deferrals_per_menu}
\end{table}

As far as Decisiveness is concerned, it has already been noted that
the proportion of subjects who violated this principle by deferring
in at least one decision problem
remained stable and in the 15\%--16\% range across all five rounds.
In addition to the percentage of \textit{deferring subjects}, however,
the total \textit{number of deferrals} also stayed practically constant---between
97 and 100---across the 5 rounds.
While the proportion of deferring subjects aligns with those seen in
deferral-permitting studies with no repeated choices, the fact that both
the intensive and extensive margin of deferrals remained constant
is a novel and, in our view, interesting finding.
First, it suggests that subjects
who are willing to incur a monetary cost to
avoid making an active choice in a difficult decision
problem are less inclined than might previously have been thought
to change this avoidance tendency after repeated exposures to the same
decision problem.
This could be thought of as evidence in support of
what \citet[pp. 763-764]{sen97} referred to as \textit{``assertive''}
rather than \textit{``tentative''} incomplete preferences:
\textit{``It is useful to consider the distinction between:
\textnormal{tentative incompleteness}, when some pairs
of alternatives are not \textnormal{yet} ranked
(though they may all get ranked with more deliberation or
information), and \textnormal{assertive incompleteness},
when some pair of alternatives is asserted to be ``non-
rankable''. Assertive incompleteness is the claim that
the failure of completeness is not provisional--waiting to
be resolved with, say, more information, or more penetrating
examination. The partial ranking, or the inexhaustive partitioning,
may simply not be ``completable'', and affirming that some $x$ may not be
rankable \textnormal{vis-\`{a}-vis} some $y$ may
be the right answer in these cases.''}

This is corroborated by the findings shown in
Table \ref{tab:deferrals_per_menu}, which
reports the deferral frequencies per menu.
In line with our hypothesis (Section \ref{s3-1}),
three of the four binary menus featuring
no SOSD relationship between the feasible lotteries
have the highest absolute deferral frequencies:
192 for \{C1, C2\}, 59 for \{B1, B2\} and 26 for \{B2, D\}.
Strikingly, moreover, the table also clarifies that
20 subjects \textit{always} deferred at menu \{C1, C2\},
which, in addition to featuring no dominance relation,
also had the most complex-looking lotteries and the lowest
expected values. It is therefore possible that
the decision to defer at this particular menu was
driven by some combination of decision difficulty and aversion
to incur the cognitive effort, given the relatively
higher complexity and lower stakes involved.
Also broadly in line with our hypothesis,
next in the list of deferral-inducing menus
are \{A1, B1, B2, D\} (25), \{B1, B2, D\} (23)
and \{A1, B1, B2\} (22).
These menus feature the ``nearly'' SOSDominant
lotteries A1, D and A1, respectively, (see Section \ref{s3-1}),
but no FOSDominance relation.
Moreover, despite the relatively low deferral frequencies
at the two menus with four lotteries, \{A1, B1, B2, D\} (25)
and \{A1, A2, C1, C2\} (13), we note the following fact that lends
support to the dominance channel in the occurrence
and alleviation of choice overload that was discussed in
Section \ref{s3-1} \citep{scheibehenne-greifeneder-todd,chernevetal15}:
\begin{highlight}
The deferral rate is lower at the 4-element
menu with a (``nearly'') FOSDominant lottery
than at the one without (0.8\% vs 1.6\%;
$p=0.071$, {\footnotesize
2-sided Fisher's exact}).
\end{highlight}
\noindent In addition, as is also clarified
in Table \ref{tab:deferrals_per_menu} (footnote),
subjects' response times, both in the first round and overall,
are significantly longer at the 4 ``difficult''
binary menus that lacked a SOSDdominant lottery
and the 3 non-binary menus that
lacked a perfectly or approximately FOSDominant lottery
than at the remaining 8 menus. It is noteworthy
that such a significant difference---of almost 3 seconds,
on average---is present in first-round response times
despite the alleviating effect brought about by the
fact that the menu \{A1,A2\} which was
seen first by all subjects in that round featured
an ``easy'' decision, which
was nevertheless associated with a long response time because
it gave subjects their very first exposure
to the experiment's lotteries and computer interface.
Taken together, these facts indicate that:
(i) Decisiveness violations are more likely at ``hard''
binary-menu decisions than at decisions in larger menus,
suggesting a bigger role of incomparability/incomplete preferences than 
traditional choice overload in our context;
(ii) other things equal, deferral is more likely at large menus
that do not have an obviously
dominant lottery than at those that do;
(iii) deferring behaviour is largely unchanged over
the course of the experiment and, in general, is associated
with longer response times.

We now take a closer look at violations of Transitivity,
focusing on the 5 triples of lotteries where such violations
may occur in our data (in O.A.\ref{a9} we extend this analysis
to the relevant lottery quadruples and quintuple).
The proportion of subjects violating the general, free-choice
version of this axiom in at least one of these five triples goes down
from 25\% initially to 12\% eventually ($p<0.001$)
in this free-choice environment.
Table \ref{tab:intransitivities} groups the total
violations across the subjects' 75 decisions by
associating each with the relevant triple where it occurred,
and clarifies the (F)(S)OSDominance structure, if any,
within each of the three pairs in the triple.
At the two extremes lie triples A1-D-A2 and A1-D-B1.
The former features one second-order and two first-order
dominance comparisons, while the other 4 triples feature no 
first-order but more (possibly approximate) second-order 
dominance comparisons.
Because FOSD comparisons 
are arguably less complex than SOSD, 
one would intuitively expect few violations in the first triple,
which is indeed what we find (19; 1.2\%).
The latter triple instead features three second-order
dominance pairwise relations.
By Proposition \ref{prp:StAR} therefore,
any expected-utility maximizing subject with risk-averse
or risk-seeking strict preferences would satisfy
Transitivity at this triple.

\begin{table}[!htbp]
\centering
\footnotesize
\caption{\centering
Frequencies of Transitivity violations and
binary cycles at the relevant triples of lottery pairs \newline
($p$-values from 2-sided Fisher's exact test).}
\setlength{\tabcolsep}{6pt} 
\renewcommand{\arraystretch}{1.3} 
\makebox[\textwidth][c]{
\begin{NiceTabular}{|l|ccccc|}
\hline
\footnotesize\qquad \textbf{Lottery triple}
& \multicolumn{1}{|c|}{A1 \;\; D \;\; A2}
& \multicolumn{1}{|c|}{A1 \;\; D \;\; B2}
& \multicolumn{1}{|c|}{A1 \;\; B1 \;\; B2}
& \multicolumn{1}{|c|}{D \;\; B2 \;\; B1}
& \multicolumn{1}{|c|}{A1 \;\; D \;\; B1}\\
\dashedline
\multirow{1.8}{*}{\footnotesize \textbf{FOSD or SOSD}}
&\multicolumn{1}{|c|}{\tiny\hspace{-20pt}A1 SOSD D}
&\multicolumn{1}{|c|}{\tiny \hspace{-14pt} A1 SOSD D}
&\multicolumn{1}{|c|}{\tiny \hspace{-20pt}A1 SOSD B1}
&\multicolumn{1}{|c|}{\tiny\hspace{-15pt} D  $\approx$SOSD B2}
&\multicolumn{1}{|c|}{\tiny\hspace{-20pt}A1 SOSD D}\\
\multirow{1.3}{*}{\footnotesize \textbf{pairwise relations}}
&\multicolumn{1}{|c|}{\tiny\hspace{15pt} D FOSD A2}
&\multicolumn{1}{|c|}{\tiny \hspace{10pt} D  $\approx$SOSD B2}
&\multicolumn{1}{|c|}{\tiny \hspace{10pt} B1 no-SOSD B2}
&\multicolumn{1}{|c|}{\tiny\hspace{10pt}B2 no-SOSD B1}
&\multicolumn{1}{|c|}{\tiny\hspace{15pt}D SOSD B1}\\
\multirow{0.8}{*}{\footnotesize \textbf{within the triple}}
&\multicolumn{1}{|c|}{\tiny A1 FOSD A2}
&\multicolumn{1}{|c|}{\tiny A1  $\approx$SOSD B2}
&\multicolumn{1}{|c|}{\tiny A1  $\approx$SOSD B2}
&\multicolumn{1}{|c|}{\tiny D SOSD B1}
&\multicolumn{1}{|c|}{\tiny A1 SOSD B1}\\
\hline
\footnotesize \textbf{Intransitivities}
& \multicolumn{1}{|c|}{19}
& \multicolumn{1}{|c|}{79}
& \multicolumn{1}{|c|}{97}
& \multicolumn{1}{|c|}{99}
& \multicolumn{1}{|c|}{136} \\
\dashedline
\multicolumn{6}{l}{\hspace{120pt}
	$p<0.001$ \hspace{24pt} $p=0.186$ \hspace{22pt} $p=0.941$
	\hspace{22pt} $p=0.014$} \\
\hline
\footnotesize \textbf{Binary cycles}
& \multicolumn{1}{|c|}{15}
& \multicolumn{1}{|c|}{77}
& \multicolumn{1}{|c|}{89}
& \multicolumn{1}{|c|}{91}
& \multicolumn{1}{|c|}{129} \\
\dashedline
 \multicolumn{6}{l}{\hspace{140pt}
	$p<0.001$ \hspace{24pt} $p=0.380$ \hspace{22pt} $p=0.938$
	\hspace{22pt} $p=0.009$} \\
\hline
\end{NiceTabular}
}
\label{tab:intransitivities}
\end{table}

Contrary to this prediction, we find that violations are
actually highest here (136; 8.8\%),
despite the fact that
triples D-B2-B1 and A1-B1-B2 each included
a pair without any dominance relation
(thereby increasing the cumulative decision difficulty
within the triple) and one with an ``approximate''
SOSD relation (denoted here by $\approx$ SOSD),
while A1-D-B2 either featured proper or approximate
SOSD relations in each of the three pairs.
Yet the violations-induced ordering between those three
triples are broadly in line with this intuition,
with the first (99; 6.4\%) followed by the second (97; 6.3\%),
which in turn is followed by the third (79; 5.1\%),
even though differences between consecutive triples
in this ranking are not significant.
Although the high incidence of intransitivities at triple
A1-D-B1 is somewhat puzzling, a possible explanation for it is
the postulated presence of risk-neutral subjects who,
by definition, are indifferent between any two lotteries
in the triple, and who might therefore reveal single-valued
choices in violation of Transitivity
(such cases are picked up by our indifference-inclusive
analysis of Section \ref{s4}). Another potential explanation, finally,
is that it emerges as a by-product of the significantly
declining yet persistently high
proportion of subjects violating StAR.\footnote{Forced-choice versions 
of Contraction Consistency and
WARP---called there \textit{Independence of Irrelevant Alternatives}
(IIA) and \textit{Consistency}---as well as of Transitivity,
Independence and FOSD, were also tested in the recent study by
\cite{nielsen-rehbeck22}. In these authors' main experiment,
subjects could first select which axioms they would like their choices to
comply with. Later, they were asked to make choices from lottery menus,
and were informed if their choices complied with the axioms they had
previously chosen. Following that, they were asked if
they wanted to retain this inconsistency or rectify it, either
by unselecting their chosen axiom or changing their chosen lotteries.
The proportions of subjects who opted for the latter
rectification were 78\%, 79\%, 66\%, 34\% and 29\% for the axioms
listed in the above order, respectively.
While the two designs differ in many important respects, one may juxtapose
the analogous---though not one-to-one so---percentages in our study.
More specifically, looking at the difference in the number of subjects
who violated each axiom in the first and fifth round as a percentage
of first-round violators, the corresponding ``rectifying'' proportions
of subjects in our feedback- and revision-free experiment are
47\%, 49,5\%, 53\%, 22\% and 57\%, respectively.
With the above caveat still in place, finally, we also remark that
broadly in line with the results from the first part of Nielsen and
Rehbeck's (\citeyear{nielsen-rehbeck22}) experiment that suggest
FOSD $\hat{\succ}$ WARP $\hat{\succ}$ Transitivity $\hat{\sim}$ Independence
$\hat{\sim}$ Contraction Consistency, our revealed preference analysis
(Table \ref{tab:first_last_15}) uncovers the---stable across
rounds---axiom-conformity
ordering FOSD $\tilde{\succ}$ Transitivity $\tilde{\succ}$ Independence
$\tilde{\succ}$ WARP $\tilde{\succ}$ Contraction Consistency.}

\begin{figure}[!htbp]
	\centering
	\caption{\centering
		Compliance gains with respect to FOSD, Contraction Consistency and WARP
		are materialized faster compared to those
		for Transitivity and StAR.\vspace{-30pt}}
	\includegraphics[width=\textwidth]{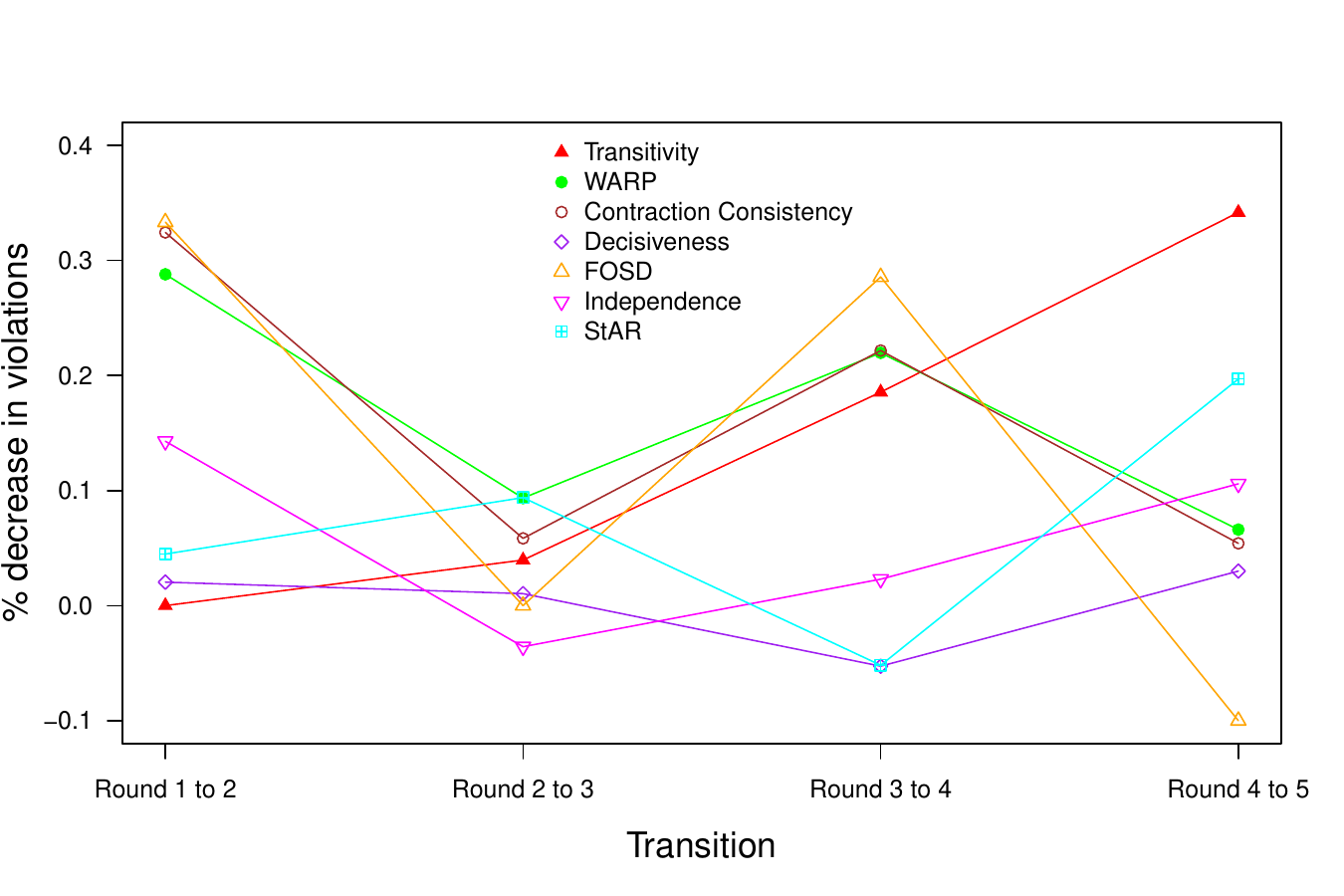}
	\label{fig:axiomViolations}
	\vspace{-30pt}
\end{figure}

We now take a closer look at the problem of understanding
which axioms subjects learn to comply with faster than others.
The last seven rows of Table \ref{tab:first_last_15} provide relevant
information from the extensive-margin perspective.
By itself, however, this is inadequate for two reasons.
First, it ignores information on the evolution of
violations' intensive margin, i.e. the absolute frequency
of each axiom's violations in each round.
Second, it does not account for the fact that different
axioms have distinct numbers of possibilities to be violated.
For example, while there is only one way in which Independence
might be violated in every round, simulations suggest that
there can be as many as 12 WARP or Contraction Consistency
violations.
While analysing neither the intensive nor the extensive margin
is helpful toward comparing which axioms' violations are
``learned away'' faster, we use the richer intensive-margin data
to make a standardized comparison by finding the
percentage decrease of each axiom's violations in the transition
from one round to the next.
Specifically, for which axioms does the sharpest such decrease
happen in the transition from the first to the second round,
and for which does it occur in the transition from the fourth
to the last round?
This analysis is summarized in Figure \ref{fig:axiomViolations}.
It reveals that FOSD, Contraction Consistency and WARP violations
see their largest percentage drops (33.3\%, 32.5\% and 29\%,
respectively) in the first transition, whereas those of
Transitivity and StAR (34\% and 20\%) in the last.
For Independence, on the other hand, the situation is less clear-cut
as there are two ``learning peaks'', occurring
in the first and last transition (14\% and 10.5\% violation decreases,
respectively). In line with the preceding analysis of subjects'
deferring behaviour, finally, there is no discernible decrease in
Decisiveness violations in any of the four transitions.

A related important question is: Which menus are more likely to be
involved in a choice inconsistency or, from the point of view of
economic rationality, a ``mistaken'' active choice?
The information presented Table \ref{tab:menusInconsistent} answers
this question by identifying 5 menus which, remarkably, were those
most frequently implicated in some pattern of choice
inconsistency in \textit{every} round.
More specifically, we used \Prests to compute, for each subject in each round,
all tuples of menus that were involved in a violation of \textit{Congruence} or,
practically synonymously, the \textit{Strong} Axiom
of Revealed Preference.\footnote{Such
tuples include those involved in a binary choice cycle;
a WARP violation; or some more general cyclical choice pattern such as
$C(\{p,q,r\})=\{p\}$; $C(\{q,r,s\})=\{q\}$; $C(\{r,s,p\})=\{s\}$).}
For each of the 15 distinct menus we then computed the number of subjects
for whom that menu was involved in at least one inconsistent choice
pattern in a given round. Table \ref{tab:menusInconsistent} summarizes
this information by ordering the---common across rounds---top-5 menus
according to its subject frequency at that round.
Menu $\{$A1, B1, B2, D$\}$ was most frequently implicated in
a choice inconsistency in rounds 2--4 (second-most in round 1).
Menu $\{$B1, B2, D$\}$ was first in round 1 and second in rounds 2-3, while
menu $\{$A1, D$\}$ was second in the last two rounds.
The other two menus on that list are $\{$A1, B1, B2$\}$ and $\{$B1, D$\}$.
These results are particularly interesting and provide further
evidence for the hypothesis that the presence or
absence of a dominant lottery in a menu determines at least partly
whether choice at that menu is easy or hard: with the
exception of $\{$B1, D$\}$, all menus on that list lack
a First- or Second-Order Stochastically Dominant lottery.
The analysis of deferring behaviour has already
demonstrated that menus of this kind are typically those where subjects
are also more likely to violate Decisiveness.
This analysis complements these findings and suggests that,
even for subjects who chose not to defer,
active choices at these menus were more likely to be fragile
and non-preference-revealing.\footnote{Although this generally
depends on the structure of a subject's active-choice graph,
a Houtman-Maks score decrease is typically seen when the menu
that is dropped from that person's dataset is one with the
highest frequency of involvement in an inconsistent choice pattern.}

\begin{table}[!htbp]
	\footnotesize
	\centering
	\caption{\centering The 5 menus that were most frequently
		implicated in a ``mistaken'' active choice per round
		(subject frequencies).}
	\setlength{\tabcolsep}{6pt} 
	\renewcommand{\arraystretch}{1.3} 
	\makebox[\textwidth][c]{
	\begin{tabular}{|lc|lc|lc|lc|lc|}
		\hline
		\multicolumn{2}{|c|}{\textbf{Round 1}} &
		\multicolumn{2}{c|}{\textbf{Round 2}} &
		\multicolumn{2}{c|}{\textbf{Round 3}} &
		\multicolumn{2}{c|}{\textbf{Round 4}} &
		\multicolumn{2}{c|}{\textbf{Round 5}}\\
		\hline
		B1,B2,D 	&159
		&A1,B1,B2,D	&133
		&A1,B1,B2,D	&124
		&A1,B1,B2,D	&109
		&A1,B1,B2,D	&90\\
		A1,B1,B2,D 	&152
		&B1,B2,D 	&124
		&B1,B2,D 	&117
		&A1,D		&94
		&A1,D 		&78\\
		B1,D 		&149
		&A1,D 		&119
		&A1,D 		&112
		&A1,B1,B2 	&92
		&B1,B2,D 	&74\\
		A1,D 		&136
		&A1,B1,B2 	&114
		&A1,B1,B2 	&105
		&B1,B2,D 	&92
		&A1,B1,B2	&70\\
		A1,B1,B2 	&135
		&B1,D 		&108
		&B1,D 		&99
		&B1,D 		&77
		&B1,D 		&62\\
		\hline
	\end{tabular}
}
\label{tab:menusInconsistent}
\end{table}

We conclude with the following finding which,
although of a different nature, resonates with and reverberates
the primary learning results presented earlier.

\begin{highlight}
Decisions in the last round are more than
twice as fast as in the first.
\end{highlight}

Indeed, the above-documented convergence to rational behaviour is accompanied
by a significant shift to the left
in the subjects' distributions of mean response times
between the first and last 15 decisions (20.7 vs 8.1 seconds per problem;
$p<0.001$).
This is an important finding because, considering also
the growing literature in support of the argument that decisions are
faster when preference comparisons are easier,\footnote{See
\citet*{alos-ferrer-fehr-netzer} and references therein}
it suggests that subjects who learned to maximize (expected)
utility in this experiment have done so while in the process of
discovering or constructing their (stable, complete and transitive)
\textit{preferences}. Moreover, the finding is important
for another reason: it suggests that the so-called
\textit{chronometric function}, which quantifies this postulated relationship
between response times and the relative easiness/difficulty of decisions, is
not stationary but changes with the agent's experience. This,
in turn, invites the analyst to exercise caution when defining and
interpreting the chronometric function in a given setting.

\section{Stability of Learning and Revealed-Preference Convergence}\label{s6}

\subsection{Learning and Strict Preferences}\label{s6-1}

The findings in the last section invite a natural follow-up question:
\textit{To what extent is learning \textnormal{stable} from one round to the next?}
More specifically, do subjects whose decisions in one round are
(E)UM-rational--after accounting for possible indifferences
in their overall behaviour---continue to be so in the next round?
The results presented in Table \ref{tab:learning-stability}
point to a positive answer. In particular, for both ordinal
and expected-utility maximization, and for each of the four possible
round transitions (i.e., from the first to the second etc.),
the majority of subjects exhibited \textit{stable learning}.
Moreover, the proportions
of such stable learners are increasing at similar rates from
the first to the fourth transition, from 80.5\% to 90\% for ordinal
and from 85\% to 90\% for expected-utility maximization.

\begin{table}[!htbp]
\centering
\footnotesize
\caption{\centering
Subjects who complied with ordinal and expected utility maximization in
one round and continued to do so in the next round*.}
\setlength{\tabcolsep}{6pt} 
\renewcommand{\arraystretch}{1.3} 
\begin{tabular}{|r|rl|rl|}
\hline
& \multicolumn{2}{c|}{\shortstack{\textbf{Ordinal Utility}\\ 
		\textbf{Maximization}}}
& \multicolumn{2}{c|}{\shortstack{\textbf{Expected Utility}\\
		\textbf{Maximization}}}\\
\hline
Round 1 to 2	
& \hspace{15pt} 91
& 80.5\%
& \hspace{20pt} 63
& 85\%\\
\hline
Round 2 to 3	
& 118
& 89\%
& 73
& 86\%\\
\hline
Round 3 to 4	
& 128
& 86.5\%
& 91
& 92\%\\
\hline
Round 4 to 5	
& 149
& 90\%
& 100
& 90\%\\
\hline
\end{tabular}

\scriptsize \*Subjects who were
(E)UM-rational throughout their 75 decisions under\\
weak preferences are also included.
\label{tab:learning-stability}
\end{table}

A closely related question is whether the transition from the first
to the fifth round, and the reported improvement in subjects' conformity
with rational choice, is also associated with learning their \textit{preferences}.
To assess this, we included an end-of-experiment question
that only showed up in subjects who had chosen different lotteries at
the same binary menu in the 1st and 5th times they were asked to choose
from it, and for every menu where such a reversal occurred.
Importantly, in order to prevent the occurrence of experimenter demand effects,
these questions and the corresponding menus were phrased and presented neutrally.
More specifically, the question stated the following:
\textit{\textit``At two different times when you saw this menu previously,
you chose different lotteries. Which one of the two lotteries
do you think you prefer more now, if any?''}
Similar to what they did in the main part of the experiment,
subjects' responded to these questions either by selecting
one of the two displayed lotteries, or by opting for
\textit{``I don't know''}
(instead of \textit{``I'm not choosing now''}).
No indication was given by experimenters on when the choice
discrepancies at those menus appeared or which lottery was
chosen earlier and which one later.

The summary of subjects' responses to these questions is
presented in Table \ref{tab:end-of-exp}, and separated
into three groups, depending on whether a FOSD, SOSD
or neither of these dominance relations exists between
the respective lotteries.
As expected, the lowest proportion of reversals occurred
at the two menus of the first type (17 in total), with
the majority (14) of affected subjects stating that they
preferred the (dominant) lottery corresponding to their
last choice (recall that they were not told which was
which).
For each of the 7 menus in the remaining 2 groups,
the majority of subjects (66 - 80\%) again stated
their last-chosen lottery as their preferred one
at that menu.
In line with our hypothesis, the highest rates of
\textit{``I don't know''} responses were seen in 3 of the 4 menus
where no stochastic-dominance relation exists (3 -- 5\%).
Two of these menus, namely \{B1,B2\}
and \{C1,C2\}, were also associated with the largest
numbers of choice reversals between those two rounds.
Across the three groups, the frequency-weighted means
for first-choice, last-choice or agnostic stated-preference
responses were 25.5\%, 71\% and 3.5\%.
These findings generally align with those
in the previous section and indicate that,
in the aggregate, the noted improvements in subjects' rationality
during the experiment were associated with their
learning of preferences along the way.	

\begin{table}[!htbp]
\centering
\footnotesize
\caption{\centering The ex-post stated preferences of
subjects who chose different lotteries at the same binary
menus in the 1st and 5th rounds.*}
\setlength{\tabcolsep}{6pt} 
\renewcommand{\arraystretch}{1} 
\begin{tabular}{|r||c|c||c|c|c||}
\hline
&\multirow{1.5}{*}{\textbf{A1,A2}}
&\multirow{1.5}{*}{\textbf{A2,D}}
&\multirow{1.5}{*}{\textbf{A1,B1}}
&\multirow{1.5}{*}{\textbf{A1,D}}
&\multirow{1.5}{*}{\textbf{B1,D}}
\\
&\tiny (FOSD) 	
&\tiny (FOSD)		
&\tiny (SOSD)		
&\tiny (SOSD)	
&\tiny (SOSD)
\\
\hline								
Lottery of first choice
&13.3\%
&50.0\%
&18.8\%
&37.5\%
&24.0\%
\\
\hline
Lottery of last choice						
&86.7\%
&50.0\%
&79.7\%
&62.5\%
&74.7\%
\\
\hline
\textit{``I don't know which one I prefer''}	
&0.00\%
& 0.0\%
& 1.5\%		
& 0.0\%	
&1.3\%	
\\
\hline
Total 1st/5th-round reversals at menu					
&15
&2
&69 		
&72
&75
\\
\hline
\end{tabular}
\label{tab:end-of-exp}
\vspace{-15pt}
\end{table}
\begin{table}[!htbp]
\centering
\footnotesize
\setlength{\tabcolsep}{6pt} 
\renewcommand{\arraystretch}{1} 
\makebox[\linewidth][c]{
\begin{tabular}{|r||c|c|c|c||}	
\hline
&\multirow{1.5}{*}{\textbf{A1,B2}}
&\multirow{1.5}{*}{\textbf{B2,D}}
&\multirow{1.5}{*}{\textbf{B1,B2}}
&\multirow{1.5}{*}{\textbf{C1,C2}}
\\
&\tiny  (``near SOSD'')	
&\tiny  (``near SOSD'')
&\tiny (SOSD-unranked)
&\tiny (SOSD-unranked)
\\
\hline
Lottery of first choice
&26.3\%	
&22.7\%				
&29.0\%				
&17.2\%
\\
\hline
Lottery of last choice
&73.7\%	
&72.7\%	
&66.3\%				
&79.6\%
\\
\hline
\textit{``I don't know which one I prefer''}
& 0.0\%	
& 4.5\%
&4.7\%				
&3.2\%
\\
\hline
Total 1st/5th-round reversals at menu
&38		
&44					
&86				
&93	
\\
\hline
\end{tabular}
}
\makebox[\textwidth][l]{\scriptsize *Relevant subjects were not
reminded which choice was made in which round.}
\end{table}	

\begin{figure}[!htbp]
\centering
\caption{\centering The most frequent strict
revealed preferences that are compatible with
EUM \linebreak (last 15 decisions).}
\begin{minipage}{0.3\textwidth}
\centering
\includegraphics[width=0.035\textheight]{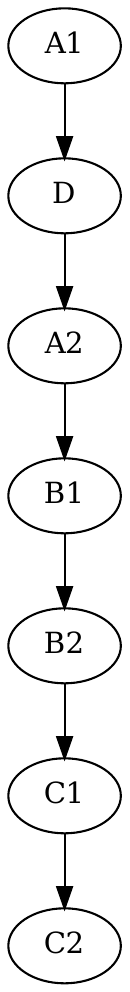}
\end{minipage}
\begin{minipage}{0.3\textwidth}
\centering
\includegraphics[width=0.035\textheight]{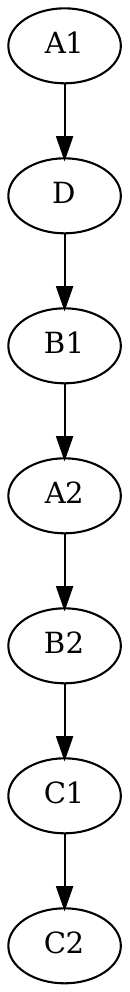}
\end{minipage}
\begin{minipage}{0.3\textwidth}
\centering
\includegraphics[width=0.035\textheight]{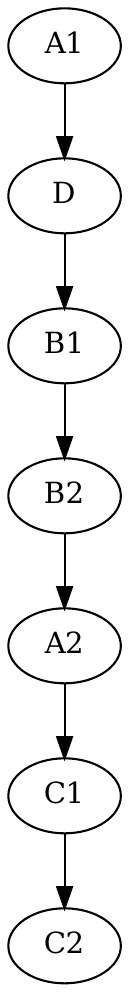}
\end{minipage}
\label{fig:pref-strict}
\end{figure}

We finally turn to the most frequent preference orderings
over lotteries that were compatible with the behaviour of those 123 subjects
who complied with expected utility maximization in their last 15 decisions.
To do this, we follow the same process that was explained in Section \ref{s4},
focusing here on strict preferences instead.
The three most common such orderings apply to 69 of the 123 EU-compliant subjects
(Figure \ref{fig:pref-strict}).
All three orders entail A1 $\succ$ D $\succ$ B1 and C1 $\succ$ C2,
differing only in how they rank B1, B2 and A2.
The first and last orderings that are displayed in Figure \ref{fig:pref-strict}
are the most and least risk-averse among them, respectively,
ranking A2---whose expected value is \pounds 11---above and below B1,
B2---both of which have expected value \pounds 12.

\section{The Role of Cognitive Ability}\label{s-ca}

The recent literature has documented a positive link between decision makers'
cognitive ability and their patience,
risk tolerance and proximity to rational behaviour, both in non-strategic
\citep{dohmen-falk-huffman-sunde10,becker-deckers-dohmen-falk-kosse12,
	dohmen-falk-huffman-sunde18,chapman_etal23,echenique-imai-saito23}
and strategic environments \citep{proto-rustichini-sofianos19,
	proto-rustichini-sofianos22,gill-prowse16,gill-rosokha23}.
Extending the investigation of such links,
we were interested to assess the potential role of
cognitive ability in our experimental subjects'
choice consistency and, additionally, their learning---or lack thereof---to 
maximize ordinal or expected utility.
We state from the outset that our analysis here is
exploratory.

To do so, as has already been noted we invited subjects to complete 
the ICAR-16 cognitive-ability questionnaire \citep{condon-revelle14}
after the main part of the experiment was over.
A cognitive ability score between 0 and 1 was constructed for every subject,
coinciding with the proportion of their correct answers.
In addition to the all-inclusive ICAR-16 score,
we also constructed in this way a variety of sub-scores that featured one
or more of the 4 blocks of questions
from the Letter-Numeric (LN), Verbal Reasoning (VR),
Matrix Reasoning (MR) and 3-Dimensional Rotation (3DR) sequences.

We find that the ICAR-16 score is significantly positively correlated with 
overall HM-consistency in subjects' merged choices over all rounds 
($\rho=-0.12$; $p=0.038$). Of the sub-measures, 
VR and LN are most predictive (VR: $\rho = 0.17$, $p = 0.002$; 
LN: $\rho = 0.11$, $p = 0.044$); 
their combination is even stronger ($\rho = 0.17$, $p = 0.003$).\footnote{More details
are available in Figure \ref{fig:ICARfigures} of O.A.\ref{a10}.}

\begin{table}[!htbp]
\footnotesize
\caption{\centering Relationships between cognitive ability and (not)
learning to maximize utility.}
\setlength{\tabcolsep}{5pt} 
\renewcommand{\arraystretch}{1.5} 
\makebox[\linewidth][c]{
\begin{tabular}{|l|c|c|c|c|c|c|c|c|}
\hline
& \multicolumn{8}{c|}{\textbf{Average ICAR Test Scores}*}\\
\cline{2-9}
&\multirow{1}{*}{\textbf{VR}}
&\multirow{2}{*}{\textbf{VR}}
&
&
&
&
&
&
\\
\multirow{3}{170pt}{\textbf{Subject groups under comparison}\\
	(group size in parenthesis)}
&\multirow{1}{*}{\textbf{LN}}
&\multirow{2}{*}{\textbf{LN}}
&\multirow{1}{*}{\textbf{VR}}
&\multirow{1}{*}{\textbf{LN}}
&\multirow{2}{*}{\textbf{VR}}
&\multirow{2}{*}{\textbf{LN}}
&\multirow{2}{*}{\textbf{3DR}}
&\multirow{2}{*}{\textbf{MR}}
\\
&\multirow{1}{*}{\textbf{3DR}}
&\multirow{2}{*}{\textbf{3DR}}
&\multirow{1}{*}{\textbf{LN}}
&\multirow{1}{*}{\textbf{3DR}}
&
&
&
& \\
&\multirow{1}{*}{\textbf{MR}}
&
&
&
&
&
&
&
\\
\hline
\multirow{3.2}{130pt}{\textbf{UM in first 15 decisions \\
		or throughout} (113)\\ \hspace{50pt} vs\\
	\textbf{UM in neither} (195)}
&\multirow{2.7}{30pt}{\centering 0.626\\ vs\\ 0.600}
&\multirow{2.7}{30pt}{\centering 0.697\\ vs\\ 0.664}
&\multirow{2.7}{30pt}{\centering 0.833\\ vs\\ 0.771}
&\multirow{2.7}{30pt}{\centering 0.580\\ vs\\ 0.557}
&\multirow{2.7}{30pt}{\centering 0.931\\ vs\\ 0.878}
&\multirow{2.7}{30pt}{\centering 0.735\\ vs\\ 0.664}
&\multirow{2.7}{30pt}{\centering 0.425\\ vs\\ 0.450}
&\multirow{2.7}{30pt}{\centering 0.412\\ vs\\ 0.409} \\
&\multirow{3.4}{*}{\tiny $p=0.271$}
&\multirow{3.4}{*}{\tiny $p=0.230$}
&\multirow{3.4}{*}{\tiny $p=0.009$}
&\multirow{3.4}{*}{\tiny $p=0.558$}
&\multirow{3.4}{*}{\tiny $p=0.004$}	
&\multirow{3.4}{*}{\tiny $p=0.103$}
&\multirow{3.4}{*}{\tiny $p=0.510$}
&\multirow{3.4}{*}{\tiny $p=0.776$} \\
&
&
&
&
&
&
&
&
\\
\hline
\multirow{3.2}{130pt}{\textbf{EUM in first 15 decisions\\
		or throughout} (87)\\
	\hspace{50pt} vs\\ \textbf{EUM in neither} (221)}
&\multirow{2.7}{30pt}{\centering 0.62\\ vs \\ 0.605}
&\multirow{2.7}{30pt}{\centering 0.685\\ vs\\ 0.673}
&\multirow{2.7}{30pt}{\centering 0.826\\ vs\\ 0.781}
&\multirow{2.7}{30pt}{\centering 0.557\\ vs\\ 0.568}
&\multirow{2.7}{30pt}{\centering 0.94\\ vs\\ 0.881}
&\multirow{2.7}{30pt}{\centering 0.713\\ vs\\ 0.681}
&\multirow{2.7}{30pt}{\centering 0.402\\ vs\\ 0.456}
&\multirow{2.7}{30pt}{\centering 0.425\\ vs\\ 0.404}\\
&\multirow{3.4}{*}{\tiny $p=0.605$}
&\multirow{3.4}{*}{\tiny $p=0.709$}
&\multirow{3.4}{*}{\tiny $p=0.067$}
&\multirow{3.4}{*}{\tiny $p=0.712$}
&\multirow{3.4}{*}{\tiny $p=0.002$}
&\multirow{3.4}{*}{\tiny $p=0.517$}
&\multirow{3.4}{*}{\tiny $p=0.247$}
&\multirow{3.4}{*}{\tiny $p=0.302$}\\
&
&
&
&
&
&
&
&
\\
\hline
\multirow{4.2}{160pt}{\textbf{Non-UM in first 15 or throughout}
	\& \textbf{UM in last 15} (78)\\ \hspace{50pt} vs\\
	\textbf{UM in neither first or last 15 or throughout} (117)}
&\multirow{3.7}{30pt}{\centering 0.632\\ vs\\ 0.579}
&\multirow{3.7}{30pt}{\centering 0.706\\ vs\\ 0.636}
&\multirow{3.7}{30pt}{\centering 0.804\\ vs\\ 0.749}
&\multirow{3.7}{30pt}{\centering 0.614\\ vs\\ 0.519}
&\multirow{3.7}{30pt}{\centering 0.891\\ vs\\ 0.870}
&\multirow{3.7}{30pt}{\centering 0.718\\ vs\\ 0.628}
&\multirow{3.7}{30pt}{\centering 0.510\\ vs\\ 0.410}
&\multirow{3.7}{30pt}{\centering 0.410\\ vs\\ 0.408}\\
&\multirow{5}{*}{\tiny $p=0.036$}
&\multirow{5}{*}{\tiny $p=0.026$}
&\multirow{5}{*}{\tiny $p=0.129$}
&\multirow{5}{*}{\tiny $p=0.021$}
&\multirow{5}{*}{\tiny $p=0.503$}
&\multirow{5}{*}{\tiny $p=0.178$}
&\multirow{5}{*}{\tiny $p=0.098$}
&\multirow{5}{*}{\tiny $p=0.810$}
\\
&
&
&
&
&
&
&
&\\
&
&
&
&
&
&
&
&\\
\hline
\multirow{4.2}{170pt}{\textbf{Non-EUM in first 15 or throughout}
	\& \textbf{EUM in last 15} (61)\\ \hspace{50pt} vs\\
	\textbf{EUM in neither first or last 15 or throughout} (160)}
&\multirow{3.7}{30pt}{\centering 0.632\\ vs\\ 0.595}
&\multirow{3.7}{30pt}{\centering 0.714\\ vs\\ 0.657}
&\multirow{3.7}{30pt}{\centering 0.816\\ vs\\ 0.768}
&\multirow{3.7}{30pt}{\centering 0.631\\ vs\\ 0.545}
&\multirow{3.7}{30pt}{\centering 0.881\\ vs\\ 0.881}
&\multirow{3.7}{30pt}{\centering 0.75\\ vs\\ 0.655}
&\multirow{3.7}{30pt}{\centering 0.512\\ vs\\ 0.434}
&\multirow{3.7}{30pt}{\centering 0.385\\ vs\\ 0.411}
\\
&\multirow{5}{*}{\tiny $p=0.183$}
&\multirow{5}{*}{\tiny $p=0.103$}
&\multirow{5}{*}{\tiny $p=0.319$}
&\multirow{5}{*}{\tiny $p=0.045$}
&\multirow{5}{*}{\tiny $p=0.643$}
&\multirow{5}{*}{\tiny $p=0.153$}
&\multirow{5}{*}{\tiny $p=0.209$}
&\multirow{5}{*}{\tiny $p=0.462$}
\\
&
&
&
&
&
&
&
&
\\
&
&
&
&
&
&
&
& \\
\hline
\multicolumn{9}{l}{\scriptsize *Each of VR, LN, 3DR and MR refers
	to the respective subset of 4 items in the ICAR-16 test
	that includes Verbal Reasoning, }\vspace{-6pt}\\
\multicolumn{9}{l}{\scriptsize \color{black} Letter-Numeric,
	3-Dimensional Rotation and Matrix Reasoning questions,
	respectively. The different columns
	report average scores}\vspace{-6pt}\\
\multicolumn{9}{l}{\scriptsize \color{black} and 2-sided
	Mann-Whitney $U$ test $p$-values for the
	corresponding combinations of questions.}
\end{tabular}
}
\label{tab:ICARlearning}
\end{table}

We also investigated potential associations between
cognitive ability and learning.
The resulting learning patterns, summarised in
Table \ref{tab:ICARlearning}, reinforce the picture that 
was outlined above. 
Subjects who either maximized utility early or consistently 
over all 75 choices score significantly higher on VR and VR-LN 
than those who never did. Late learners, i.e. those who were 
initially inconsistent but utility maximizing by the final 15 choices, 
also have higher ICAR-16, VR-LN-3DR and LN-3DR scores, suggesting 
a link between ability and eventual convergence. 
These patterns are qualitatively the same using expected-utility maximization, 
though differences are not always statistically significant.

\begin{highlight}
	Subjects who are ordinal or expected-utility maximizers
	in the first round or throughout tend
	to have a higher cognitive ability than those who are not,
	particularly in the (individual or combined)
	verbal reasoning and letter-numeric scores.
 	Moreover, subjects who learn to be so by the
 	last round tend to have a higher cognitive
 	ability than those who do not, particularly in the combined
 	letter-numeric and 3-D rotation score.
 \end{highlight}

Estimates from probit models, reported in Table \ref{tab:um_determinants}, 
confirm the result that more cognitively able individuals are more likely to 
be utility- or expected-utility maximizers in every round. 
In particular, the verbal-reasoning scores are significant. 
The regression results also reveal that relative to humanities students 
(base category in the probit models), those studying economics and business, 
computer science, mathematics, psychology, or law are more likely to be 
utility maximizers of any kind 
(see Table \ref{tab:um_determinants_all_controls} in Online Appendix \ref{a2} 
for the full set of coefficient estimates).
All of these results are robust to restricting the sample to subjects
who never deferred a choice at any menu [columns (2) and (4)], 
and also to not controlling for level of study and field of study 
(see Table \ref{tab:um_determinants_basic_controls} in Online Appendix \ref{a2}). 
Males tend to be more likely to make choices consistent with utility 
maximization than females, corroborating the evidence from the raw data 
that shows that males have slightly lower HM scores than females
in each round (1.24 vs 1.55; 1.02 vs 1.3; 0.89 vs 1.23; 
0.73 vs 1.12; 0.63 vs 1)\footnote{All round-per-round differences
except the first one are significant at the 5\% level.} and overall 
(1.74 vs 2.13; $p=0.079$).
In addition, a higher proportion of males were ordinal or
expected-utility maximizers
(30\% vs 24\% and 20\% vs 16\%), but insignificantly so
($p=0.286$ and $p=0.356$, respectively).

Importantly, and consistent with Table \ref{tab:first_last_15}, 
the number of utility maximizers and expected utility maximizers 
increases significantly from round to round. 
An analysis of response times---summarized in Table \ref{tab:response_times}
in O.A.B---aligns with, and bolsters, the findings on learning 
from round to round.
In particular, response times become shorter from round to round, and they
are shorter for individuals who have a higher verbal reasoning score.
Response times are also significantly shorter for utility maximizers---and
particularly shorter for expected utility maximizers.
Notably, response times are longer at more difficult and more complex menus.
Deferring a choice also significantly reduces response times at that menu,
indicating that individuals who defer spend less time and effort to think
about the choice. Interestingly, while this is not the case
in the first 15 decisions, during which the average response times were
similar for deferrals and active choices
(21.6 vs 20.6 seconds, respectively; $p=0.125$ from 2-sided Mann-Whitney test),
it does become so in the last 15 decisions (5.57 vs 8.2 seconds; $p<0.001$).
These facts suggests that subjects spend a similar amount of time at both
kinds of decisions initially, but quickly learn which menus are hard
for them and then spend little time before opting for the deferral
option when they see those menus again. Also interestingly,
this response-time reduction when deferring a choice
is higher for more cognitively able individuals, suggesting that they use
deferring more efficiently and decide to defer a choice more quickly to
save cognitive effort.

\begin{table}[!htbp]
	\centering
	\caption{Utility Maximizers at all menus:
		Probit marginal effects}\label{tab:um_determinants}
	\footnotesize
	\begin{tabular}{lcccc} \hline\hline
		& \multicolumn{4}{c}{Dependent variable:}\\
		\cline{2-5}
		& \multicolumn{2}{c}{1 if U-maximizer across 15 menus} &
		\multicolumn{2}{c}{1 if EU-maximizer across 15 menus}\\
		& (1) & (2) & (3) & (4) \\
		\cline{2-5}
		1 if round 2 & 0.082*** & 0.090** & 0.042 & 0.041 \\
		& (0.031) & (0.036) & (0.032) & (0.041) \\
		1 if round 3 & 0.132*** & 0.127*** & 0.093*** & 0.081** \\
		& (0.033) & (0.038) & (0.033) & (0.041) \\
		1 if round 4 & 0.194*** & 0.209*** & 0.135*** & 0.156*** \\
		& (0.033) & (0.036) & (0.035) & (0.043) \\
		1 if round 5 & 0.247*** & 0.270*** & 0.176*** & 0.194*** \\
		& (0.031) & (0.032) & (0.034) & (0.041) \\
		Verbal reasoning score & 0.463*** & 0.521*** & 0.269* & 0.284 \\
		& (0.162) & (0.182) & (0.149) & (0.180) \\
		Letter-numeric sequence score & 0.062 & 0.149 & 0.045 & 0.084 \\
		& (0.081) & (0.096) & (0.074) & (0.091) \\
		Matrix reasoning score & 0.085 & 0.011 & -0.014 & -0.056 \\
		& (0.123) & (0.158) & (0.105) & (0.139) \\
		3-dimensional rotation score & -0.045 & -0.095 & -0.054 & -0.100 \\
		& (0.066) & (0.076) & (0.059) & (0.069) \\
	
        Joint significance (ICAR) & 0.0315 & 0.0165 & 0.302 & 0.223 \\

	     Additional controls & yes & yes & yes & yes \\\hline

		&  &  &  &  \\
		Observations & 1,540 & 1,130 & 1,540 & 1,130 \\ \hline
		\hline
	\end{tabular}
	\begin{minipage}[t]{\textwidth}\scriptsize \raggedright
		\textbf{Notes:} Marginal effect estimates of probit models
		evaluated at the means of independent variables.
		Robust standard errors clustered at the subject level in parentheses.
		*** $p<0.01$, ** $p<0.05$.\\
		In the first two columns, the dependent variable is an indicator
		variable that equals one if a subject made choices in the 15 menus
		of a round that are consistent with ordinal utility maximization.
		In column (1) all subjects are included, in column (2) the sample
		is restricted to subjects who never deferred at any of the 75 menus.
		In columns (3) and (4), the dependent variable is an indicator variable
		that equals one if a subject made choices in the 15 menus of a round
		that are consistent with expected utility maximization.
		In column (3) all subjects are included, in column (4) the sample
		is restricted to subjects who never deferred at any of the 75 menus. 
		Indicator variables for gender (female, non-binary), level of study 
		(undergraduate, master, PhD) and field of study are included 
		in all regressions.
	\end{minipage}
\end{table}

To explore how cognitive ability is related to particular violations 
of utility maximization, we estimated a set of six probit models in which 
the dependent variable is an indicator variable that equals 1 if 
the choices in a round of 15 menus violate FOSD, Independence,
WARP, Contraction Consistency, and Decisiveness, respectively.
Table \ref{tab:violations} reports marginal effects evaluated at the 
means of independent variables, and reveals that higher cognitive ability, 
and specifically better verbal reasoning ability, reduces WARP and 
Contraction Consistency violations in particular, as the $p$-values 
of a Wald-Test for the joint significance of the four ICAR4 scores 
and the Romano-Wolf $p$-values that correct for multiple hypothesis 
testing of the six outcomes and across the four ICAR4 variables confirm.\footnote{See 
Table \ref{tab:violations_all_controls} in Appendix \ref{a2} for coefficient 
estimates of the additional controls in the probit models reported in 
Table \ref{tab:violations}.} These results are robust do not controlling for level of study, 
and field of study (see Table \ref{tab:violations_basic_controls} in Appendix \ref{a2}).

The estimates also clearly demonstrate that the probability of any violation is lower
in rounds 2 to 5 than in round 1.
First-Order Stochastic Dominance violations
become particularly less likely from round 1 to round 2,
while violations of Independence become particularly less
likely in the last round.
WARP and Contraction Consistency violations, on the other hand,
become steadily less likely from round to round.
Consistent with the findings in Table \ref{tab:first_last_15},
violations of Decisiveness are not statistically significantly
different across rounds.

\begin{table}[!htbp]
\centering
\caption{Violations of Utility Maximization at All Menus:
	Probit Marginal Effects}
\footnotesize
\begin{tabular}{lcccccc} \hline\hline
	& \multicolumn{6}{c}{Dependent variable: 1 if violating (across 15 menus)}\\
	& FOSD  & Indepen- & StAR  & WARP & Contraction & Decisive- \\
	&       & dence    &            &      & consistency & ness \\
	& (1) & (2) & (3) & (4) & (5) & (6) \\
	\cline{2-7}
	&  &  &  &  &  &  \\
	1 if round 2 & -0.015** & -0.043 & -0.021 & -0.120*** & -0.127*** & -0.013 \\
	& (0.007) & (0.032) & (0.034) & (0.030) & (0.030) & (0.018) \\
	1 if round 3 & -0.014** & -0.034 & -0.052* & -0.165*** & -0.150*** & -0.011 \\
	& (0.006) & (0.032) & (0.031) & (0.031) & (0.032) & (0.018) \\
	1 if round 4 & -0.021*** & -0.040 & -0.037 & -0.222*** & -0.227*** & -0.008 \\
	& (0.006) & (0.034) & (0.032) & (0.030) & (0.030) & (0.018) \\
	1 if round 5 & -0.019*** & -0.069** & -0.103*** & -0.291*** & -0.293*** & -0.014 \\
	& (0.006) & (0.031) & (0.030) & (0.027) & (0.027) & (0.018) \\
Verbal reasoning score & -0.005 & 0.053 & -0.232* & -0.532*** & -0.499*** & -0.031 \\
 & (0.025) & (0.110) & (0.125) & (0.156) & (0.157) & (0.112) \\
 & [0.965] & [0.965] & [0.370] & [0.025] & [0.040] & [0.965] \\
Letter-numeric sequence score & -0.027** & -0.041 & -0.013 & -0.067 & -0.057 & 0.038 \\
 & (0.013) & (0.057) & (0.067) & (0.081) & (0.080) & (0.056) \\
 & [0.215] & [0.895] & [0.895] & [0.855] & [0.895] & [0.895] \\
Matrix reasoning score & 0.002 & 0.029 & 0.133 & 0.001 & -0.020 & -0.102 \\
 & (0.022) & (0.080) & (0.090) & (0.121) & (0.118) & (0.089) \\
 & [1.000] & [1.000] & [0.580] & [1.000] & [1.000] & [0.580] \\
3-dimensional rotation score & 0.008 & 0.055 & 0.080 & 0.079 & 0.052 & -0.058 \\
 & (0.014) & (0.046) & (0.052) & (0.065) & (0.065) & (0.047) \\
 & [0.705] & [0.695] & [0.515] & [0.695] & [0.705] & [0.695] \\
Joint significance (ICAR) & 0.346 & 0.715 & 0.0639 & 0.00861 & 0.0225 & 0.364 \\ \hline
 &  &  &  &  &  &  \\
Additional controls	&   yes & yes & yes  &  yes & yes & yes   \\
\hline
	&  &  &  &  &  &  \\
	Observations & 1,375 & 1,540 & 1,540 & 1,540 & 1,540 & 1,540 \\
	\hline\hline
\end{tabular}
\begin{minipage}[t]{\textwidth}\scriptsize \raggedright
	\textbf{Notes:} Marginal effect estimates of probit models evaluated
	at the means of independent variables.
	Robust standard errors clustered at the subject level in parentheses.
	*** $p<0.01$, ** $p<0.05$. Romano-Wolf p-values corrected for multiple 
	hypothesis testing in brackets.\\
	The dependent variable in column (1) is an indicator variable that
	equals one if a subject makes at least one first-order
	stochastically dominated choice. Likewise the dependent variables
	in columns (2) to (6) are indicator variables that equal 1 if a
	subject's choices in a round of 15 menus violate Independence, StAR,
	WARP, Contraction Consistency, and Decisiveness, respectively.
\end{minipage}
\label{tab:violations}
\end{table}

\section{Related Literature}\label{s7}

Although, to our knowledge, no previous study has raised a similar
set of questions or reported an analogous set of results,
we note that \cite{hey01}, \cite{vandekuilen-wakker06},
\cite{birnbaum-schmidt15}, \cite*{nicholls-romm-zimper15}, 
and \cite*{spicer-mullett-sanborn24}\footnote{The latter
two works were pointed out to us during
the paper's peer-reviewed process.}
have also tested aspects of the learning question that constitutes
the main focus of our paper.
Hey's (\citeyear{hey01}) experiment included
53 subjects who, over the course of 5 experimental
sessions, were shown five times the same 100 binary menus
of lotteries with two outcomes.
The five sessions were conducted on different days and
with no less than two days between them, thereby enabling subjects to
acquire information and experience outside the lab environment.
Leaving aside the significant differences in motivation,
design and sample sizes between that study and ours,
no evidence that subjects learned to behave rationally over time
was provided in that paper.
\cite{vandekuilen-wakker06} reported on a repeated-choice experiment under
risk with two treatments and 52 student subjects of various levels
and fields, who could either learn ``by experience and by thought''
(in this treatment subjects' played their chosen lottery after each
decision\footnote{Recent surveys of the literature
on learning by experience in risky choice are \cite{hertwig-erev09}
and \cite{erev-haruvy16}.})
or ``only by thought''. The 26 participants in each treatment made
decisions in two trial and fifteen actual rounds from two binary menus
of money lotteries with two outcomes that featured ``common-ratio''
types of tests of Independence.
Importantly, and unlike our study, the lotteries in the two menus
differed in each round.
The authors found that the aggregate behaviour resulting from
subjects' two decisions tended to converge to expected utility
maximization in the dual-learning treatment but not in the
``only by thought'' one.

Following an approach which, in their own words,
is a synthesis of \cite{hey01} and \cite{vandekuilen-wakker06},
\cite{birnbaum-schmidt15} recruited 54 mainly economics and business
undergraduate student subjects and presented them
four times with the same 20 binary menus of money lotteries.
These menus were designed to test \textit{Coalescing}
(splitting vs non-splitting an outcome's probability
should not alter choices between otherwise identical lotteries),
Independence (in their case, the ``common-ratio'' and ``common-consequence''
implications thereof), and risk-attitude inconsistencies
(manifested in that study when choices between the same risky
and safe lotteries are reversed). The authors found evidence of ``by thought''
learning in all three dimensions, as evidenced by the significant
decrease in the respective total violations.\footnote{A similar
result concerning reduced violations of the Sure-Thing Principle
	of preferences under uncertainty is the main finding in the study by
	\cite{nicholls-romm-zimper15}.}
In the adaptation of Hey's (\citeyear{hey01}) experiment by
	\cite{spicer-mullett-sanborn24}, finally,
	the authors found that their 166 online subjects' choices
	from 50 pairs of lotteries, repeated four times without feedback,
	reduce choice noise but not adherence to expected utility,
	possibly because of subjects' increasing reliance on simplistic heuristics
	as the experiment progressed.

Compared to these earlier studies, ours differs in several important ways:
(i) it features \textit{free choices}; (ii) presents both binary and non-binary menus;
(iii) was designed to test several implications of deterministic ordinal-
and expected-utility maximization that
go beyond the Independence axiom;
(iv) uses rather involved computational methods to assess with precision
each subject's conformity with ordinal- and expected-utility maximization,
as well as with each one of several behavioural axioms implied
by them;
(v) has 6 times as large a sample than the lab-based studies
	by \cite{hey01,vandekuilen-wakker06,birnbaum-schmidt15} and nearly twice
	as large as the online one by \cite{spicer-mullett-sanborn24};
(vi) analyses the data both at the individual and aggregate levels; (vii)
accounts for potential indifferences in the former type of analysis,
and quantifies the importance of doing so; (viii)
finds significant evidence of introspective learning
in what we believe were considerably more challenging decision environments;
and
(ix) relates subjects' overall consistency and (non-) learning to cognitive ability.

Our paper also differs significantly from the recent studies by
\cite{ert-haruvy17},
\cite{charness-chemaya-trujano-ochoa23,charness-chemaya23},
\cite{nielsen-rehbeck22}, \cite{benjamin-fontana-kimball21},
and \cite{breig-feldman24}. A common feature in the latter three
works is the finding that subjects show a higher conformity with principles of
rational choice under risk when they are asked if they want to
revise their previous choices, with or without receiving
relevant feedback prior to the revision opportunity.
Unlike our experiment, the ones in these studies were
not designed to test for the existence and evolution
of subjects' neutral learning (i.e., only ``by-thought'')
to be (expected-)utility maximizers, either with or without
also accounting for the possibility of them being occasionally indifferent.
On the other hand, \cite{ert-haruvy17} (60 subjects)
and, concurrent to our own study,
\cite{charness-chemaya-trujano-ochoa23} and \cite{charness-chemaya23} (99 subjects)
studied stability of risk preferences
using several rounds---with feedback and trialling between rounds---of
the multiple price list format due to \cite{holt-laury02}
and a variation of the one-menu-with-six-lotteries method due to
\cite{eckel-grossman02}, respectively. All lotteries in these studies
featured two monetary outcomes. The first study found generally
unstable preferences between rounds, and subjects' tendency to become more risk
neutral with experience. \cite{charness-chemaya23}, analysing the data from
\cite{charness-chemaya-trujano-ochoa23}, also found more than 50\% of subjects
changing their choices between the first and last decisions that preceded
and succeeded the intermediating trial/feedback rounds.
Consistent with the main message that comes from our own analysis that builds on
substantially different experimental procedures and data-analytic methods,
no feedback, free choices,
and more than three times their sample size, \cite{charness-chemaya23}
highlight the importance of providing participants with experience
in risk-elicitation tasks towards reducing preference-measurement errors.

Finally, although this is outwith the focus of our study,
we note that there is an extensive body of research that
documents subjects' learning to conform with various types of
equilibrium predictions in strategic games as they become
more experienced. In ultimatum bargaining, for example,
\cite{slonim-roth98}, followed by \cite{list-cherry00} and
\cite{grimm-mengel11}, among others, find evidence
in favour of such learning by proposers and responders, respectively.

\section{Summary and Discussion}\label{s8}

Our aim in this study has been to conduct a detailed and targeted
investigation of behaviour in experimental
choices under risk when all decision problems are presented to subjects
multiple times (in our case, five) and in a carefully structured
semi-random order.
The lotteries and decision problems were designed
to test subjects' conformity with a variety of ordinal-
and expected-utility maximization principles, including ones that
pertain to binary menus only, as well as ones that require revealed-preference
consistency across binary and/or non-binary menus.
The data collected from 308 subjects who participated in our
free-/non-forced choice experiment in the UK and in Germany
allowed us to carry out rather comprehensive and, we hope,
illuminating tests of some important questions on choice under risk:
\begin{enumerate}
\item Can proper accounting for the possibility of subjects' being
indifferent between some lotteries partly explain the often observed
choice reversals across different instances where the same menu
was presented to them?
\item Do subjects learn to maximize ordinal or expected utility
as they progress in the experiment, without receiving any feedback
or invitations to revise their previous choices,
and without being forced to make active choices?
\item Which principles of utility maximization appear to be the
hardest for subjects to conform with?
\item What is the relation between cognitive ability
and utility maximization or learning?
\end{enumerate}

Our analysis pertaining to the first question uncovers
that choice reversals between different presentations
of the same decision problem may often stem from the
decision makers' rational indifference
between the respective choice alternatives.
More specifically, 15\% of all subjects in our sample
exhibited behaviour across their 75 decisions that is
perfectly consistent with indifference-revealing ordinal
utility maximization, with half of them also being
perfectly consistent with expected-utility maximization.
Without testing for the indifference hypothesis, this sizeable
proportion of subjects would have been discarded as non-rational.
While this possibility had been acknowledged in the related
literature as early as \cite{davidson-marschak59},
our study appears to be the first to document
and quantify it, and does so using a sophisticated
combinatorial-optimization method that is freely accessible.

In response to the second question, our analysis suggests that
a substantial fraction of participants in choice experiments
can learn introspectively, i.e.
without any feedback or other interventions,
to conform in a strict sense with the benchmark models of
economic rationality when they are repeatedly
exposed to the same decision problems,
even when several of these problems involve relatively high
degrees of decision difficulty.
In our data, those fractions were 23\% and 12\% for
ordinal- and expected-utility maximization between the first
and last decision round, respectively, representing highly
significant increases in theory-abiding subjects.
This conclusion is in contrast to those
in some---and in line to those in others---pre-existing
studies that explored similar themes using more constrained
analytical methods and experimental designs, smaller sample sizes
and, typically, fewer repetitions of the same menus.

As far as specific rationality principles are concerned,
subjects' violations of all axioms except Decisiveness exhibited a steady
decline over the course of the experiment. Notably, however,
despite there being only 1 and 3 pairs of menus, respectively, where
violations of the Independence and Stability of Attitudes to Risk
(StAR) axioms could be observed in any one round (as opposed to 20 such
pairs for the Contraction Consistency axiom, for example), 25\% and 26\%
of all subjects still violated the strict-preference versions
of these axioms in the fifth round.
Unlike many tests of Independence in the literature that revolve
around patterns inspired by the Allais-paradox, ours
does not involve probabilities near or equal to one or zero
in any of the lotteries involved.
Instead, it features the novelty whereby each of the relevant
two binary menus contains lotteries not ranked
by SOSD, hence representing
a decision with difficult trade-offs.

On the other hand, analysing subjects' deferring behaviour
(equivalently, conformity with the Decisiveness axiom) 
reveals that:
(i) such decisions are made by a sizeable minority of
15-16\% participants; (ii) they are persistent across
rounds and relatively predictable in their occurrence; 
(iii) in line with intuition and the decision-difficulty
theoretic channel to choice avoidance/deferral that
has been suggested and documented in the existing literature,
they tend to occur at menus that are more complex than others,
either because they lack a (first-/second-order) stochastically
dominant lottery or because they present information in a complicated way.
Considering that deferring in this experiment comes with a positive
expected cost for subjects, the ensuing violations of the Decisiveness
axiom cannot be attributed to indifference but would be better thought of
as being caused by incomplete or imprecise preferences,
or complexity-aversion considerations.

Introducing and testing StAR (Proposition \ref{prp:StAR}) in our
specific environment, moreover, enables us to uncover patterns
of risk-attitude reversals that go beyond
those in the classic ``reflection effect'' \citep{tversky-kahneman81}
where the said reversals are mediated
by the framing of decision in terms of gains or losses.
More specifically, testing StAR in our framework amounts to checking whether
a decision maker always opts for the second-order stochastically
dominant (risk-averse) or dominated (risk-seeking) lottery
at every collection of pairs that contain such a lottery.
The vast majority of subjects respect this axiom, and do so
in the direction of exhibiting consistent risk aversion.
We acknowledge the possibility that those who violate it
in any given round do so because they are risk-neutral,
a hypothesis that cannot be definitively rejected in an experiment
with single-valued choice data from individual rounds.
Moreover, the relatively limited variation in the expected values
of the lotteries in our specific experiment poses an obstacle toward testing
for the possibility that some subjects start comparing
lotteries by looking at their expected values before turning
to other criteria.
The risk-neutrality hypothesis, however, \textit{can} be tested in the
indifference-permitting analysis where the possibly distinct
single choices per menu across different rounds are merged.
There, risk neutrality would be detected by subjects'
revealed indifference in \textit{every} binary menu that contained
SOSD-ranked lotteries.
Yet only one expected-utility maximizing subject 
was revealed to be risk-neutral.
In light of this fact, StAR violations suggest
that context-dependent risk attitudes exist even
without gain/loss framing.
In particular, they also emerge when SOSD-ranked
lotteries feature the same or similar three or more outcomes.
This, in conjunction with our other findings,
points to a potentially relevant descriptive role for the development
of theories of choice under risk that allow for similarity-based reversals
in attitudes to risk while simultaneously respecting FOSD 
and the basic axioms of ordinal utility theory.

Finally, concerning the role of cognitive ability,
our results indicate that it is not only related
to overall choice consistency,
but also plays a predictive role in individuals'
capacity to autonomously adapt towards rational
decision-making over the course of the experiment.
In particular, we find an important
association between choice consistency and learning
on the one hand and the verbal-reasoning and letter-numeric tasks
of the administered ICAR-16 on the other.
Assuming that the latter causes the former, which is something that
we are clearly unable to test, this suggests that efforts to improve
people's decision-making quality in the real-world, for example
their financial literacy, can benefit from the inclusion of
not only numerical problems but also of verbal ones.

The clear presence of feedback-independent learning in our data carries
important implications for experimental design, theory testing
and preference elicitation. Indeed, it suggests that
in conducting choice experiments
or surveys where participants encounter the same scenarios repeatedly,
focusing on subjects' decisions in the
final instance of these scenarios,
and properly accounting for the possibility
of indifferences across all their decisions,
could yield significantly
more accurate information about
the subjects' underlying decision process and preferences,
whether these were ``discovered'' \citep{plott96}
or ``constructed'' \citep{kahneman96}.
This postulated more accurate elicitation could
in turn---as in our study--- paint a relatively more
favourable picture of the baseline models
of economic rationality as descriptive theories of
choice under risk than what is often inferred.
Indeed, combining our answers to the first two questions above,
as these are summarized in the relevant entries of Tables \ref{tab:merged}
and \ref{tab:first_last_15}, leads to the conclusion
that 51\% and 36\%, of all subjects, respectively,
are revealed to be ordinal or expected-utility maximizers
either across all 75 decisions or in their last 15 ones,
with 6.5\% and 3\% of the indifference-revealing subjects in these
groups breaking their indifference ties in the last round.
Of course, a natural question is, which mechanisms underlie 
the observed reduction in violations of axioms across repeated presentations? 
Our design does not provide feedback and 
thus does not support learning in the sense of belief updating or reinforcement. 
Instead, several non-feedback mechanisms may plausibly contribute to increased 
consistency over time. After encountering the same lotteries multiple times, 
some subjects may adopt simplified decision rules that reduce the likelihood 
of violating axioms. Our interpretation of convergence to behaviour 
consistent with utility maximization, i.e. what we refer to as learning, 
is intentionally mechanism-agnostic: the evidence shows that behaviour 
becomes more consistent with models of rational choice, but we do not claim 
that this arises from a single source or that it reflects preference discovery 
in a strong sense.

While our experiment is not directly comparable to any pre-existing
one that we are aware of, it is probably fair to say
that such degrees of conformity with those two models 
is considerably higher than what one might
have anticipated in the decision environment of our experiment.
One may wonder if this is could be partly
driven by subjects' possible desire to be consistent
once they start to recognize patterns across menus.
Although such an interpretation might be relevant
for the results from the online experiment by \cite{spicer-mullett-sanborn24}
discussed earlier, where choice noise was reduced
without subjects' eventual adherence to expected-utility maximization,
such a desire could not possibly be the main explanation behind
the primary findings of our study. Indeed,
if a preference for consistency was significant among our subjects,
then we should not observe the kinds of UM- and EUM-learning behaviour
that we documented above in the transition from the first to the fifth round.
A more plausible explanation for our learning findings and, simultaneously, for
how these seemingly go against those of some earlier studies may be traced in the
fact that our experiment featured only 15 menus that were derived from 7 lotteries.
Although, as we discussed in Section \ref{s3}, these do indeed generate difficult
decisions and are in some respects more complex than those in
\cite{hey01,vandekuilen-wakker06,birnbaum-schmidt15,spicer-mullett-sanborn24},
their smaller number may have enabled many subjects' introspection processes to converge
to (E)UM-compatible revealed preferences.
In any case, our analysis and findings suggest that
learning without feedback or other exogenous interventions---e.g.
choice-revision queries---in such relatively more complex but smaller
domains raises the possibility that an additional, neutral, and cost-effective approach to
achieving a meaningful reduction of measurement error for risk-preference elicitation,
either in the lab or in the field
\citep{schildberg-horisch18,gillen-snowberg-yariv19,dohmen-jagelka24},
could be the wider use of appropriately structured
repeated-choice experiments or surveys.

{\color{magenta}

}

The substantial evidence in favour of the learning
hypothesis notwithstanding, non-trivial proportions of subjects in our
study still deviated from expected
or ordinal utility maximization by
the end of the experiment, even after accounting
for the possibility of indifference.
This fact reinforces the view that favours the development
of bounded-rational choice models for decisions under risk.
It is outside this paper's scope to explore which
of the numerous existing such models
might explain those subjects' behaviour better
or to provide detailed outlines of potentially new models that
might do so. Our approximate-rationality
analysis for the ordinal model does suggest that the vast
majority of subjects make up to one ``mistaken'' choice
by the fifth time they are asked to decide from the same menus.
This fact and our remarks above about the potential role
of similarity-driven risk-attitude reversals
point towards a potentially promising
avenue in this respect.
We hope that the rich new dataset
that we are introducing with this study will
facilitate further empirical and theoretical exploration
of these important questions.

{

\linespread{1}

\footnotesize

\bibliographystyle{ecta} 
\bibliography{Learning}  

}
\newpage


\appendix

\newpage

\centerline{\huge \textbf{Online Appendix}}

\counterwithin{figure}{section}
\counterwithin{table}{section}
\setcounter{page}{0}
\thispagestyle{empty}

\newpage

\clearpage

\section{Additional Tables Cited in the Main Text}\label{a2}
\subsection{Utility Maximization}

\begin{table}[!htbp]
	\centering
	\caption{Utility Maximizers at all menus:
		Probit marginal effects}\label{tab:um_determinants_all_controls}
\resizebox{!}{0.55\textwidth}{
	\begin{tabular}{lcccc} \hline\hline
		& \multicolumn{4}{c}{Dependent variable:}\\
		\cline{2-5}
		& \multicolumn{2}{c}{1 if U-maximizer across 15 menus} &
		\multicolumn{2}{c}{1 if EU-maximizer across 15 menus}\\
		& (1) & (2) & (3) & (4) \\
		\cline{2-5}
		1 if round 2 & 0.082*** & 0.090** & 0.042 & 0.041 \\
		& (0.031) & (0.036) & (0.032) & (0.041) \\
		1 if round 3 & 0.132*** & 0.127*** & 0.093*** & 0.081** \\
		& (0.033) & (0.038) & (0.033) & (0.041) \\
		1 if round 4 & 0.194*** & 0.209*** & 0.135*** & 0.156*** \\
		& (0.033) & (0.036) & (0.035) & (0.043) \\
		1 if round 5 & 0.247*** & 0.270*** & 0.176*** & 0.194*** \\
		& (0.031) & (0.032) & (0.034) & (0.041) \\
		Verbal reasoning score & 0.463*** & 0.521*** & 0.269* & 0.284 \\
		& (0.162) & (0.182) & (0.149) & (0.180) \\
		Letter-numeric sequence score & 0.062 & 0.149 & 0.045 & 0.084 \\
		& (0.081) & (0.096) & (0.074) & (0.091) \\
		Matrix reasoning score & 0.085 & 0.011 & -0.014 & -0.056 \\
		& (0.123) & (0.158) & (0.105) & (0.139) \\
		3-dimensional rotation score & -0.045 & -0.095 & -0.054 & -0.100 \\
		& (0.066) & (0.076) & (0.059) & (0.069) \\
        1 if female & -0.102** & -0.101* & -0.029 & 0.007 \\
         & (0.050) & (0.055) & (0.044) & (0.053) \\
        1 if gender non-binary & -0.251* & -0.310* & -0.074 & -0.077 \\
         & (0.139) & (0.178) & (0.132) & (0.162) \\
        1 if undergraduate student & 0.073 & -0.133 & -0.089 & -0.327 \\
         & (0.149) & (0.209) & (0.154) & (0.218) \\
        1 if Master student & 0.045 & -0.107 & -0.122 & -0.292* \\
         & (0.157) & (0.237) & (0.128) & (0.153) \\
        1 if PhD student & 0.252 & 0.152 & 0.161 & 0.045 \\
         & (0.156) & (0.217) & (0.186) & (0.253) \\
        Economics and business & 0.147** & 0.193*** & 0.143** & 0.202** \\
         & (0.074) & (0.075) & (0.071) & (0.082) \\
        Earth Sciences and agriculture & 0.120 & 0.244*** & 0.058 & 0.145 \\
         & (0.107) & (0.093) & (0.111) & (0.144) \\
        Physics and chemistry & 0.022 & -0.060 & 0.048 & -0.031 \\
         & (0.089) & (0.099) & (0.078) & (0.089) \\
        Life science & -0.044 & -0.030 & -0.073 & -0.051 \\
         & (0.080) & (0.104) & (0.068) & (0.095) \\
        Computer science & 0.217*** & 0.261*** & 0.151** & 0.208*** \\
         & (0.084) & (0.071) & (0.077) & (0.080) \\
        Mathematics & 0.222*** & 0.177** & 0.157** & 0.141* \\
         & (0.076) & (0.078) & (0.078) & (0.085) \\
        Languages & 0.070 & 0.079 & -0.024 & -0.019 \\
         & (0.107) & (0.119) & (0.083) & (0.099) \\
        Psychology & 0.266*** & 0.286*** & 0.184* & 0.190* \\
         & (0.096) & (0.080) & (0.101) & (0.110) \\
        Law & 0.245** & 0.330*** & 0.168 & 0.256** \\
         & (0.103) & (0.069) & (0.107) & (0.124) \\
        Other field of study & 0.133 & 0.059 & 0.091 & 0.049 \\
         & (0.102) & (0.108) & (0.095) & (0.105) \\
        Observations & 1,540 & 1,130 & 1,540 & 1,130 \\ \hline
        		\hline
	\end{tabular}
    }
	\begin{minipage}[t]{\textwidth}\scriptsize \raggedright
			\footnotesize \textbf{Notes:} Marginal effect estimates of probit models
		evaluated at the means of independent variables.
		Robust standard errors clustered at the subject level in parentheses.
		*** $p<0.01$, ** $p<0.05$.\\
		In the first two columns, the dependent variable is an indicator
		variable that equals one if a subject made choices in the 15 menus
		of a round that are consistent with ordinal utility maximization.
		In column (1) all subjects are included, in column (2) the sample
		is restricted to subjects who never deferred at any of the 75 menus.
		In columns (3) and (4), the dependent variable is an indicator variable
		that equals one if a subject made choices in the 15 menus of a round
		that are consistent with expected utility maximization.
		In column (3) all subjects are included, in column (4) the sample
		is restricted to subjects who never deferred at any of the 75 menus.
	\end{minipage}
\end{table}

\newpage

\begin{table}[!htbp]
	\centering
	\caption{Utility Maximizers at all menus:
		Probit marginal effects---basic controls only}\label{tab:um_determinants_basic_controls}
\resizebox{!}{0.3\textwidth}{
	\begin{tabular}{lcccc} \hline\hline
		& \multicolumn{4}{c}{Dependent variable:}\\
		\cline{2-5}
		& \multicolumn{2}{c}{1 if U-maximizer across 15 menus} &
		\multicolumn{2}{c}{1 if EU-maximizer across 15 menus}\\
		& (1) & (2) & (3) & (4) \\
		\cline{2-5}
        Verbal reasoning score & 0.420*** & 0.419** & 0.246* & 0.214 \\
 & (0.161) & (0.183) & (0.147) & (0.177) \\
Letter-numeric sequence score & 0.089 & 0.138 & 0.079 & 0.116 \\
 & (0.077) & (0.086) & (0.070) & (0.083) \\
Matrix reasoning score & 0.053 & -0.049 & -0.049 & -0.135 \\
 & (0.119) & (0.146) & (0.102) & (0.130) \\
3-dimensional rotation score & -0.015 & -0.064 & -0.032 & -0.080 \\
 & (0.063) & (0.070) & (0.056) & (0.066) \\
1 if round 2 & 0.080*** & 0.084** & 0.042 & 0.040 \\
 & (0.030) & (0.034) & (0.031) & (0.039) \\
1 if round 3 & 0.127*** & 0.119*** & 0.090*** & 0.077** \\
 & (0.032) & (0.036) & (0.032) & (0.038) \\
1 if round 4 & 0.186*** & 0.196*** & 0.131*** & 0.149*** \\
 & (0.032) & (0.035) & (0.034) & (0.041) \\
1 if round 5 & 0.238*** & 0.255*** & 0.171*** & 0.185*** \\
 & (0.030) & (0.031) & (0.033) & (0.039) \\
1 if female & -0.105** & -0.109** & -0.043 & -0.017 \\
 & (0.047) & (0.053) & (0.042) & (0.050) \\
1 if gender non-binary & -0.270** & -0.310** & -0.144 & -0.150 \\
 & (0.114) & (0.122) & (0.109) & (0.130) \\
 &  &  &  &  \\
Observations & 1,540 & 1,130 & 1,540 & 1,130 \\
 Joint significance (ICAR) & 0.0350 & 0.0572 & 0.247 & 0.189 \\ \hline
      		\hline
	\end{tabular}
    }
	\begin{minipage}[t]{\textwidth}\scriptsize \raggedright
			\footnotesize \textbf{Notes:} Marginal effect estimates of probit models
		evaluated at the means of independent variables.
		Robust standard errors clustered at the subject level in parentheses.
		*** $p<0.01$, ** $p<0.05$.\\
		In the first two columns, the dependent variable is an indicator
		variable that equals one if a subject made choices in the 15 menus
		of a round that are consistent with ordinal utility maximization.
		In column (1) all subjects are included, in column (2) the sample
		is restricted to subjects who never deferred at any of the 75 menus.
		In columns (3) and (4), the dependent variable is an indicator variable
		that equals one if a subject made choices in the 15 menus of a round
		that are consistent with expected utility maximization.
		In column (3) all subjects are included, in column (4) the sample
		is restricted to subjects who never deferred at any of the 75 menus.
	\end{minipage}
\end{table}

\newpage

\subsection{Violations of Axioms}
\begin{table}[!htbp]
\centering
\caption{Violations of Utility Maximization at All Menus:
	Probit Marginal Effects}
 \resizebox{!}{0.6\textwidth}{

\begin{tabular}{lcccccc} \hline\hline
	& \multicolumn{6}{c}{Dependent variable: 1 if violating (across 15 menus)}\\
	& FOSD  & Indepen- & StAR  & WARP & Contraction & Decisive- \\
	&       & dence    &            &      & consistency & ness \\
	& (1) & (2) & (3) & (4) & (5) & (6) \\
	\cline{2-7}
	&  &  &  &  &  &  \\
	1 if round 2 & -0.015** & -0.043 & -0.021 & -0.120*** & -0.127*** & -0.013 \\
	& (0.007) & (0.032) & (0.034) & (0.030) & (0.030) & (0.018) \\
	1 if round 3 & -0.014** & -0.034 & -0.052* & -0.165*** & -0.150*** & -0.011 \\
	& (0.006) & (0.032) & (0.031) & (0.031) & (0.032) & (0.018) \\
	1 if round 4 & -0.021*** & -0.040 & -0.037 & -0.222*** & -0.227*** & -0.008 \\
	& (0.006) & (0.034) & (0.032) & (0.030) & (0.030) & (0.018) \\
	1 if round 5 & -0.019*** & -0.069** & -0.103*** & -0.291*** & -0.293*** & -0.014 \\
	& (0.006) & (0.031) & (0.030) & (0.027) & (0.027) & (0.018) \\
Verbal reasoning score & -0.005 & 0.053 & -0.232* & -0.532*** & -0.499*** & -0.031 \\
 & (0.025) & (0.110) & (0.125) & (0.156) & (0.157) & (0.112) \\
 & [0.965] & [0.965] & [0.370] & [0.025] & [0.040] & [0.965] \\
Letter-numeric sequence score & -0.027** & -0.041 & -0.013 & -0.067 & -0.057 & 0.038 \\
 & (0.013) & (0.057) & (0.067) & (0.081) & (0.080) & (0.056) \\
 & [0.215] & [0.895] & [0.895] & [0.855] & [0.895] & [0.895] \\
Matrix reasoning score & 0.002 & 0.029 & 0.133 & 0.001 & -0.020 & -0.102 \\
 & (0.022) & (0.080) & (0.090) & (0.121) & (0.118) & (0.089) \\
 & [1.000] & [1.000] & [0.580] & [1.000] & [1.000] & [0.580] \\
3-dimensional rotation score & 0.008 & 0.055 & 0.080 & 0.079 & 0.052 & -0.058 \\
 & (0.014) & (0.046) & (0.052) & (0.065) & (0.065) & (0.047) \\
 & [0.705] & [0.695] & [0.515] & [0.695] & [0.705] & [0.695] \\
1 if female & 0.019* & 0.059* & -0.009 & 0.078 & 0.079 & 0.018 \\
 & (0.010) & (0.035) & (0.041) & (0.049) & (0.049) & (0.036) \\
1 if gender non-binary & 0.278 & 0.010 & -0.014 & 0.325** & 0.309** & -0.079 \\
 & (0.173) & (0.113) & (0.164) & (0.142) & (0.143) & (0.055) \\
1 if undergraduate student & 0.219*** & 0.085 & -0.058 & -0.049 & -0.033 & -0.091 \\
 & (0.038) & (0.124) & (0.126) & (0.134) & (0.132) & (0.121) \\
1 if Master student & 0.976*** & 0.153 & -0.055 & -0.108 & -0.089 & -0.006 \\
 & (0.005) & (0.156) & (0.121) & (0.135) & (0.136) & (0.100) \\
1 if PhD student & 0.986*** & 0.019 & -0.216*** & -0.224* & -0.216 & -0.051 \\
 & (0.005) & (0.157) & (0.080) & (0.127) & (0.132) & (0.083) \\
Economics and business & -0.014 & 0.033 & -0.088 & -0.101 & -0.111 & -0.050 \\
 & (0.011) & (0.053) & (0.055) & (0.070) & (0.070) & (0.046) \\
Earth Sciences and agriculture & 0.006 & -0.035 & -0.051 & -0.132 & -0.105 & 0.015 \\
 & (0.022) & (0.068) & (0.077) & (0.090) & (0.096) & (0.072) \\
Physics and chemistry & -0.026*** & 0.008 & -0.044 & 0.043 & 0.037 & -0.045 \\
 & (0.007) & (0.059) & (0.064) & (0.091) & (0.090) & (0.048) \\
Life science & -0.015* & -0.007 & 0.049 & 0.054 & 0.044 & -0.006 \\
 & (0.009) & (0.052) & (0.063) & (0.082) & (0.081) & (0.051) \\
Computer science & -0.012 & 0.083 & -0.128** & -0.185** & -0.183** & -0.060 \\
 & (0.011) & (0.066) & (0.065) & (0.075) & (0.077) & (0.045) \\
Mathematics &  & -0.017 & -0.109* & -0.145** & -0.146** & -0.096*** \\
 &  & (0.066) & (0.058) & (0.073) & (0.074) & (0.034) \\
Languages & -0.027*** & 0.070 & -0.003 & 0.033 & 0.005 & -0.054 \\
 & (0.007) & (0.072) & (0.090) & (0.106) & (0.106) & (0.050) \\
Psychology & -0.016 & -0.018 & -0.175*** & -0.194** & -0.185** & -0.085** \\
 & (0.011) & (0.068) & (0.062) & (0.089) & (0.091) & (0.040) \\
Law & -0.008 & 0.043 & -0.135* & -0.271*** & -0.265*** & 0.001 \\
 & (0.015) & (0.080) & (0.078) & (0.074) & (0.077) & (0.077) \\
Other field of study & -0.027*** & 0.020 & -0.069 & 0.001 & -0.019 & -0.153*** \\
 & (0.007) & (0.068) & (0.071) & (0.104) & (0.103) & (0.020) \\

\hline
	&  &  &  &  &  &  \\
	Observations & 1,375 & 1,540 & 1,540 & 1,540 & 1,540 & 1,540 \\
	\hline\hline
\end{tabular}
}
\begin{minipage}[t]{\textwidth}\scriptsize \raggedright
	 \footnotesize \textbf{Notes:} Marginal effect estimates of probit models evaluated
	at the means of independent variables.
	Robust standard errors clustered at the subject level in parentheses.
	*** $p<0.01$, ** $p<0.05$.  Romano-Wolf p-values corrected for multiple hypothesis testing in brackets.\\
	*** $p<0.01$, ** $p<0.05$. \\
	The dependent variable in column (1) is an indicator variable that
	equals one if a subject makes at least one first-order
	stochastically dominated choice. Likewise the dependent variables
	in columns (2) to (6) are indicator variables that equal 1 if a
	subject's choices in a round of 15 menus violate Independence, StAR,
	WARP, Contraction Consistency, and Decisiveness, respectively.
\end{minipage}
\label{tab:violations_all_controls}
\end{table}

\newpage

\begin{table}[!htbp]
\centering
\caption{Violations of Utility Maximization at All Menus:
	Probit Marginal Effects}
 \resizebox{!}{0.3\textwidth}{

\begin{tabular}{lcccccc} \hline\hline
	& \multicolumn{6}{c}{Dependent variable: 1 if violating (across 15 menus)}\\
	& FOSD  & Indepen- & StAR  & WARP & Contraction & Decisive- \\
	&       & dence    &            &      & consistency & ness \\
	& (1) & (2) & (3) & (4) & (5) & (6) \\
	\cline{2-7}
	&  &  &  &  &  &  \\
1 if round 2 & -0.018** & -0.043 & -0.020 & -0.115*** & -0.122*** & -0.013 \\
 & (0.008) & (0.032) & (0.033) & (0.028) & (0.029) & (0.018) \\
1 if round 3 & -0.018** & -0.034 & -0.051* & -0.159*** & -0.145*** & -0.011 \\
 & (0.008) & (0.032) & (0.030) & (0.030) & (0.031) & (0.018) \\
1 if round 4 & -0.026*** & -0.040 & -0.035 & -0.214*** & -0.219*** & -0.007 \\
 & (0.007) & (0.034) & (0.031) & (0.029) & (0.030) & (0.019) \\
1 if round 5 & -0.024*** & -0.068** & -0.102*** & -0.281*** & -0.283*** & -0.013 \\
Verbal reasoning score & -0.003 & 0.029 & -0.221* & -0.479*** & -0.451*** & -0.020 \\
 & (0.034) & (0.108) & (0.122) & (0.153) & (0.155) & (0.119) \\
 & [0.990] & [0.985] & [0.295] & [0.035] & [0.060] & [0.990] \\
Letter-numeric sequence score & -0.042** & -0.043 & -0.028 & -0.078 & -0.071 & 0.020 \\
 & (0.019) & (0.051) & (0.065) & (0.077) & (0.076) & (0.058) \\
 & [0.135] & [0.765] & [0.890] & [0.725] & [0.765] & [0.890] \\
Matrix reasoning score & 0.001 & 0.009 & 0.157* & 0.030 & 0.010 & -0.101 \\
 & (0.026) & (0.079) & (0.089) & (0.117) & (0.115) & (0.094) \\
 & [1.000] & [1.000] & [0.345] & [0.995] & [1.000] & [0.715] \\
3-dimensional rotation score & 0.009 & 0.045 & 0.052 & 0.041 & 0.019 & -0.066 \\
 & (0.019) & (0.044) & (0.051) & (0.062) & (0.062) & (0.049) \\
 & [0.880] & [0.770] & [0.770] & [0.795] & [0.880] & [0.600] \\
1 if gender non-binary & 0.180 & 0.006 & 0.028 & 0.356*** & 0.339*** & -0.099** \\
 & (0.110) & (0.117) & (0.170) & (0.117) & (0.118) & (0.050) \\

\hline
	&  &  &  &  &  &  \\
	Observations & 1,375 & 1,540 & 1,540 & 1,540 & 1,540 & 1,540 \\
	\hline\hline
\end{tabular}
}
\begin{minipage}[t]{\textwidth}\scriptsize \raggedright
	 \footnotesize \textbf{Notes:} Marginal effect estimates of probit models evaluated
	at the means of independent variables.
	Robust standard errors clustered at the subject level in parentheses.
	*** $p<0.01$, ** $p<0.05$.  Romano-Wolf p-values corrected for multiple hypothesis testing in brackets.\\
	*** $p<0.01$, ** $p<0.05$. \\
	The dependent variable in column (1) is an indicator variable that
	equals one if a subject makes at least one first-order
	stochastically dominated choice. Likewise the dependent variables
	in columns (2) to (6) are indicator variables that equal 1 if a
	subject's choices in a round of 15 menus violate Independence, StAR,
	WARP, Contraction Consistency, and Decisiveness, respectively.
\end{minipage}
\label{tab:violations_basic_controls}
\end{table}

\newpage

\subsection{Analysis of Response Times}
\vspace{-20pt}

\begin{table}[!htbp]
\centering
\vspace{0pt}	
\caption{Determinants of Response Times}\label{tab:response_times}
\footnotesize
\begin{tabular}{lccc}\hline\hline
& \multicolumn{3}{c}{Dependent variable: }\\
\cline{2-4}
& \multicolumn{3}{c}{Response time at choice menu in seconds}\\
& (1) & (2) & (3) \\
\cline{2-4}
&  &  &  \\
1 if EU maximizer &  &  & -1.298** \\
&  &  & (0.643) \\
1 if utility maximizer &  &  & -1.774** \\
&  &  & (0.685) \\
1 if choice deferred &  & -2.976* & -4.331*** \\
&  & (1.519) & (1.522) \\
1 if round 2 & -7.392*** & -7.394*** & -7.215*** \\
& (0.339) & (0.339) & (0.343) \\
1 if round 3 & -9.924*** & -9.926*** & -9.608*** \\
& (0.427) & (0.427) & (0.426) \\
1 if round 4 & -11.227*** & -11.226*** & -10.752*** \\
& (0.466) & (0.467) & (0.468) \\
1 if round 5 & -12.514*** & -12.514*** & -11.899*** \\
& (0.497) & (0.497) & (0.497) \\
1 if menu A1,A2 & -4.671*** & -5.017*** & -5.175*** \\
& (0.516) & (0.534) & (0.533) \\
1 if menu A1,A2,C1 & -3.703*** & -4.047*** & -4.203*** \\
& (0.508) & (0.534) & (0.533) \\
1 if menu A1,A2,C1,C2 & -4.705*** & -5.051*** & -5.208*** \\
& (0.525) & (0.552) & (0.551) \\
1 if menu A1,A2,C2 & -5.748*** & -6.088*** & -6.243*** \\
& (0.539) & (0.568) & (0.568) \\
1 if menu A1,B1 & -6.464*** & -6.804*** & -6.959*** \\
& (0.500) & (0.530) & (0.530) \\
1 if menu A1,B1,B2 & -2.412*** & -2.740*** & -2.890*** \\
& (0.491) & (0.536) & (0.539) \\
1 if menu A1,B1,B2,D & 0.753 & 0.430 & 0.283 \\
& (0.557) & (0.590) & (0.594) \\
1 if menu A1,B2 & -7.736*** & -8.076*** & -8.231*** \\
& (0.523) & (0.550) & (0.548) \\
1 if menu A1,D & -6.633*** & -6.979*** & -7.136*** \\
& (0.512) & (0.546) & (0.545) \\
1 if menu A2,D & -7.111*** & -7.445*** & -7.598*** \\
& (0.526) & (0.554) & (0.554) \\
1 if menu B1,B2 & -1.580*** & -1.837*** & -1.954*** \\
& (0.436) & (0.458) & (0.458)
\end{tabular}
\end{table}

\begin{table}[!htbp]
\centering
\footnotesize
\caption*{}
\begin{tabular}{lccc}
1 if menu B1,B2,D & -2.185*** & -2.512*** & -2.660*** \\
& (0.494) & (0.530) & (0.530) \\
1 if menu B1,D & -4.978*** & -5.312*** & -5.465*** \\
& (0.468) & (0.498) & (0.495) \\
1 if menu B2,D & -6.461*** & -6.782*** & -6.928*** \\
& (0.519) & (0.553) & (0.552)\\
Verbal reasoning score & -4.371** & -4.270** & -3.198* \\
& (1.967) & (1.947) & (1.901) \\
Letter-numeric sequence score & 0.033 & 0.069 & 0.228 \\
& (1.000) & (0.995) & (0.959) \\
Matrix reasoning score & 0.332 & 0.305 & 0.393 \\
& (1.706) & (1.706) & (1.653) \\
3-dimensional rotation score & 0.867 & 0.868 & 0.727 \\
& (0.814) & (0.813) & (0.768) \\
1 if female & -1.376** & -1.354** & -1.555** \\
& (0.687) & (0.683) & (0.669) \\
1 if gender non-binary & -3.218*** & -3.212*** & -3.743*** \\
& (1.195) & (1.195) & (1.197) \\
1 if undergraduate student & -1.784 & -1.798 & -1.806 \\
& (1.901) & (1.885) & (1.791) \\
1 if Master student & -2.466 & -2.372 & -2.428 \\
& (2.040) & (2.020) & (1.938) \\
1 if PhD student & -1.601 & -1.584 & -0.978 \\
& (2.342) & (2.329) & (2.207) \\
Economics and business & 1.246 & 1.270 & 1.677 \\
& (1.065) & (1.063) & (1.031) \\
Earth Sciences and agriculture & -1.075 & -1.049 & -0.781 \\
& (1.086) & (1.084) & (1.064) \\
Physics and chemistry & 0.094 & 0.070 & 0.143 \\
& (1.218) & (1.216) & (1.198) \\
Life science & -1.248 & -1.252 & -1.408 \\
& (0.967) & (0.961) & (0.946) \\
Computer science & -1.262 & -1.273 & -0.760 \\
& (1.049) & (1.045) & (1.074) \\
Mathematics & -0.053 & -0.085 & 0.448 \\
& (1.148) & (1.153) & (1.111) \\
Languages & 1.335 & 1.304 & 1.383 \\
& (1.590) & (1.590) & (1.513) \\
Psychology & 1.172 & 1.131 & 1.773 \\
& (1.293) & (1.295) & (1.224) \\
Law & -1.870* & -1.770* & -1.127 \\
& (1.061) & (1.037) & (1.029) \\
Other field of study & -0.822 & -0.876 & -0.605 \\
& (1.201) & (1.199) & (1.107) \\
Constant & 31.059*** & 31.307*** & 31.136*** \\
& (3.146) & (3.130) & (3.053) \\
&  &  &  \\
Observations & 23,100 & 23,100 & 23,100 \\
R-squared & 0.191 & 0.193 & 0.204 \\ \hline\hline
\end{tabular}
\begin{minipage}[t]{\textwidth}\scriptsize \raggedright
\textbf{Notes:} OLS estimates.
Robust standard errors clustered at the subject
level in parentheses. Menu $\{C1,C2\}$ is the reference category.
*** $p<0.01$, ** $p<0.05$. The dependent variable
is the response time in seconds at a choice menu.
\end{minipage}
\end{table}

\newpage

\newpage

\section{The 7 Lotteries and 15 Menus}\label{a3}

\begin{figure}[!htbp]
\centering
\caption{\centering The 7 lotteries.}
\label{fig:lotteries}
\begin{subfigure}[b]{0.3\textwidth}
\caption*{\vspace{-5pt}A1}
\centering
\includegraphics[width=1\textwidth]{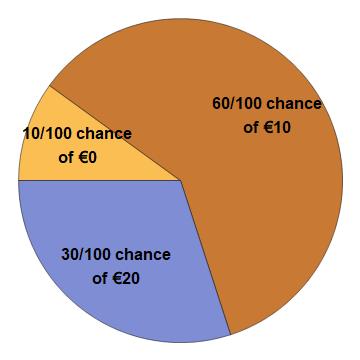}
\end{subfigure}
\begin{subfigure}[b]{0.3\textwidth}
\caption*{\vspace{-5pt}A2}
\centering
\includegraphics[width=1\textwidth]{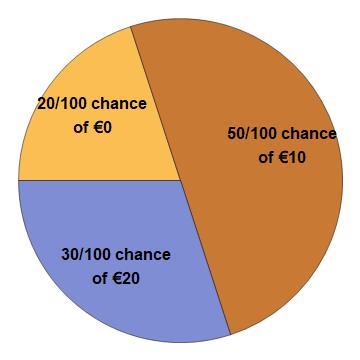}
\end{subfigure}\hspace{100pt}
\begin{subfigure}[b]{0.3\textwidth}
\centering
\caption*{\vspace{-5pt}B1}
\includegraphics[width=1\textwidth]{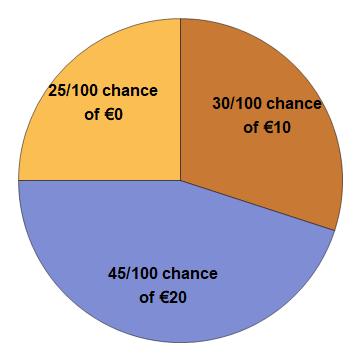}
\end{subfigure}
\begin{subfigure}[b]{0.3\textwidth}
\caption*{\vspace{-5pt}B2}
\centering
\includegraphics[width=1\textwidth]{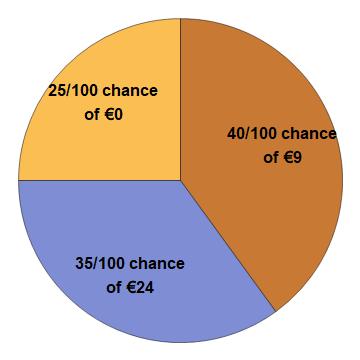}
\end{subfigure}\hspace{100pt}
\begin{subfigure}[b]{0.3\textwidth}
\caption*{\vspace{-5pt}C1}
\centering
\includegraphics[width=1\textwidth]{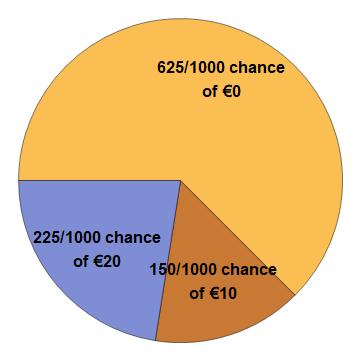}
\end{subfigure}
\begin{subfigure}[b]{0.3\textwidth}
\caption*{\vspace{-5pt}C2}
\includegraphics[width=1\textwidth]{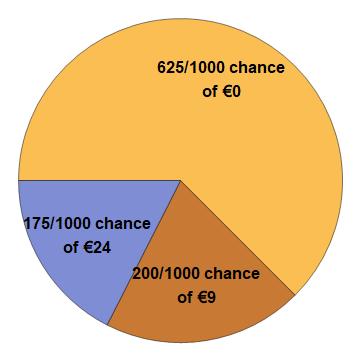}
\end{subfigure}\hspace{20pt}
\begin{subfigure}[b]{0.3\textwidth}
\caption*{\vspace{-5pt}D}
\centering
\includegraphics[width=1\textwidth]{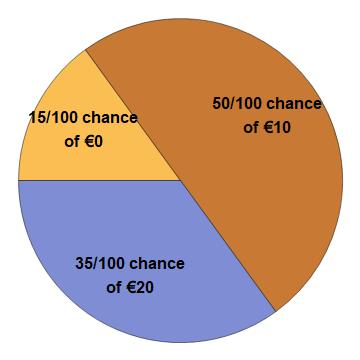}
\end{subfigure}
\caption*{\raggedright \scriptsize }
\end{figure}

\begin{figure}
\centering
\caption{\centering The 15 distinct menus.}
\label{fig:menus}
\begin{subfigure}[b]{0.35\textwidth}
\caption*{\vspace{-5pt}A1 \& A2}
\centering
\includegraphics[width=0.4\textwidth]{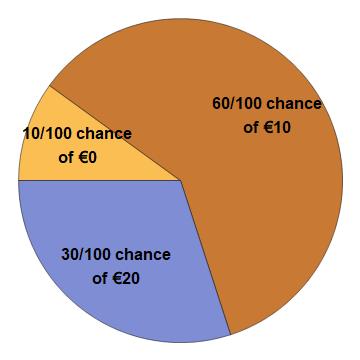}
\includegraphics[width=0.4\textwidth]{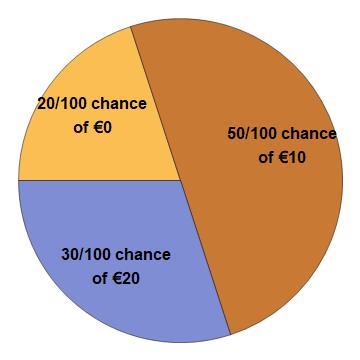}
\end{subfigure}
\begin{subfigure}[b]{0.35\textwidth}
\caption*{\vspace{-5pt}B1 \& B2}
\centering
\includegraphics[width=0.4\textwidth]{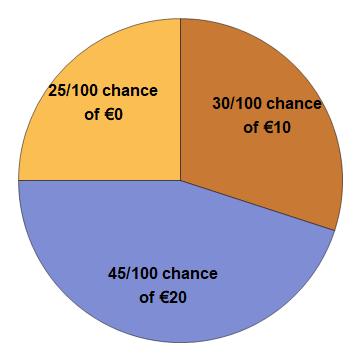}
\includegraphics[width=0.4\textwidth]{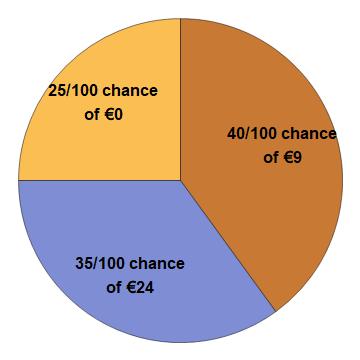}
\end{subfigure}
\begin{subfigure}[b]{0.35\textwidth}
\caption*{\vspace{-5pt}C1 \& C2}
\centering
\includegraphics[width=0.4\textwidth]{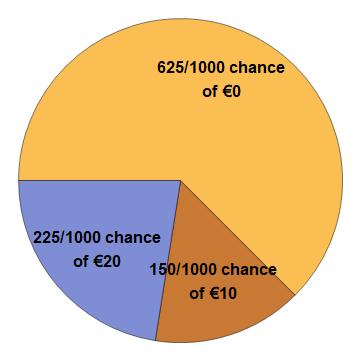}
\includegraphics[width=0.4\textwidth]{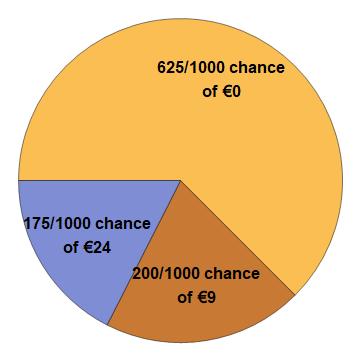}
\end{subfigure}
\begin{subfigure}[b]{0.35\textwidth}
\caption*{\vspace{-5pt}B1 \& D}
\centering
\includegraphics[width=0.4\textwidth]{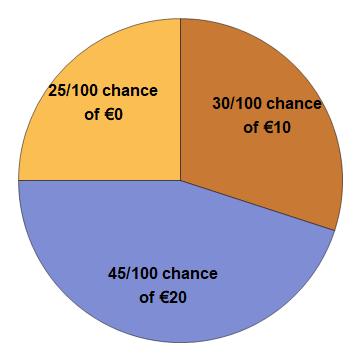}
\includegraphics[width=0.4\textwidth]{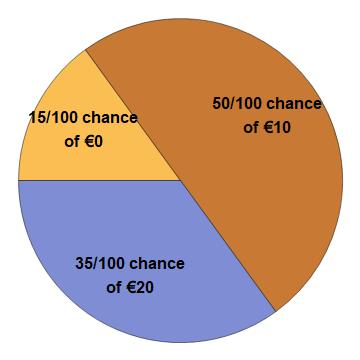}
\end{subfigure}
\begin{subfigure}[b]{0.35\textwidth}
\caption*{\vspace{-5pt}B2 \& D}
\centering
\includegraphics[width=0.4\textwidth]{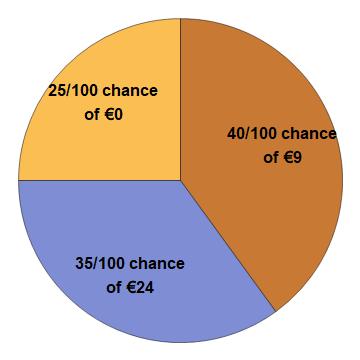}
\includegraphics[width=0.4\textwidth]{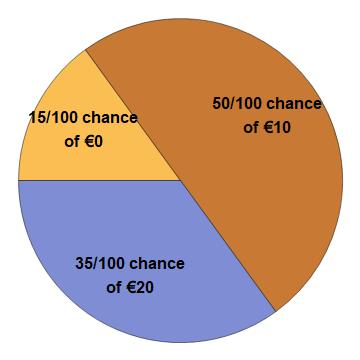}
\end{subfigure}
\begin{subfigure}[b]{0.35\textwidth}
\caption*{\vspace{-5pt}A1 \& B1}
\centering
\includegraphics[width=0.4\textwidth]{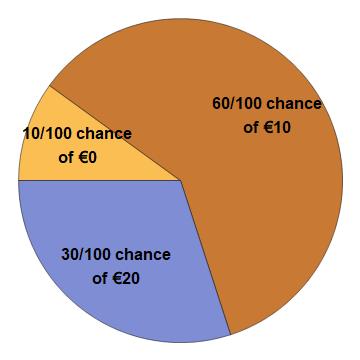}
\includegraphics[width=0.4\textwidth]{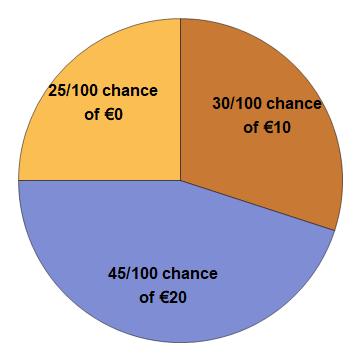}
\end{subfigure}
\begin{subfigure}[b]{0.35\textwidth}
\caption*{\vspace{-5pt}A1 \& B2}
\centering
\includegraphics[width=0.4\textwidth]{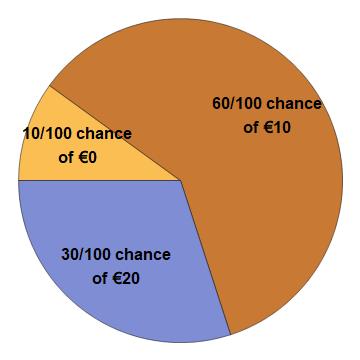}
\includegraphics[width=0.4\textwidth]{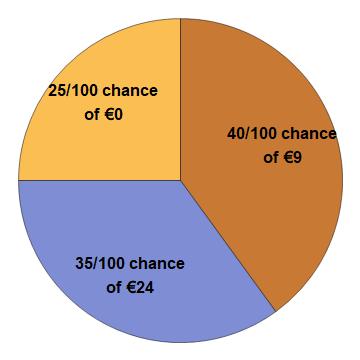}
\end{subfigure}
\begin{subfigure}[b]{0.35\textwidth}
\caption*{\vspace{-5pt}A2 \& D}
\centering
\includegraphics[width=0.4\textwidth]{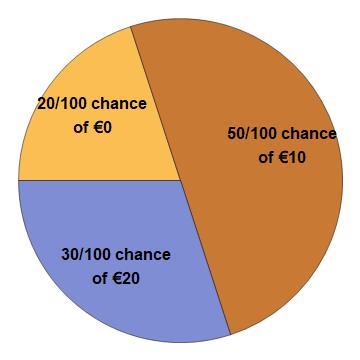}
\includegraphics[width=0.4\textwidth]{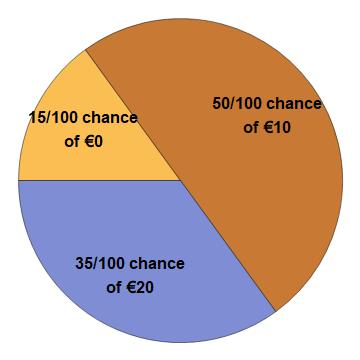}
\end{subfigure}
\begin{subfigure}[b]{0.35\textwidth}
\caption*{\vspace{-5pt}A1 \& D}
\centering
\includegraphics[width=0.4\textwidth]{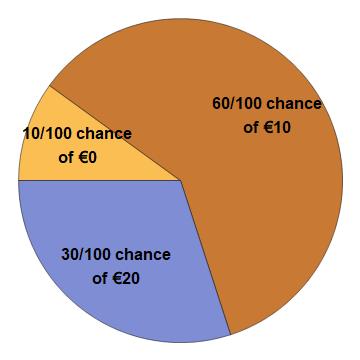}
\includegraphics[width=0.4\textwidth]{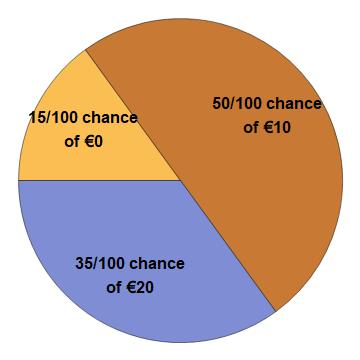}
\end{subfigure}\hspace{200pt}
\begin{subfigure}[b]{0.4\textwidth}
\caption*{\vspace{-5pt}A1 \& A2 \& C1}
\centering
\includegraphics[width=0.32\textwidth]{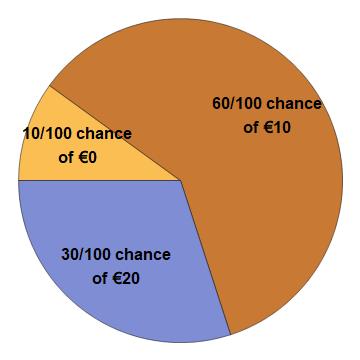}
\includegraphics[width=0.32\textwidth]{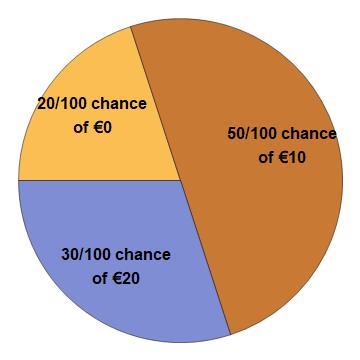}
\includegraphics[width=0.32\textwidth]{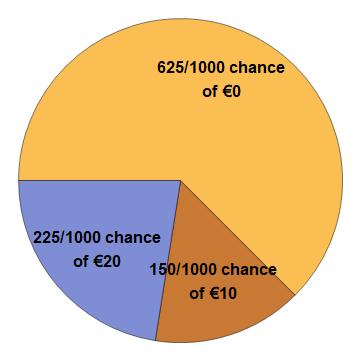}
\end{subfigure}\hspace{30pt}
\begin{subfigure}[b]{0.4\textwidth}
\caption*{\vspace{-5pt}A1 \& A2 \& C2}
\centering
\includegraphics[width=0.32\textwidth]{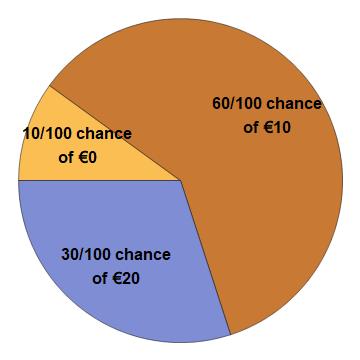}
\includegraphics[width=0.32\textwidth]{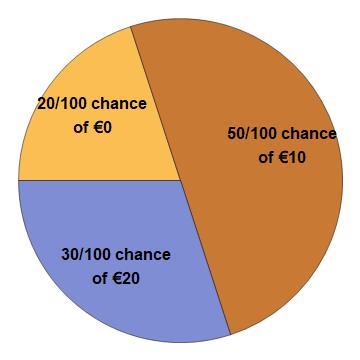}
\includegraphics[width=0.32\textwidth]{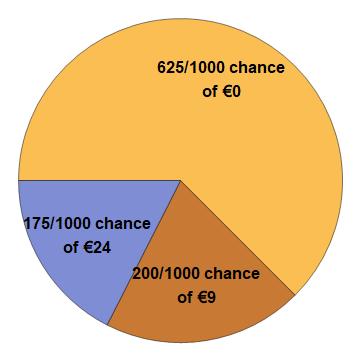}
\end{subfigure}
\begin{subfigure}[b]{0.4\textwidth}
\caption*{\vspace{-5pt}A1 \& B1 \& B2}
\centering
\includegraphics[width=0.32\textwidth]{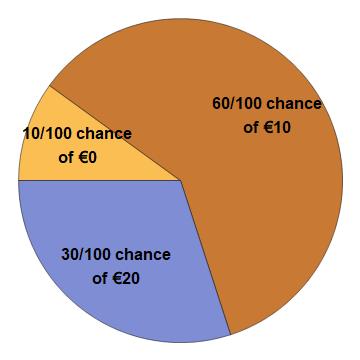}
\includegraphics[width=0.32\textwidth]{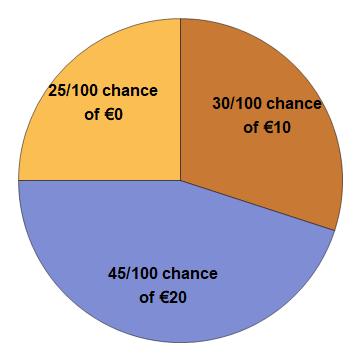}
\includegraphics[width=0.32\textwidth]{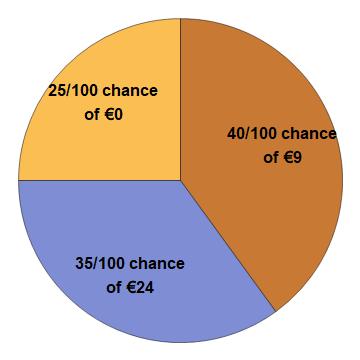}
\end{subfigure}\hspace{30pt}
\begin{subfigure}[b]{0.4\textwidth}
\caption*{\vspace{-5pt}B1 \& B2 \& D}
\centering
\includegraphics[width=0.32\textwidth]{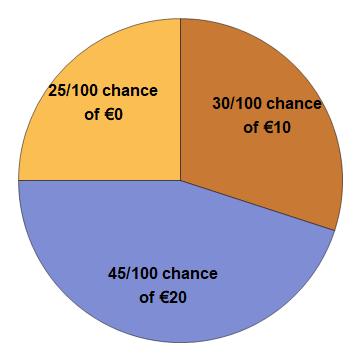}
\includegraphics[width=0.32\textwidth]{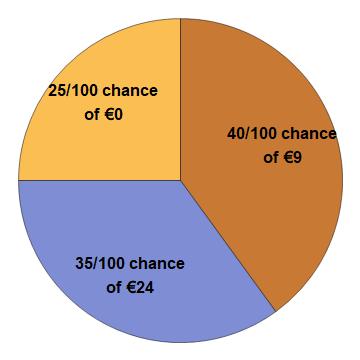}
\includegraphics[width=0.32\textwidth]{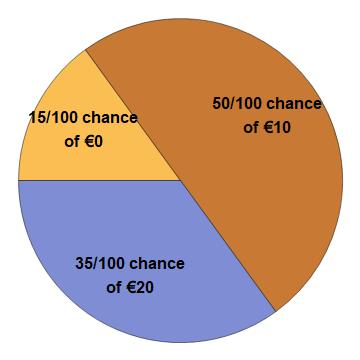}
\end{subfigure}\hspace{20pt}
\begin{subfigure}[b]{0.45\textwidth}
\caption*{\vspace{-5pt}A1 \& B1 \& B2 \& D}
\centering
\includegraphics[width=0.3\textwidth]{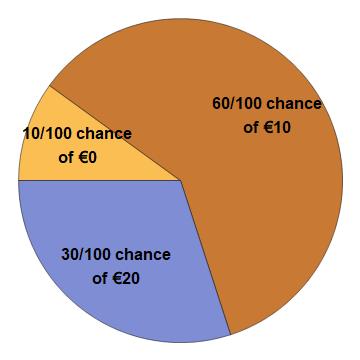}
\includegraphics[width=0.3\textwidth]{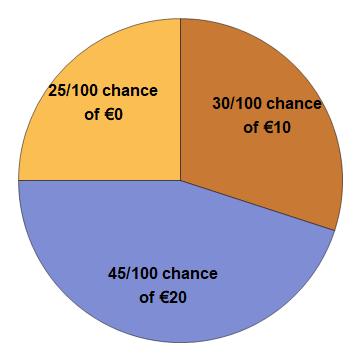}\hspace{100pt}
\includegraphics[width=0.3\textwidth]{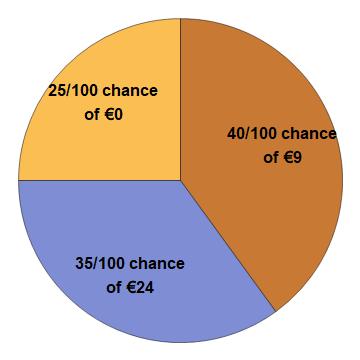}
\includegraphics[width=0.3\textwidth]{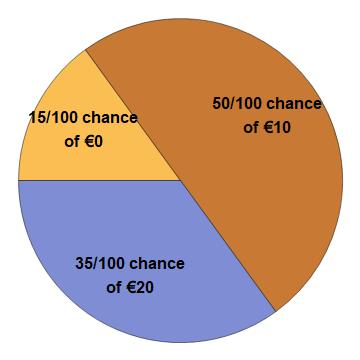}
\end{subfigure}
\begin{subfigure}[b]{0.45\textwidth}
\caption*{\vspace{-5pt}A1 \& A2 \& C1 \& C2}
\centering
\includegraphics[width=0.3\textwidth]{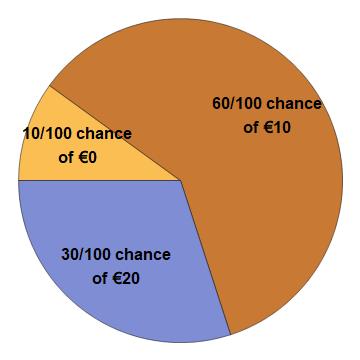}
\includegraphics[width=0.3\textwidth]{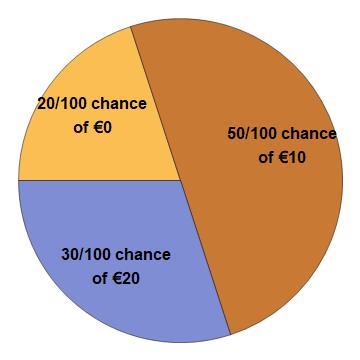}\hspace{100pt}
\includegraphics[width=0.3\textwidth]{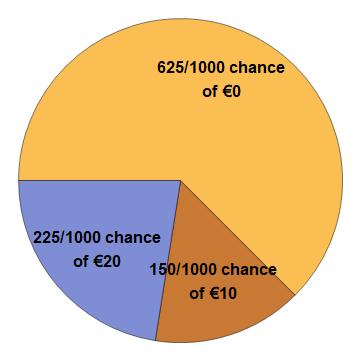}
\includegraphics[width=0.3\textwidth]{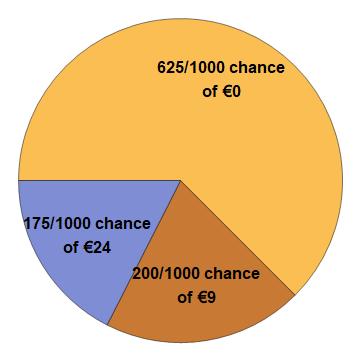}
\end{subfigure}
\end{figure}

\newpage

\section{Instructions}\label{a4}

\includegraphics[width=0.9\textwidth]{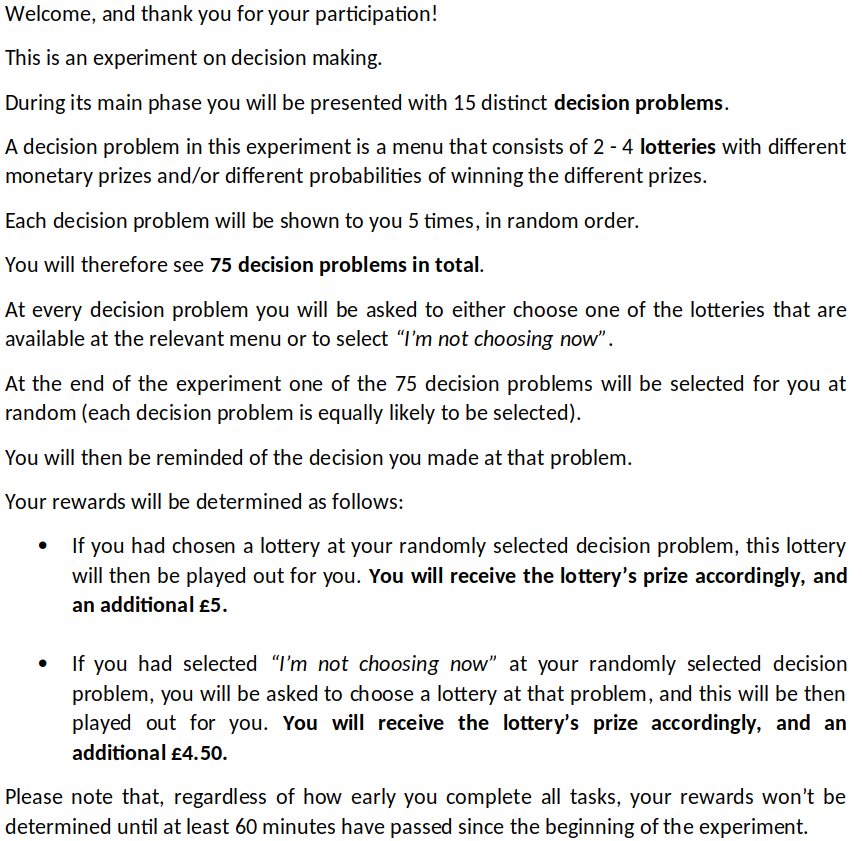}

\newpage

\section{Experimental Interface: Screenshots}\label{a5}

\subsection{Understanding Quiz}

{

\centering

\includegraphics[width=0.9\textwidth]{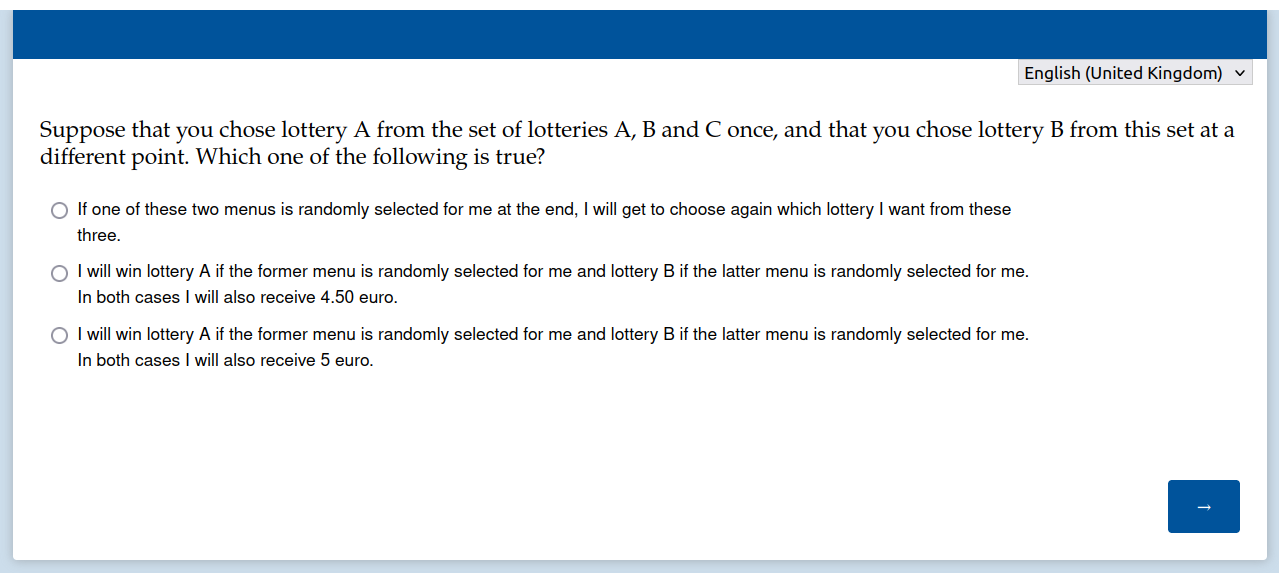}

\includegraphics[width=0.9\textwidth]{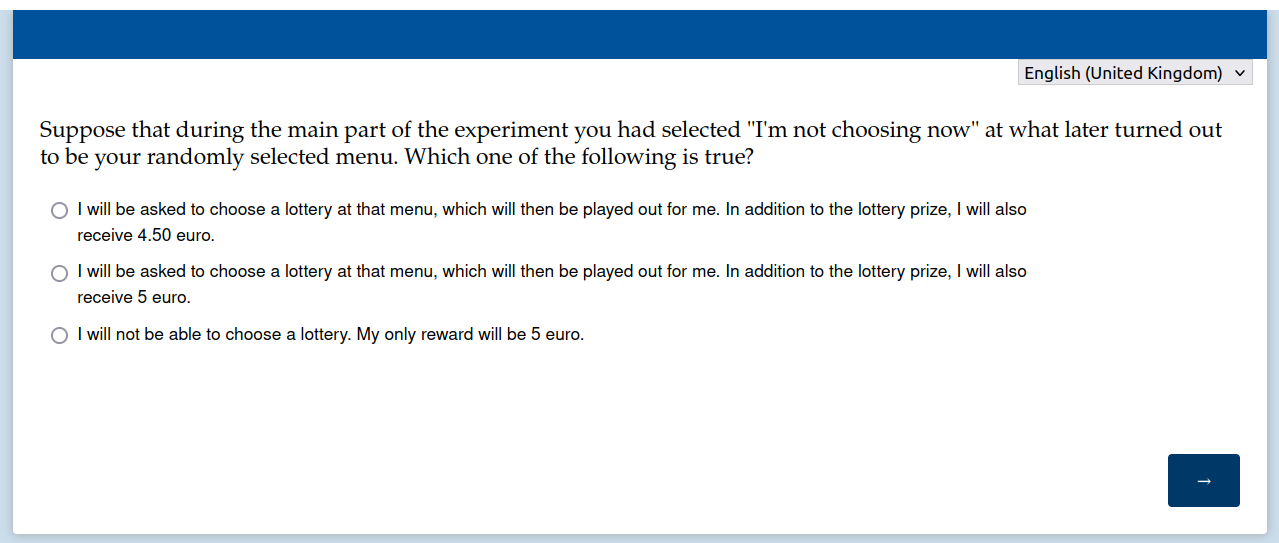}

\includegraphics[width=0.9\textwidth]{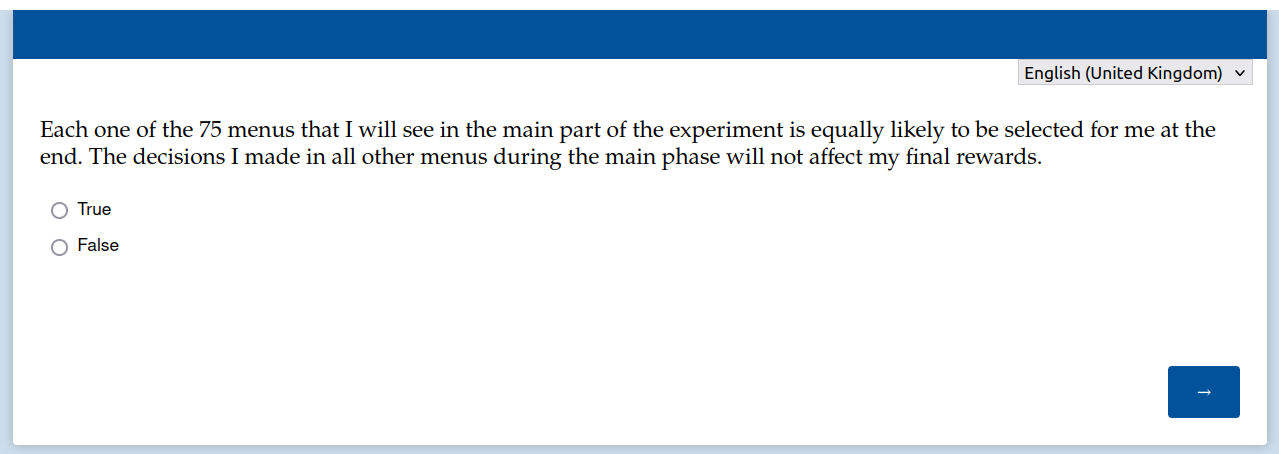}

}

\subsection{Main Part}

{

\centering

\includegraphics[width=0.85\textwidth]{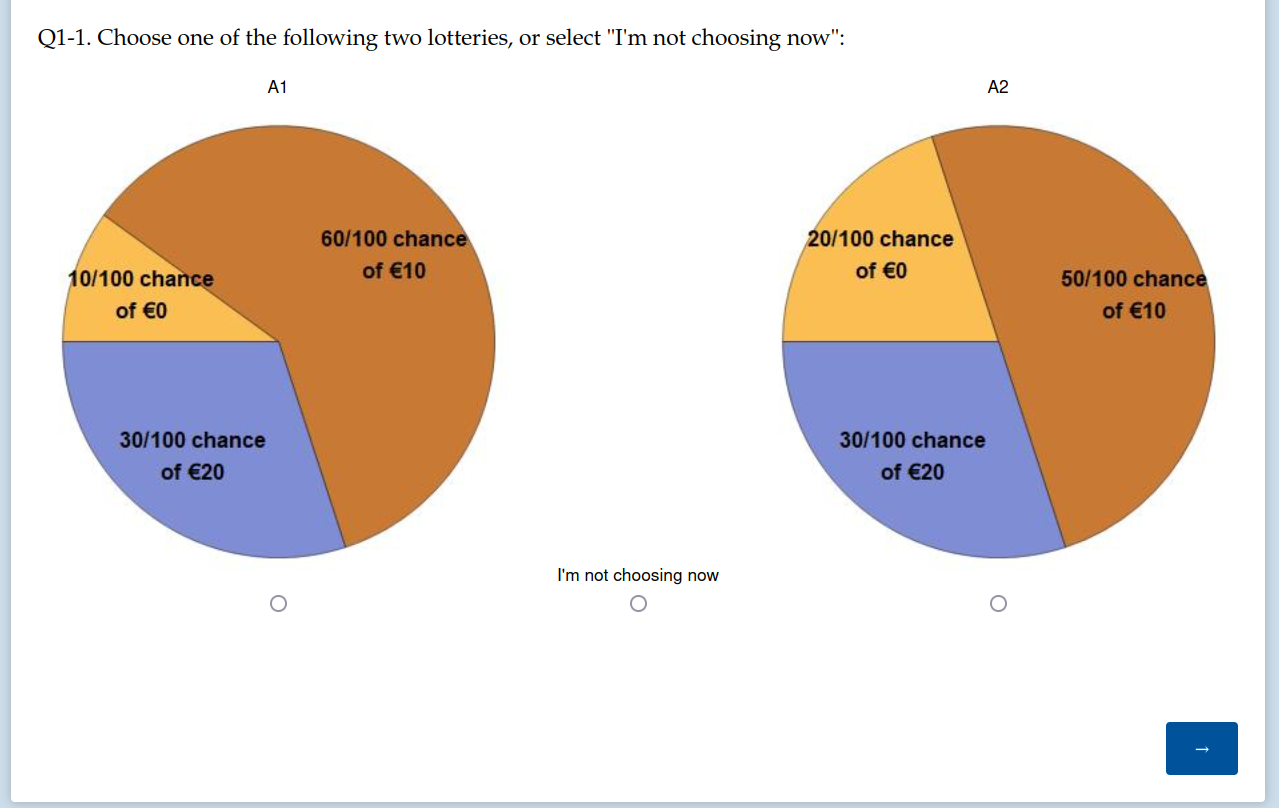}

\includegraphics[width=0.85\textwidth]{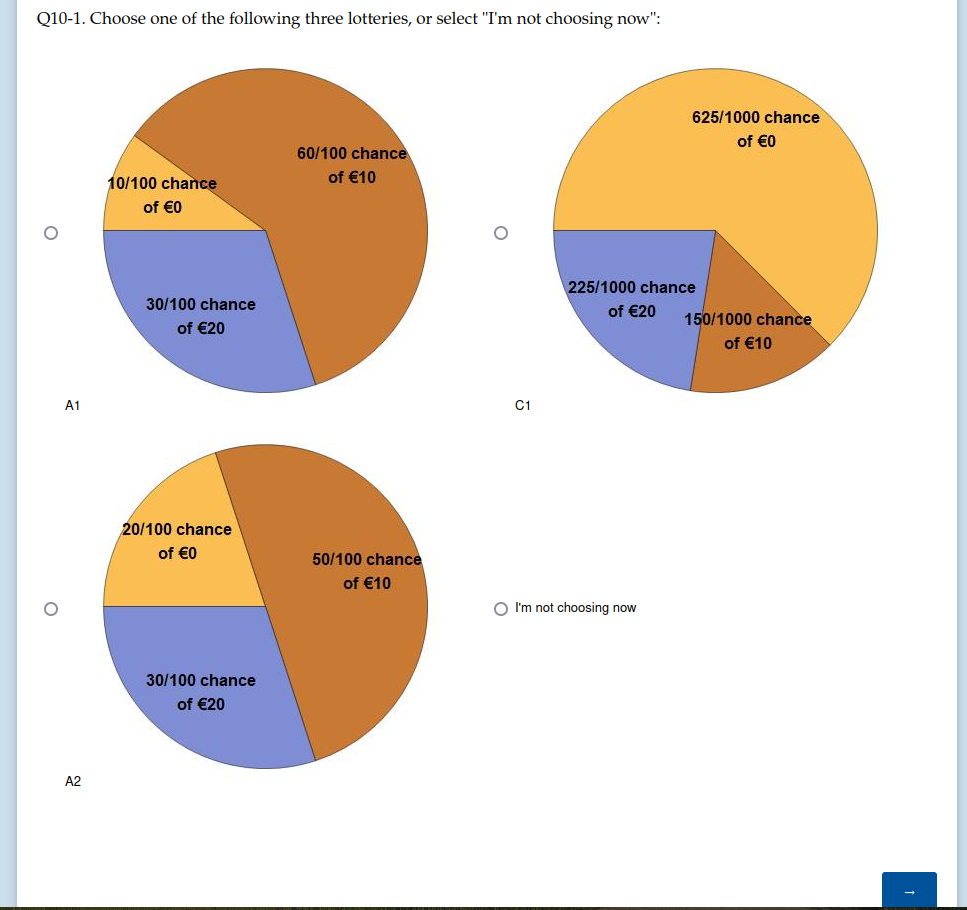}

\includegraphics[width=0.85\textwidth]{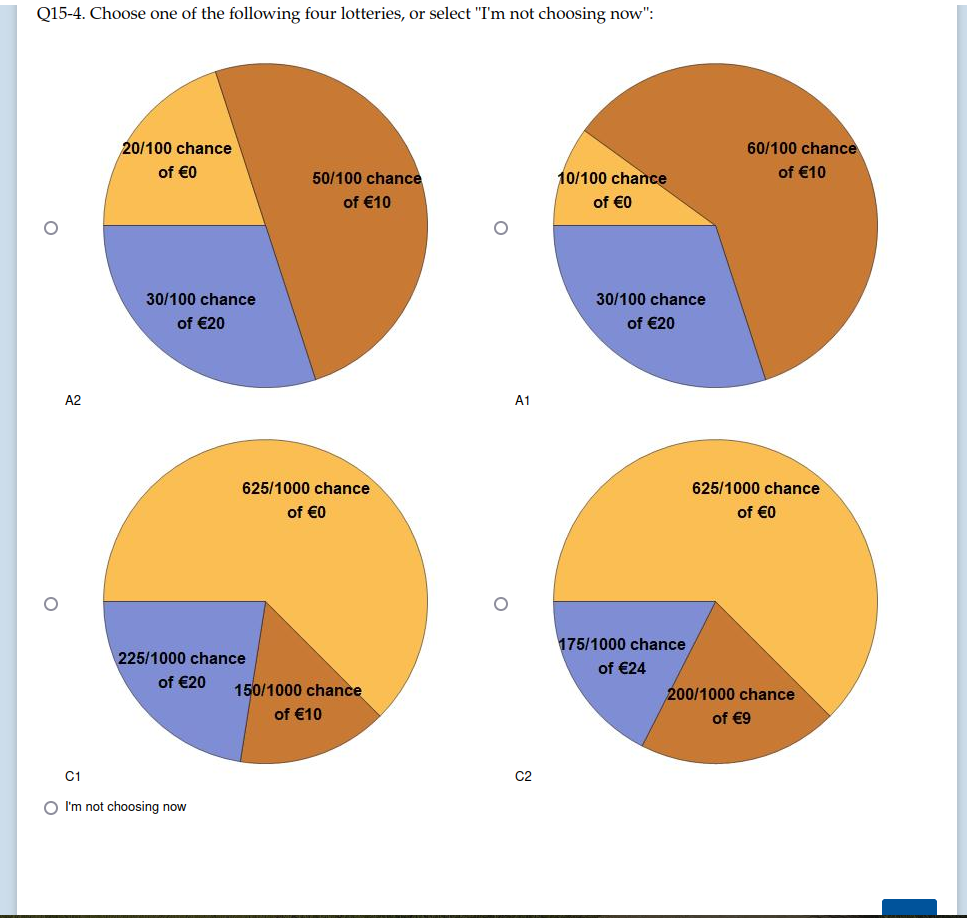}

}
\subsection{Randomly Selected Menu}

\begin{center}

{

\centering

\includegraphics[width=0.9\textwidth]{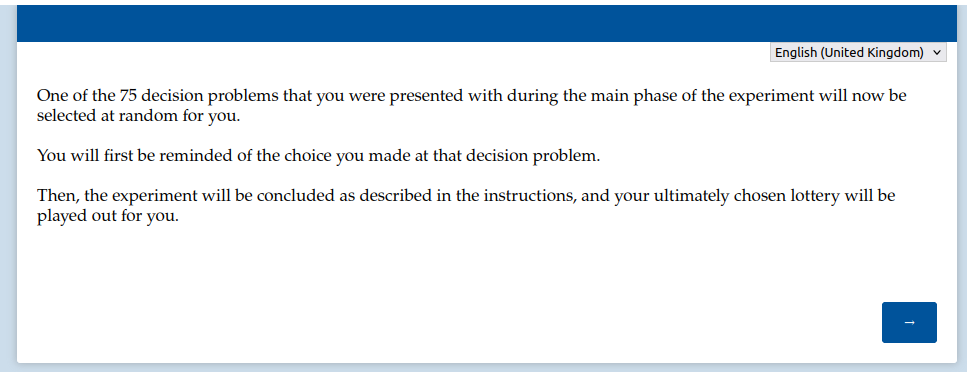}

\includegraphics[width=0.9\textwidth]{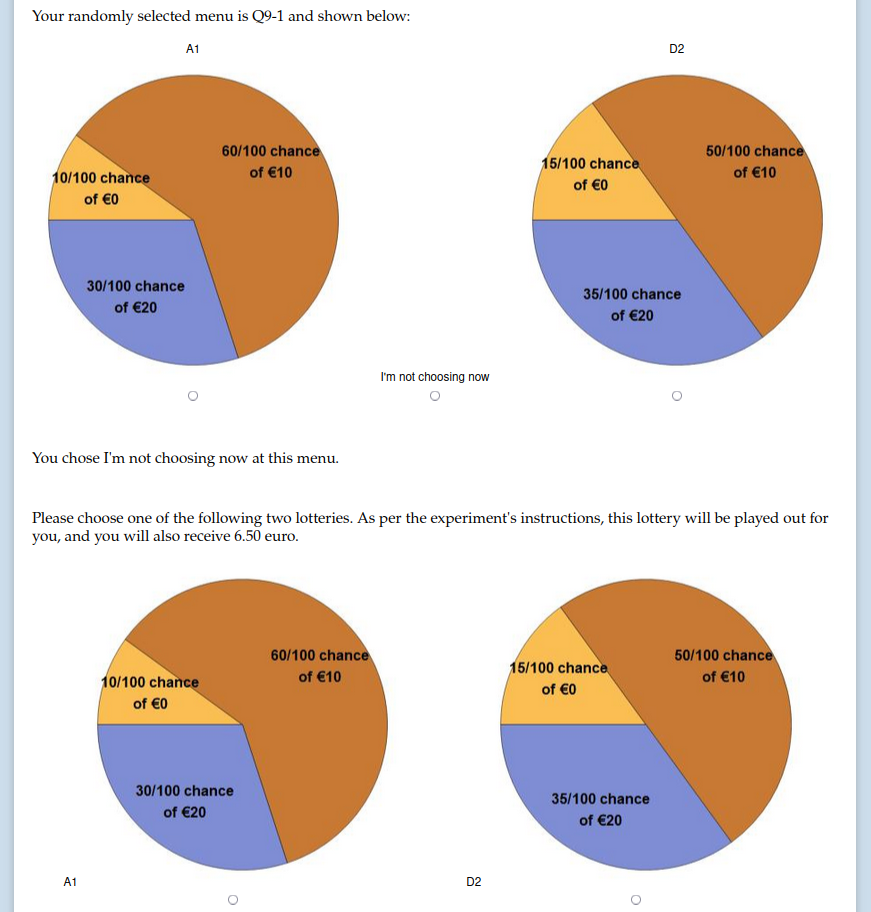}

}

\end{center}

\newpage

\section{Stochastic Dominance (Non-)Relations}\label{a6}

\paragraph*{Definitions.} Recall that, for money lotteries $p$ and $q$ that are defined over
a finite set $X$ and have cumulative distributions denoted by $F_p$ and $F_q$,
$p$ is said to FOSD $q$ if $F_p(x)\leq F_q(x)$
for all $x\in X$, with strict inequality at
some $x$. Furthermore,
$p$ is said to SOSD $q$ if $\int_a^x[F_p(t)-F_q(t)]dt\leq 0$
for all $x\in [a,b]$, with strict inequality at some $x$.

\begin{table}[!htbp]
\small
\centering
\caption{\centering The cumulative density functions and
expected values \linebreak of lotteries involved in a FOSD claim.}
\setlength{\tabcolsep}{6pt} 
\renewcommand{\arraystretch}{1.3} 
\begin{tabular}{|r|c|c|c|c|c|c|}
\hline
\multirow{2}{*}{\hspace{20pt} $x\in $}
&\multirow{2}{*}{$[0,9)$}&\multirow{2}{*}{$[9,10)$}
&\multirow{2}{*}{$[10,20)$}&\multirow{2}{*}{$[20,24)$}
&\multirow{2}{*}{$[24,\infty)$} & \multirow{2}{*}{$\mathbb{E}(x)$} \\
\multirow{1}{*}{Lottery \hspace{30pt}} &&&&&& \\	
\hline	
\textbf{A1} & 0.1	& 0.1	& 0.7	& 1		& 1	& 12 \\
\hline
\textbf{A2} & 0.2 	& 0.2 	& 0.7 	& 1 	& 1	& 11 \\
\hline
\textbf{C1} & 0.625 & 0.625 & 0.775 & 1 	& 1 & 6	\\
\hline
\textbf{C2} & 0.625 & 0.825 & 0.825 & 0.825 & 1 & 6	\\
\hline
\textbf{D} 	& 0.15 	& 0.15 	& 0.65 	& 1 	& 1 & 12 \\
\hline		
\end{tabular}
\label{tab:fosd}
\end{table}

\begin{enumerate}

\small	
	
\item A1 \textit{FOSD} A2.

\item A1 \textit{FOSD} C1.

\item A1 \textit{``nearly'' FOSD} C2:
A1 dominates C2 in $[0,20)$ with a weighted-average difference
in probability mass of 0.335;
C2 dominates A1 in $[20,24]$ with a probability-mass difference of 0.175;
for $x\in [0,24]$, the net difference in weighted-average probability mass
is 0.25 in favour of A1;
finally $\mathbb{E}_{\textnormal{A1}}(x)=12=2\, \mathbb{E}_{\textnormal{C2}}(x)$.

\item A2 \textit{``nearly'' FOSD} C2:
A2 dominates C2 in $[0,20)$ with a weighted-average difference
in probability mass of 0.285;
C2 dominates A2 in $[20,24]$ with a probability-mass difference
of 0.175; for $x\in [0,24]$, the net
difference in weighted-average probability mass is
$\approx0.21$ in favour of A2;
finally $\mathbb{E}_{\textnormal{A2}}(x)=11>6=\mathbb{E}_{\textnormal{C2}}(x)$.

\item A1 and D are FOSD-unranked: compare their cdf values at $[0,10)$
and $[10,20)$. 

\end{enumerate}

\pagebreak

\begin{table}[!htbp]
\centering
\small
\caption{\centering The area under the
cumulative density function of each lottery involved in
a SOSD claim, evaluated in the range $[0,x]$ for
each integer $x\in\{1,2,\ldots,24\}$.}
\setlength{\tabcolsep}{6pt} 
\renewcommand{\arraystretch}{1.3} 
\begin{tabular}{|c|c|c|c|c|}
\hline	
\multirow{3}{*}{\hspace{20pt} Lottery}
&\multirow{4}{*}{\textbf{A1}}&\multirow{4}{*}{\textbf{B1}}
&\multirow{4}{*}{\textbf{B2}}&\multirow{4}{*}{\textbf{D}} \\
\multirow{3}{*}{$x=$ \hspace{30pt}} &&&& \\
&&&&\\
\hline
1	& 0.1		& 0.25		& 0.25		& 0.15	\\
\hline
2	& 0.2		& 0.50		& 0.50		& 0.30	\\
\hline
3	& 0.3		& 0.75		& 0.75		& 0.45	\\
\hline
4	& 0.4		& 1.00		& 1.00		& 0.60	\\
\hline
5	& 0.5		& 1.25		& 1.25		& 0.75	\\
\hline
6	& 0.6		& 1.50		& 1.50		& 0.90	\\
\hline
7	& 0.7		& 1.75		& 1.75		& 1.05	\\
\hline
8	& 0.8		& 2.00		& 2.00		& 1.20	\\
\hline
9	& 0.9		& 2.25		& 2.25		& 1.35	\\
\hline
10	& 1.0		& 2.50		& 2.25		& 1.50	\\
\hline
11	& 1.7		& 3.05		& 2.90		& 2.15	\\
\hline
12	& 2.4		& 3.60		& 3.55		& 2.80	\\
\hline
13	& 3.1		& 4.15		& 4.20		& 3.45	\\
\hline
14	& 3.8		& 4.70		& 4.85		& 4.10	\\
\hline
15	& 4.5		& 5.25		& 5.50		& 4.75	\\
\hline
16	& 5.2		& 5.80		& 6.15		& 5.40	\\
\hline
17	& 5.9		& 6.35		& 6.80		& 6.05	\\
\hline
18	& 6.6		& 6.90		& 7.45		& 6.70	\\
\hline
19	& 7.3		& 7.45		& 8.10		& 7.35	\\
\hline
20	& 8.0		& 8.0		& 8.75		& 8.0	\\
\hline
21	& 9.0		& 9.0		& 9.40		& 9.0	\\
\hline
22	& 10.0		& 10.0		& 10.05		& 10.0	\\
\hline
23	& 11.0		& 11.0		& 10.70		& 11.0	\\
\hline
24	& 12.0		& 12.0		& 11.35		& 12.0	\\
\hline
\end{tabular}
\label{tab:sosd}
\end{table}

\begin{enumerate}
	
\small
	
\item \textit{A1 SOSD D}: The claim can be established with a
line-by-line inspection of the second and fifth columns of
Table \ref{tab:sosd}.

\item \textit{D SOSD B1}: -//- third and fifth columns.

\item \textit{A1 SOSD B1}: -//- second and third columns.

\item \textit{A1 ``nearly'' SOSD B2}:
A1 dominates B2 in this sense
for $x\in[0,22.143]$.

\item \textit{D ``nearly'' SOSD B2}:
D dominates B2 in this sense for
$x\in [0,22.143]$.

\item \textit{B1 \& B2 are SOSD-unranked}: Compare values at $x=12$ vs $x=13$
and at $x=23$ vs $x=24$.

\item \textit{C1 \& C2 are SOSD-unranked}:
The areas under the cumulative density functions
of C1, C2 are half those of B1, B2.
Hence, the claim follows from the test of B1 vs B2 above.

\end{enumerate}

\section{Additional Information on Experimental Procedures}\label{a7}

In all sessions the image describing each lottery was identical: its description
was in English and the rewards were expressed in Euro (Figure \ref{fig:menus}
in O.A. \ref{a3}).
St Andrews subjects were told that the Euro amounts in the lotteries
would be converted to
Pounds Sterling at parity (one-for-one). After the choice part of the experiment
--and before administering the additional questionnaires-- the 107
subjects from Bonn and the 115 subjects from the St Andrews May '23 sessions
were told that they would receive an additional 2 \euro/\pounds to respond
to a few more questions.
This extra payment was first introduced in the Bonn sessions
to bring the total expected hourly payment
of every subject in line with that lab's guidelines for the hourly rate in Euro.
Compared to the '22 St Andrews sessions, those conducted in '23
in both locations also contained the following
technical improvements in the online administration of the experiment:
(i) a fixed data-recording bug which had led to a few missing
choice and questionnaire responses from 14 subjects in the '22 sessions
(we discarded those participants' datasets);
(ii) inclusion of the instructions that were missing from
the 4 cognitive-ability questions
that pertained to 3-dimensional rotation tasks.
The implementation of our design across all sessions
was identical in all other respects.

Upon entering the lab, subjects were asked to keep
their phones switched off and be silent throughout the experiment.
As soon as subjects finished all tasks, their randomly selected menu
showed up on their screens, together with the reminder of the decision
they had made at this menu. As an additional incentive for subjects to make
deliberated and non-rushed decisions, they were told from the beginning
that no participant would be able
to receive their rewards and leave the lab in the first 60 minutes of
the session.
After such time, an experimenter went to the desk of each subject who
had finished,
had their chosen lottery played out for
them using the random-number generating website \url{https://random.org}
(St Andrews)
or using an urn with pieces
of paper numbered from 1 to 1000 (Bonn).
A three-question understanding quiz was administered at
the beginning of all sessions.
Subjects could not proceed until they answered all questions correctly.

\section{Proof of Proposition \ref{prp:StAR}}\label{a8}

Consider an expected-utility maximizer with preferences over lotteries
$\succsim$ that are represented by the Bernoulli utility index $u$ on $Z$.
This may be normalized without loss so that $u(0)\equiv 0$.
Suppose $C(\{\textnormal{A1, D}\})=\{\textnormal{A1}\}$.
This is equivalent to A1 $\succ$ D and
$0.1u(10)>0.05u(20)$. Suppose to the contrary that
$\{\textnormal{B1}\}\in C(\{\textnormal{A1, B1}\})$.
This is equivalent to $B1\succsim A1$ and $0.15u(20)\geq 0.3u(10)$ or,
equivalently, $0.05u(20)\geq 0.1u(10)$. This contradicts the above inequality.
Since, by the expected-utility hypothesis,
$C(\{\textnormal{A1,B1}\})\neq\emptyset$, it follows that
$\{\textnormal{A1}\}=C(\{\textnormal{A1, B1}\})$.
Now suppose to the contrary that
$\textnormal{B1} \in C(\{\textnormal{B1, D}\})$.
This is equivalent to $B1\succsim D$ and $0.1u(20)\geq 0.2u(1)$ or,
equivalently, $0.05u(20)\geq 0.1u(10)$. This, too, is a contradiction.
Thus, \eqref{star-ra} is true.
The argument for \eqref{star-rs} and \eqref{star-rn}
is the same but with the inequalities `$>$', `$\leq$' above
replaced by `$<$', `$\geq$' and by equalities, respectively.
\hfill $\blacksquare$

\section{Analysis of Binary Cycles of Length Greater than Three}\label{a9}

\begin{table}[!htbp]
	\centering
	\footnotesize
	\caption{\centering
		Frequencies of binary cycles at the relevant quadruples
		and quintuple of lottery pairs \newline
		($p$-values from 2-sided Fisher's exact test).}
	\setlength{\tabcolsep}{6pt} 
	\renewcommand{\arraystretch}{1.3} 
	\makebox[\textwidth][c]{
		\begin{NiceTabular}{|l|c|c|c|}
			\hline
			\footnotesize \hspace{45pt} \textbf{Lottery quadruple}
			& A1 \;\; B2 \;\; D\;\; A2
			& A1 \;\; B1 \;\; D\;\; A2
			& A1 \;\; B2 \;\; D\;\; B1\\
			\dashedline
			\multirow{1.8}{*}{\footnotesize \textbf{Possible cyclic-inducing}}
			&\multirow{3.5}{*}{\tiny\shortstack{
					$\{$A1,B2$\}$
					\\
					$\{$B2,D$\}$
					\\
					$\{$A2,D$\}$
					\\
					$\{$A1,A2$\}$
			}}
			&\multirow{3.5}{*}{\tiny\shortstack{
					\hspace{5pt} $\{$A1,A2$\}$
					\hspace{5pt} $\{$A1,D$\}$
					\\
					\hspace{5pt} $\{$A2,D$\}$
					\hspace{5pt} $\{$A1,B1$\}$
					\\
					\hspace{5pt} $\{$B1,D$\}$
					\hspace{5pt} $\{$B1,B2$\}$
					\\
					\hspace{5pt} $\{$A1,B1$\}$
					\hspace{5pt} $\{$B2,D$\}$
			}}	
			&\multirow{3.5}{*}{\tiny\shortstack{
					$\{$A1,B2$\}$
					\hspace{5pt} $\{$A1,B1$\}$
					\hspace{5pt} $\{$A1,D$\}$
					\\
					$\{$B2,D$\}$
					\hspace{5pt} $\{$B1,B2$\}$
					\hspace{5pt} $\{$B1,D$\}$
					\\
					$\{$B1,D$\}$
					\hspace{5pt} $\{$B2,D$\}$
					\hspace{5pt} $\{$B1,B2$\}$
					\\			$\{$A1,B1$\}$
					\hspace{5pt} $\{$A1,D$\}$
					\hspace{5pt} $\{$A1,B2$\}$
			}}
			\\
			\multirow{1.5}{*}{\footnotesize \textbf{structures within quadruple}}
			&
			&
			&\\
			\multirow{1.2}{*}{\footnotesize }
			&
			&
			&\\
			\hline
			\footnotesize \textbf{Binary cycles}
			& 5
			& 48
			& 100\\
			\dashedline
			\multicolumn{4}{|c|}{\hspace{110pt}
				$p<0.001$ \hspace{45pt}
				$p<0.001$}
			\\
			\hline
			\footnotesize \textbf{Normalized binary cycles*}
			& 5
			& 24
			& 33\\
			\dashedline
			\multicolumn{4}{|c|}{\hspace{110pt}
				$p<0.001$ \hspace{45pt}
				$p=0.284$}
			\\
			\hline
			\hline
			\footnotesize \hspace{45pt} \textbf{Lottery quintuple}
			& \multicolumn{3}{|c|}{A1 \;\; A2 \;\; B1\;\; B2\;\; D}\\
			\dashedline
			\multirow{1.3}{*}{\footnotesize \textbf{Possible  cyclic-inducing}}
			&\multicolumn{3}{c}{\tiny $\{$A1,B1$\}$\;\; $\{$B1,B2$\}$\;\; $\{$B2,D$\}$\;\;
				$\{$A2,D$\}$\;\; $\{$A1,A2$\}$} \\
			\multirow{0.8}{*}{\footnotesize \textbf{structures within quintuple}}
			&\multicolumn{3}{c}{\tiny  $\{$A1,A2$\}$\;\; $\{$A2,D$\}$\;\; $\{$B1,D$\}$\;\;
				$\{$B1,B2$\}$\;\; $\{$A1,B2$\}$} \\
			\hline
			\footnotesize \textbf{Binary cycles}
			& \multicolumn{3}{c}{5}\\
			\hline
		\end{NiceTabular}
	}
	\makebox[\textwidth][l]{\scriptsize *Binary cycles divided by the respective
		number of cyclic-inducing structures.}
\end{table}

\section{Consistency in Merged-15 Decisions and Cognitive Ability}\label{a10}

\begin{figure}[!htbp]
	\centering
	\caption{\centering Associations between cognitive ability and choice
		consistency in subjects' merged decisions.\vspace{-10pt}}
	\label{fig:ConsistencyIntelligence}
	\begin{subfigure}[b]{1\textwidth}
		\caption*{\vspace{-5pt}}
		\centering
		\includegraphics[width=0.47\textwidth]{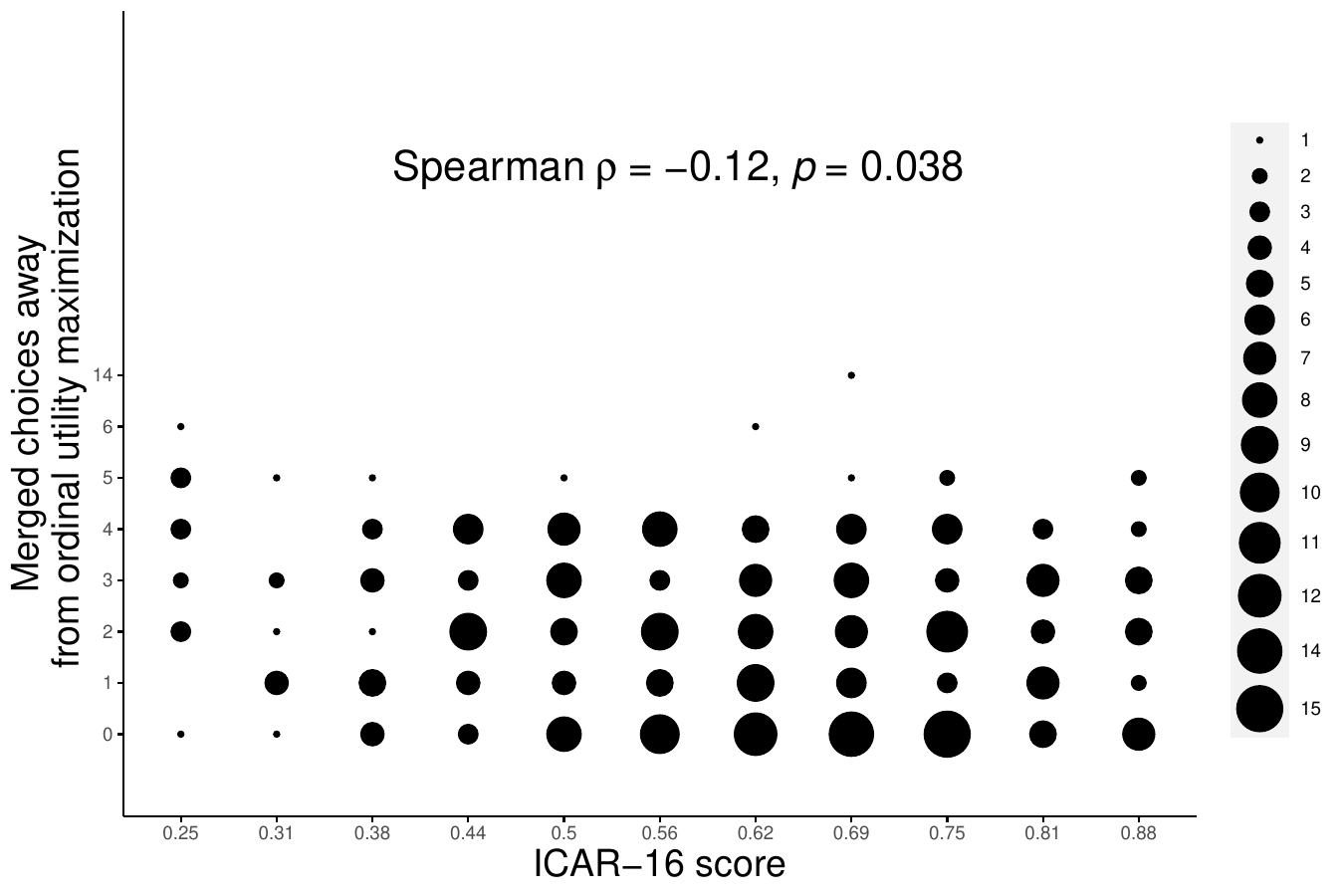}
		\hspace{20pt}
		\includegraphics[width=0.47\textwidth]{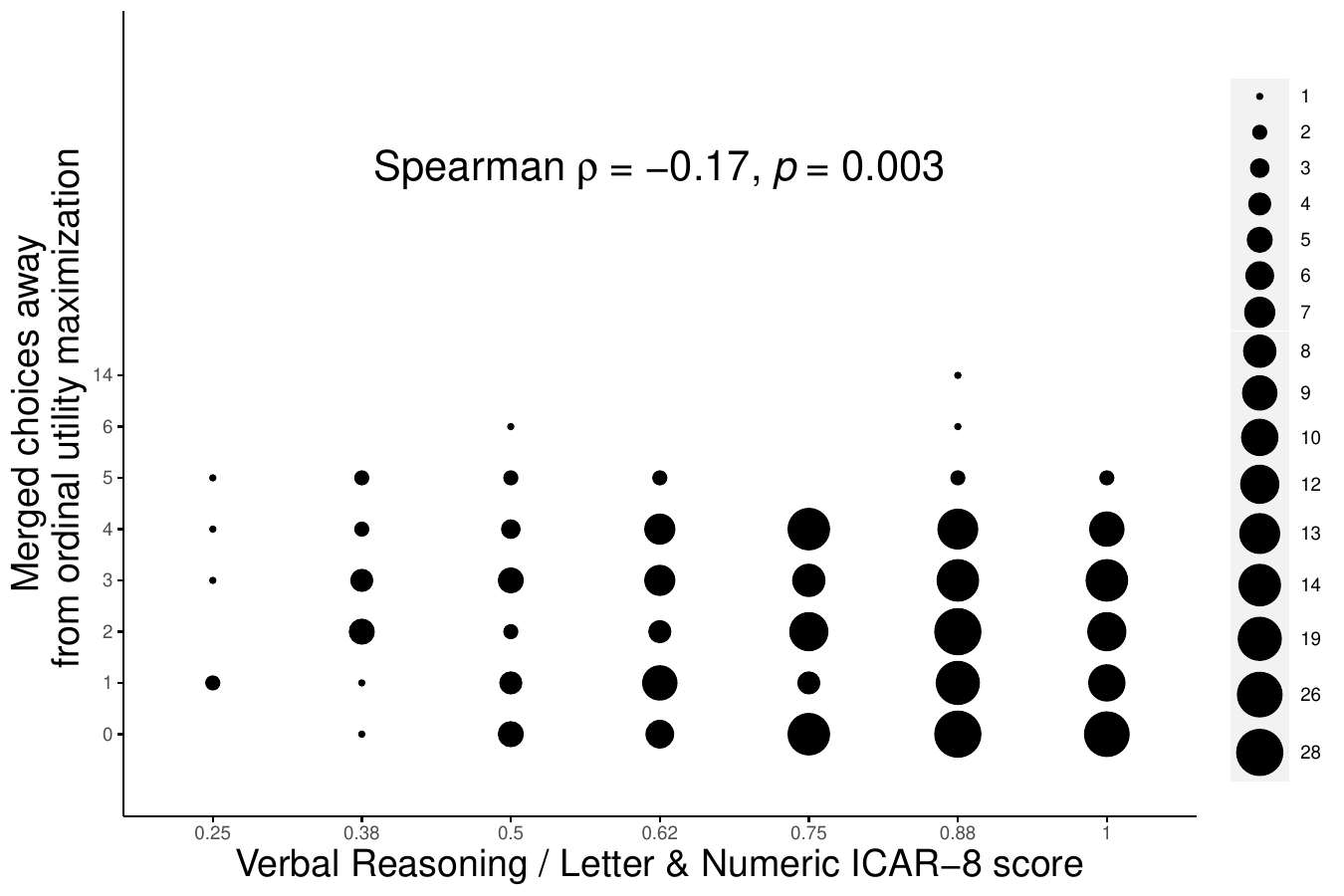}
	\end{subfigure}
	\begin{subfigure}[b]{1\textwidth}
		\caption*{\vspace{-5pt}}
		\centering
		\includegraphics[width=0.47\textwidth]{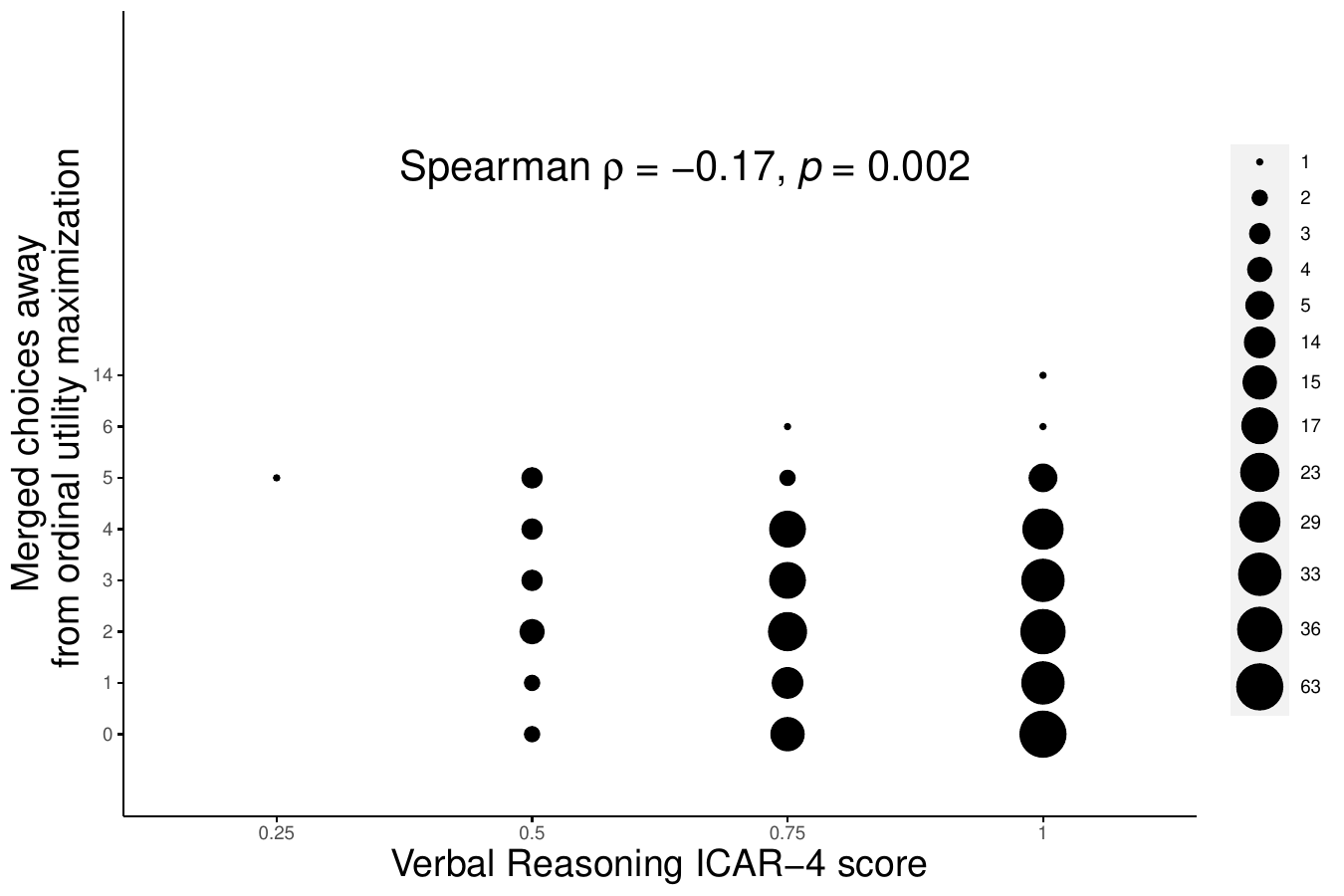}
		\hspace{20pt}
		\includegraphics[width=0.47\textwidth]{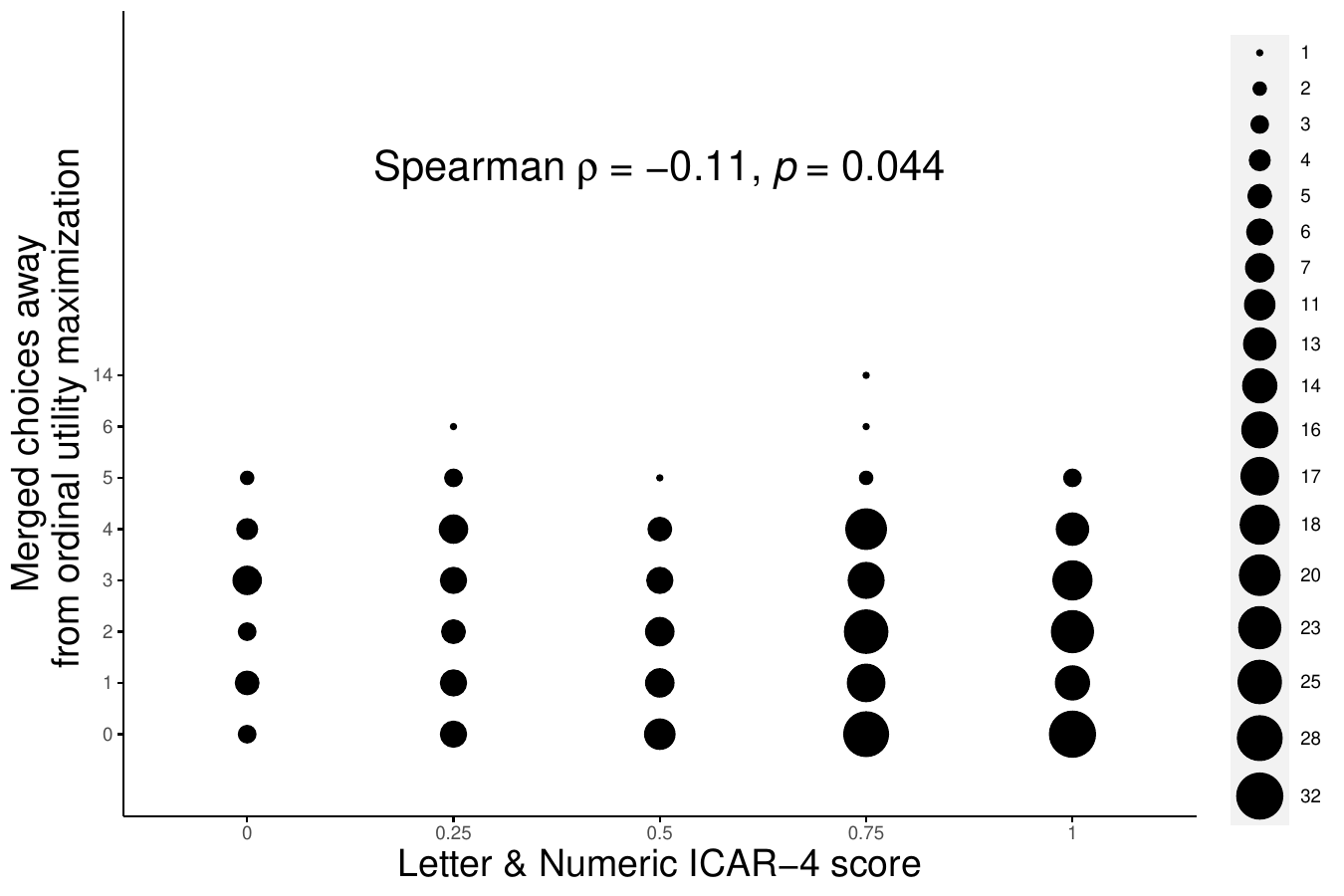}
	\end{subfigure}
	\begin{subfigure}[b]{1\textwidth}
		\caption*{\vspace{-5pt}}
		\centering
		\includegraphics[width=0.47\textwidth]{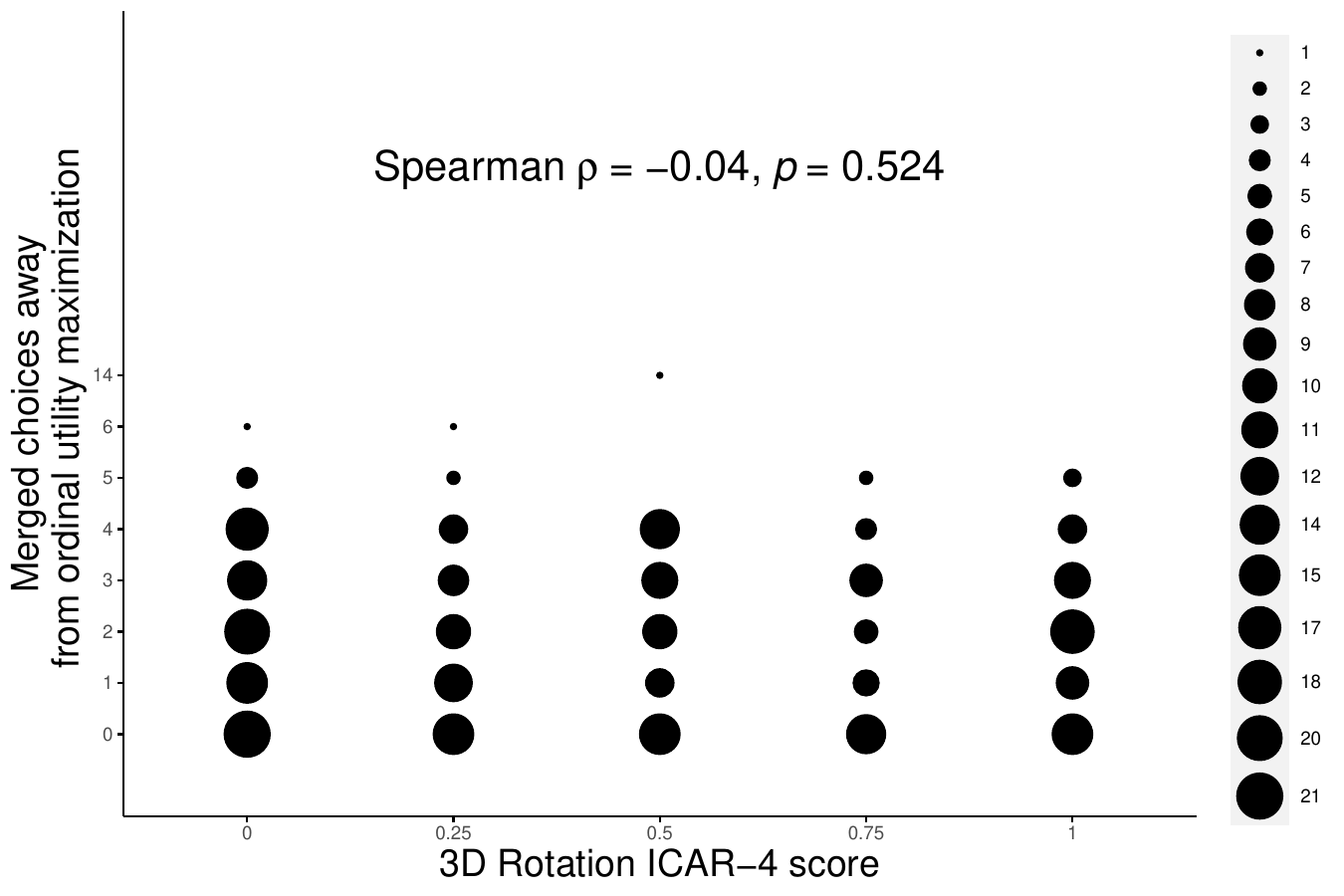}
		\hspace{20pt}
		\includegraphics[width=0.47\textwidth]{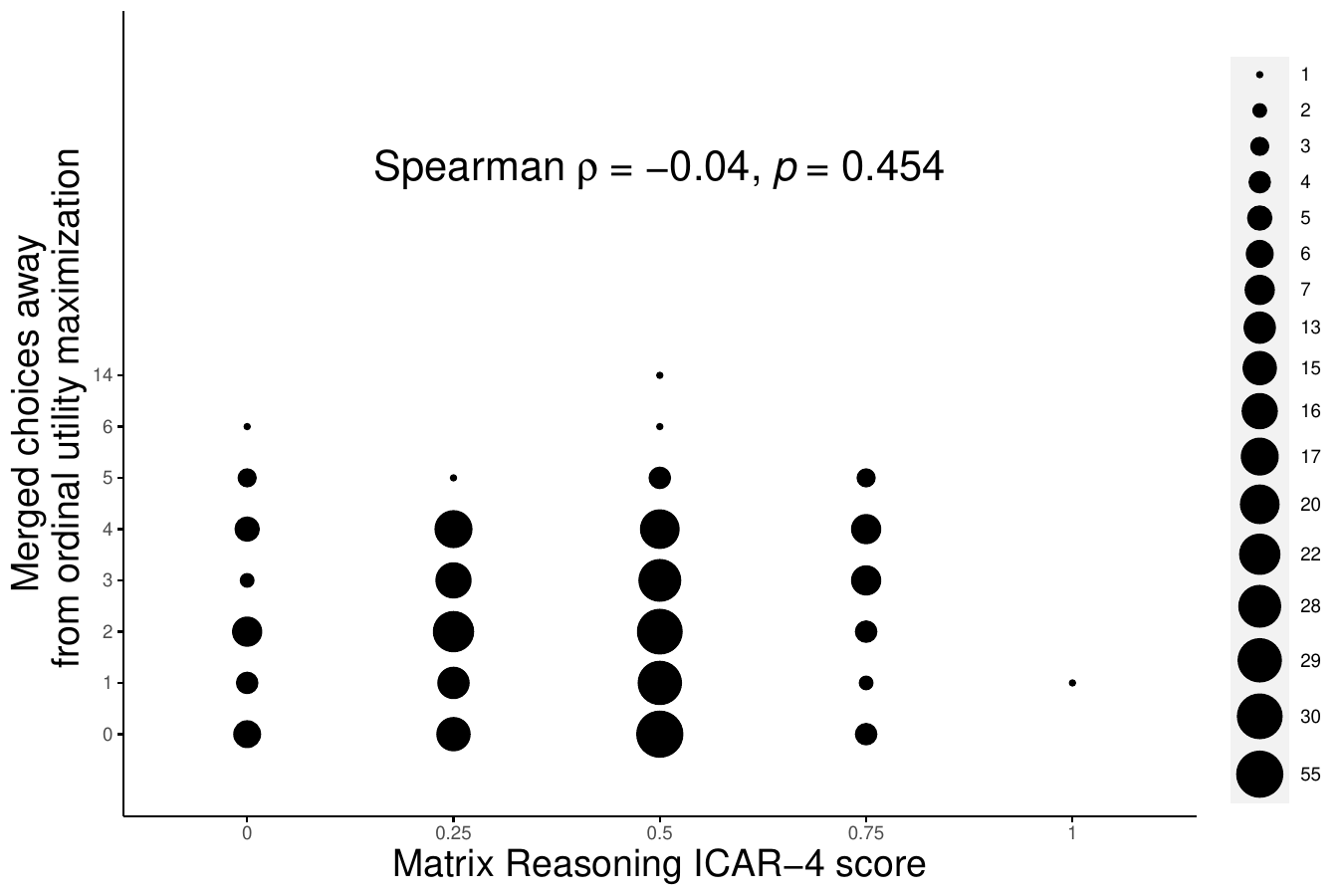}
	\end{subfigure}
	\label{fig:ICARfigures}
\end{figure}

Figure \ref{fig:ConsistencyIntelligence} presents density-inclusive correlograms
between subjects' HM scores in their merged 15 decisions
(where relevant, also penalizing deferrals) and their ICAR-16 scores,
as well as a variety of more theme-focused subscores.

\end{document}